\documentclass[useAMS,usenatbib]{mn2e}
\usepackage{graphicx} 
\usepackage{lscape} 
\usepackage{longtable} 
\usepackage{afterpage} 
\usepackage{multirow} 
\usepackage{definitions}
\begin{document}

\title[Searching for Planets Around White Dwarfs]{The DODO Survey II: A Gemini 
Direct Imaging Search for Substellar and Planetary Mass Companions around 
Nearby Equatorial and Northern Hemisphere White Dwarfs}

\author[E. Hogan et al.]{E. Hogan,\textsuperscript{1,2} M.R. 
Burleigh,\textsuperscript{1} and F.J. Clarke\textsuperscript{3}\\
\textsuperscript{1}Department of Physics and Astronomy, University of
Leicester, University Road, Leicester, LE1 7RH, UK\\ 
\textsuperscript{2}Gemini Observatory, Casilla 603, La Serena, Chile\\ 
\textsuperscript{3}Department of Astrophysics, Denys Wilkinson Building, 
University of Oxford, Keble Road, Oxford, OX1 3RH, UK}

\maketitle

\begin{abstract} 
\label{abstract} 
The aim of the Degenerate Objects around Degenerate Objects (DODO) survey is 
to search for very low mass brown dwarfs and extrasolar planets in wide orbits 
around white dwarfs via direct imaging. The direct detection of such 
companions would allow the spectroscopic investigation of objects with 
temperatures much lower ($<500$~K) than the coolest brown dwarfs currently 
observed. These ultra--low mass substellar objects would have spectral types 
$>$T8.5 and so could belong to the proposed Y dwarf spectral sequence. The 
detection of a planet around a white dwarf would prove that such objects can 
survive the final stages of stellar evolution and place constraints on the 
frequency of planetary systems around their progenitors (with masses between 
$1.5-8\,M_{\odot}$, i.e., early B to mid F). This paper presents the results 
of a multi--epoch $J$ band common proper motion survey of 23 nearby equatorial 
and northern hemisphere white dwarfs. We rule out the presence of any common 
proper motion companions, with limiting masses determined from the 
completeness limit of each observation, to 18 white dwarfs. For the remaining 
five targets, the motion of the white dwarf is not sufficiently separated from 
the non--moving background objects in each field. These targets require 
additional observations to conclusively rule out the presence of any common 
proper motion companions. From our completeness limits, we tentatively suggest 
that $\la5\%$ of white dwarfs have substellar companions with 
$T_{\rm{eff}}\ga500$~K between projected physical separations of $60-200$~AU.
\end{abstract}


\section{Introduction}
\label{introduction}
Directly imaging the extrasolar planets found in orbit around solar type stars 
is difficult as these faint companions are too close to their bright parent 
stars. As this paper was being finalised, \citet{kgc2008} announced the 
discovery of a directly imaged $\sim3\,M_{\rm{Jup}}$ extrasolar planet with a 
projected physical separation of $119$~AU in orbit around the A-type star 
Fomalhaut. On the same day, \citet{mmb2008} announced the discovery of three 
directly imaged companions around the A-type star HR$8799$ with likely masses 
between $5-13\,M_{\rm{Jup}}$ and projected physical separations of $24$, $38$ 
and $68$~AU. However, coronagraphy and adaptive optics were needed to detect 
these faint extrasolar planets. Another, perhaps simpler, solution to the 
problems of contrast and resolution is to instead target intrinsically faint 
stars. For example, many groups are already searching for planetary mass 
companions\footnote{We make the distinction between very low mass brown dwarfs 
and massive extrasolar planets by formation mechanism rather than mass, since 
the mass distributions of these two types of object likely overlap. For 
example, the $9\,M_{\rm{Jup}}$ transiting object HAT-P-2b is too dense to be a 
brown dwarf \citep{bcb2008}, while the free floating objects with masses of 
the order of a few Jupiter masses that have been found in young clusters 
possibly formed in the same manner as stars. Indeed, some authors insist that 
the IAU distinction between these two populations, based on mass alone, has no 
valid foundation \citep{cbl2008}. Therefore, throughout this paper we prefer 
to use the term "planetary mass object" to refer to any body at or below the 
deuterium burning limit ($13-14\,M_{\rm{Jup}}$), since the evolutionary 
history of any companions discovered with these masses is uncertain.} in orbit 
around young, low mass stars and brown dwarfs 
(e.g., \citealt{ctd2003,nga2003}). Any planetary mass companions found in 
orbit around these young stars will have relatively high luminosities, since 
planetary mass objects cool continuously from the moment they form. Famously, 
a $\sim4\pm1\,M_{\rm{Jup}}$ \citep{dtc2008} companion to the 
$\sim25\,M_{\rm{Jup}}$ brown dwarf member of the TW~Hydrae association 
2MASSW~J$1207334-393254$ ($2$M$1207$) was imaged by \citet{cld2004,cld2005}. 
However, \citet{ldc2005} argue that $2$M$1207$Ab more likely formed as a 
binary brown dwarf system, since the core accretion model, thought to be the 
most likely formation mechanism for gas giants like those in the Solar System, 
is unable to account for the formation of $2$M$1207$b.

An alternative approach, rather than looking at the bright, early part of a 
planet's life, is to look at the faint, late part of a star's life. White 
dwarfs are intrinsically faint stars and can be up to $10,000$ times less 
luminous than their main sequence progenitors, significantly enhancing the 
contrast between any companion and the white dwarf. In addition, any companion 
that avoids direct contact with the red giant envelope as the main sequence 
progenitor evolves into a white dwarf will migrate outwards as mass is lost 
from the central star by a maximum factor of $M_{\rm{MS}}/M_{\rm{WD}}$ 
\citep{j1924}. This increases the projected physical separation between the 
companion and the white dwarf, substantially increasing the probability of 
obtaining a ground based direct image of a planetary mass companion. The 
evolution of planetary systems during the post--main sequence phase is 
discussed in more detail by \citet{dl1998}, \citet*{bch2002}, \citet{ds2002} 
and \citet{vl2007}.

The direct detection of a planetary mass companion in orbit around a white 
dwarf would allow the spectroscopic investigation of very low mass objects 
much cooler ($<500$~K) than previously found. The coolest known brown dwarfs, 
ULAS~J$003402.77-005206.7$ \citep{wml2007} and CFBDS~J$005910.90-011401.3$ 
\citep{dda2008}, have effective temperatures of $600<T_{\rm{eff}}<700$~K and 
spectral types of T8.5. The letter Y has been suggested for the next, cooler, 
spectral type \citep{k2005}. Any planetary mass companions directly detectable 
around old ($>2$~Gyr) white dwarfs could well belong to this class, regardless 
of formation mechanism \citep{zs2008}. Such a discovery would help provide 
constraints on models for the evolution of planets and planetary systems 
during the final stages of stellar evolution. In addition, the age of any 
substellar and planetary mass companions discovered in such a system can be 
estimated using the white dwarf cooling age and the mass and the lifetime of 
the main sequence progenitor, providing model-free benchmark estimates of 
their mass and luminosity, which could be used to test evolutionary models 
\citep{pjl2006}. Radial velocity searches have concentrated mainly on stars 
with spectral types between mid F and M, since the faster rotation and 
increased activity of early B, A and mid F type stars broadens the low number 
of absorption lines in their spectra. As a result, it is difficult to 
accurately measure the Doppler shift of stars with these earlier spectral 
types. However, new methods in manipulating the measurements acquired when 
using the radial velocity technique has allowed planetary mass companions to 
be found around stars with spectral types of A and F (e.g., 
\citealt{glu2005}). Nevertheless, as the $1.5-8\,M_{\odot}$ progenitor stars 
of white dwarfs have spectral types of early B, A and mid F, searching for 
planetary mass companions in orbit around white dwarfs allows the examination 
of a currently inadequately explored region of parameter space, supplying new 
information on the frequency and mass distribution of extrasolar planets 
around intermediate mass main sequence stars. 

A number of extrasolar planets have been discovered around evolved giant stars 
using the radial velocity technique, e.g., HD~$11977$ (G5~III; 
\citealt{srd2005}), HD~$13189$ (K2~II; \citealt{hge2005}) and $\beta$~Gem 
(K0~III; \citealt{hce2006,rqm2006}). These stars have entered the red giant 
phase of stellar evolution, proving that planets can survive the early stages 
of the RGB phase. Evolved giant stars are significantly more massive than the 
Sun, so their progenitors were likely to be A or B type stars (see Table~6 of 
\citealt{hce2006}). The companions all have masses significantly larger than 
Jupiter, implying that significantly more massive planets are formed around 
these intermediate mass stars than around lower mass stars \citep{lm2007}. In 
fact, both \citet{lm2007} and \citet{jbm2007} suggest that intermediate mass 
stars are more likely to host extrasolar planets of all masses than solar mass 
stars. 

Up to three white dwarfs are known to be wide companions to stars hosting 
extrasolar planets. The bright ($V=11$~mag), well studied white dwarf 
WD~$1620-391$ (CD$-38^{\circ}10980$) was known to be a common proper motion 
companion to the solar type star HD~$147513$ \citep{w1973} before a planet, 
with a minimum mass, $M_{\rm{p}}\,{\rm{sin}}\,i=1.21\,M_{\rm{Jup}}$ and an 
orbital radius of $1.32$~AU, was discovered in orbit around the latter 
\citep{mun2004}. Since WD~$1620-391$ and the parent star are separated by 
$\sim5360$~AU, it is highly unlikely that the evolution of the main sequence 
progenitor of the white dwarf affected the mass or the orbit of HD~$147513$b. 
It has been suggested that the faint companions to two other planet hosting 
stars are also white dwarfs. The planet in orbit around Gliese~$86$ has a 
minimum mass, $M_{\rm{p}}\,{\rm{sin}}\,i=4.01\,M_{\rm{Jup}}$ and an orbital 
period of $15.8$~days \citep{qmw2000}, while the likely white dwarf companion 
has a mass between $0.48\leq M\leq0.62\,M_{\odot}$ and an orbital radius of 
$\sim20$~AU \citep{mn2005,lbu2006}. The possible effects of the evolution of 
the main sequence progenitor on the mass and the orbit of the planet are 
discussed by \citet{db2007}. The planet in orbit around HD~$27442$ 
($\epsilon$~Ret) has a minimum mass, 
$M_{\rm{p}}\,{\rm{sin}}\,i=1.28\,M_{\rm{Jup}}$ and an orbital radius of 
$1.18$~AU \citep{btm2001}. The white dwarf companion \citep{rhm2006,clu2006}, 
currently separated from HD~$27442$ by $\sim240$~AU, was recently confirmed 
from an analysis of its optical and infrared (IR) spectrum 
\citep{mnm2007b,clu2007}. 

The discovery of metal rich dust disks in close orbits around white dwarfs may 
indicate the existence of old, rocky planetary systems, suggesting that even 
terrestrial planets and asteroids can survive the final stages of stellar 
evolution. The first dust disk was discovered around the DAZ white dwarf 
G~$29-38$ (WD~$2326+049$) from the identification of a large IR excess in its 
spectrum at wavelengths between $2-5\mu$m \citep{zb1987b}. This IR excess was 
initially attributed to a spatially unresolved, $T_{\rm{eff}}=1200\pm200$~K 
brown dwarf companion to the white dwarf. However, subsequent measurements 
showed that the blackbody--like IR excess was due to a dust disk rather than a 
brown dwarf \citep*{tbz1990}. A mid--IR (MIR) spectrum of G~$29-38$, obtained 
by the Spitzer Space Telescope, shows a strong emission feature in the 
spectrum between $9-11\mu$m, which indicates the presence of silicates 
(SiO$_{4}$) in the dust disk \citep{rkv2005}. Eight additional white dwarfs 
are now known to have dust disks in orbit around them: GD~$362$ 
($T_{\rm{eff}}=9740$~K; \citealt{bfj2005,kvl2005}), GD~$56$ 
($T_{\rm{eff}}=14400$~K; \citealt{kvl2006}), WD~$1150-153$ 
($T_{\rm{eff}}=12800$~K; \citealt{kr2007}), WD~$2115-560$ 
($T_{\rm{eff}}=9700$~K; \citealt{vkk2007}), GD~$40$ ($T_{\rm{eff}}=15200$~K; 
\citealt*{jfz2007}), GD~$133$ ($T_{\rm{eff}}=12200$~K; \citealt{jfz2007}), 
PG~$1015+161$ ($T_{\rm{eff}}=19300$~K; \citealt{jfz2007}) and G~$166-58$ 
($T_{\rm{eff}}=7390$~K; \citealt{fzb2008}). The generally accepted origin of 
the material in these dust disks is from the tidal disruption of an asteroid 
that had strayed within the Roche lobe radius of the white dwarf after its 
orbital radius was altered during the AGB phase of stellar evolution 
\citep{gmn1990,ds2002,j2003}. In addition to these dust disks, there have been 
metal rich gas disks found in orbit around 2 hotter DAZ white dwarfs; 
SDSS~J$122859.93+104032.0$ ($T_{\rm{eff}}=22292$~K; \citealt{gms2006}) and 
SDSS~J$104341.53+085558.2$ ($T_{\rm{eff}}=18330$~K; \citealt*{gms2007}). These 
gas disks could also indicate the presence of old planetary systems, since it 
is likely that the hot white dwarfs caused the dust in the disk to sublimate. 
The fraction of known single DAZ white dwarfs with IR excesses, which can be 
attributed to a dust disk, is $14\%$ \citep{kvl2006}. In addition, 
\citet{j2006} argues that $>7\%$ of white dwarfs possess asteroid belts 
similar to that of the Solar System, and by implication, remnant planetary 
systems.

The first search for low mass substellar companions to white dwarfs was 
conducted by \citet{p1983}, who measured the IR magnitudes of $\sim100$ white 
dwarfs to determine whether any excess emission was present. No companions 
were found during this survey. Subsequently, a number of groups unsuccessfully 
attempted to detect substellar companions to white dwarfs using the same 
method (e.g.,~\citealt{s1986,zb1987a}). The first confirmed subtellar 
companion to a white dwarf was discovered in 1988 around the DA white dwarf 
GD~$165$ \citep{bz1988}. Over 15 years later, a second brown dwarf companion 
was found in orbit around GD~$1400$ \citep{fc2004,dbl2005}. More recently, 
radial velocity measurements revealed a brown dwarf companion in a close 
($\sim116$~minutes) orbit around WD~$0137-349$ \citep{mnd2006}, while its 
spectral type (L8) was later determined from near--IR (NIR) spectroscopy 
\citep{bhd2006}. The L dwarf companion fraction, determined from a wide field, 
proper motion, NIR search for wide substellar companions to 347 white dwarfs, 
is estimated to be $<0.5\%$ \citep{fbz2005}.

The recent discovery of an extrasolar planet around the post--red giant star 
V~$391$~Pegasi proves that planetary mass companions with an initial orbital 
radius outside the maximum radius of the red giant envelope can survive the 
RGB phase of stellar evolution. The planet was discovered from the periodic 
variation in the precise timing measurements of V~$391$~Pegasi's extremely 
stable, short period pulsations \citep{ssj2007}. It has a minimum mass, 
$M_{\rm{p}}\,{\rm{sin}}\,i\sim3.2\,M_{\rm{Jup}}$, an orbital radius of 
$\sim1.7$~AU and an estimated age of $\sim10$~Gyr. Strong evidence for the 
existence of a planetary mass companion to the DAV white dwarf GD~$66$ has 
been recently found from the periodic variation in the precise timing 
measurements of GD~$66$'s extremely stable non--radial pulsations 
\citep{mwd2008}. While current measurements suggest that this companion has a 
minimum mass, $M_{\rm{p}}\,{\rm{sin}}\,i\sim2.11\,M_{\rm{Jup}}$ and an orbital 
radius of $\sim2.356$~AU, further observations, to cover the entire orbit of 
the companion, are now required. 

In \citet{bch2008} we reported limits on planetary mass companions to the 
nearest single white dwarf, vMa~$2$. Preliminary results and progress reports 
from the Degenerate Objects around Degenerate Objects (DODO) survey have been 
published previously \citep*{cb2004,bhc2006,hbc2007}. In this paper we report 
further extensive results of the DODO survey; a NIR direct imaging search for 
substellar and planetary mass common proper motion companions in wide orbits 
around nearby white dwarfs. 

\section{Target Selection}
\label{targetselection}
The ability to directly image an extrasolar planet in orbit around a white 
dwarf will depend on the apparent magnitude of the planet, which in turn 
depends on its absolute magnitude and distance from the Earth. The absolute 
magnitude of the planet is determined from its intrinsic luminosity, which is 
dependant primarily on the age and the mass of the planet, since it will cool 
continuously from the moment it formed. The age of an extrasolar planet found 
in orbit around a white dwarf equals the sum of the main sequence progenitor 
lifetime and the white dwarf cooling age. Using this age and the distance to 
the white dwarf, \citet{bch2002} used evolutionary models for cool brown 
dwarfs and extrasolar planets \citep{bmh1997} to make initial predictions of 
the IR magnitudes of planetary mass companions around white dwarfs. The 
results published in this paper use the more recent ``COND'' evolutionary 
models of \citet{bcb2003}. These models assume irradiation effects from the 
parent star on the planet are negligible and predict that a $5\,M_{\rm{Jup}}$ 
planet with an age of $\sim2$~Gyr will have an apparent magnitude of 
$J\sim24$~mag at $10$~pc. This magnitude is comparable with the expected 
sensitivity of a one hour exposure acquired using an $8$m telescope. Brighter 
companions such as massive brown dwarfs and M dwarfs should easily be 
detected. Indeed, these more luminous objects would have already been detected 
in previous IR studies (e.g., \citealt{fbz2005}) and in 2MASS data.

Lower mass planets could be more easily detected around younger white dwarfs, 
since these companions would be brighter than their older counterparts. In 
addition, fainter, and therefore lower mass, companions could be more easily 
detected around nearby white dwarfs compared to more distant targets. Nearby 
white dwarfs are also more likely to have large proper motions, which require 
a smaller baseline between the observations of the two epochs. An initial 
sample of $\sim40$ targets, with total ages (main sequence progenitor lifetime 
plus the white dwarf cooling age) $<4$~Gyr and distances within $\sim20$~pc, 
were selected from the catalogue of white dwarfs within 20pc \citet{hos2002}. 
Of these targets, 23 equatorial and northern hemisphere white dwarfs 
(Table~\ref{parameters}) are presented in this paper. The remaining southern 
hemisphere white dwarfs will be presented in an upcoming paper. 
\begin{table*}
\caption{Parameters of the 23 equatorial and northern hemisphere white dwarfs in the DODO survey\label{parameters}}
\begin{center}
\begin{tabular}{clcrrrrcllcccr}
\hline
\hline
\multicolumn{1}{c}{White} & \multicolumn{1}{c}{Name} & \multicolumn{1}{c}{Sp.} & \multicolumn{1}{c}{$\mu$} & \multicolumn{1}{c}{$\theta$} & \multicolumn{1}{c}{d} & \multicolumn{1}{c}{$T_{\rm{eff}}$} & \multicolumn{1}{c}{log $g$} & \multicolumn{1}{c}{$M_{\rm{WD}}$} & \multicolumn{1}{c}{$t_{\rm{WD}}$} & \multicolumn{1}{c}{$M_{\rm{MS}}$} & \multicolumn{1}{c}{$t_{\rm{MS}}$} & \multicolumn{1}{c}{$t_{\rm{tot}}$} & R\\
\multicolumn{1}{c}{Dwarf} & & \multicolumn{1}{c}{Class} & \multicolumn{1}{c}{[m$^{\prime\prime}$/yr]} & \multicolumn{1}{c}{[m$^{\prime\prime}$/yr]} & \multicolumn{1}{c}{[pc]} & \multicolumn{1}{c}{[K]} & & \multicolumn{1}{c}{[$M_{\odot}$]} & \multicolumn{1}{c}{[Gyr]} & \multicolumn{1}{c}{[$M_{\odot}$]} & \multicolumn{1}{c}{[Gyr]} & \multicolumn{1}{c}{[Gyr]} &\\
\hline
0115$\,+\,$159 & LHS 1227 & DQ & --20$^{1}$ & --647$^{1}$ & 15.41$^{2}$ & 9050 & 8.19 & 0.69 & 1.02$^{*}$ & 3.0 & 0.63 & 1.7 & 3\\
0148$\,+\,$467 & GD 279 & DA & 1$^{4}$ & 124$^{4}$ & 15.85$^{4}$ & 13990 & 7.89 & 0.53 & 0.21$^{*}$ & 1.8 & 2.26 & 2.5 & 5\\
0208$\,+\,$396 & G 74--7 & DAZ & 1031$^{1}$ & --497$^{1}$ & 16.72$^{2}$ & 7310 & 8.01 & 0.60 & 1.38 & 2.3 & 1.20 & 2.6 & 6\\
0341$\,+\,$182 & Wolf 219 & DQ & 415$^{1}$ & --1125$^{1}$ & 19.01$^{2}$ & 6510 & 7.99 & 0.57 & 1.79$^{*}$ & 2.1 & 1.54 & 3.3 & 3\\
0435$\,-\,$088 & L 879--14 & DQ & 243$^{1}$ & --1555$^{1}$ & 9.51$^{2}$ & 6300 & 7.93 & 0.53 & 1.79$^{*}$ & 1.8 & 2.26 & 4.1 & 3\\
0644$\,+\,$375 & G 87--7 & DA & --226$^{4}$ & --936$^{4}$ & 15.41$^{4}$ & 21060 & 8.10 & 0.54$^{7}$ & 0.07$^{*}$ & 1.9 & 2.04 & 2.1 & 5\\
0738$\,-\,$172 & L 745--46A & DZ & 1147$^{1}$ & --538$^{1}$ & 8.90$^{2}$ & 7710 & 8.09 & 0.63 & 1.45 & 2.6 & 0.95 & 2.4 & 6\\
0912$\,+\,$536 & G 195--19 & DCP$^a$ & --1086$^{1}$ & --1125$^{1}$ & 10.28$^{2}$ & 7160 & 8.28 & 0.75 & 2.54 & 3.5 & 0.45 & 3.0 & 6\\
1055$\,-\,$072 & LHS 2333 & DC & --822$^{1}$ & 91$^{1}$ & 12.15$^{2}$ & 7420 & 8.42 & 0.85 & 3.01 & 4.2 & 0.27 & 3.3 & 6\\
1121$\,+\,$216 & Ross 627 & DA & --1040$^{1}$ & --14$^{1}$ & 13.44$^{2}$ & 7490 & 8.20 & 0.72 & 1.76 & 3.2 & 0.53 & 2.3 & 6\\
1134$\,+\,$300 & GD 140 & DA & --147$^{4}$ & --6$^{4}$ & 15.32$^{4}$ & 21280 & 8.55 & 0.96 & 0.20 & 5.0 & 0.17 & 0.37 & 8\\
1344$\,+\,$106 & LHS 2800 & DAZ & --871$^{9}$ & --181$^{9}$ & 20.04$^{2}$ & 7110 & 8.10 & 0.65 & 1.67 & 2.7 & 0.82 & 2.5 & 6\\
1609$\,+\,$135 & LHS 3163 & DA & 14$^{1}$ & --551$^{1}$ & 18.35$^{2}$ & 9080 & 8.75 & 1.07 & 2.71 & 5.9 & 0.12 & 2.8 & 6\\
1626$\,+\,$368 & Ross 640 & DZ & --469$^{9}$ & 709$^{9}$ & 15.95$^{2}$ & 8640 & 8.03 & 0.60 & 1.02 & 2.3 & 1.20 & 2.2 & 6\\
1633$\,+\,$433 & G 180--63 & DAZ & 218$^{1}$ & --302$^{1}$ & 15.11$^{2}$ & 6650 & 8.14 & 0.68 & 2.28 & 2.9 & 0.67 & 3.0 & 6\\
1647$\,+\,$591 & G 226--29 & DAV & 139$^{4}$ & --292$^{4}$ & 10.97$^{4}$ & 12260 & 8.31 & 0.80 & 0.56$^{*}$ & 3.8 & 0.35 & 0.91 & 10\\
1900$\,+\,$705 & G 260--15 & DAP$^b$ & 105$^{9}$ & 479$^{9}$ & 12.99$^{2}$ & 12070 & 8.58 & 0.95 & 0.94 & 5.0 & 0.18 & 1.1 & 6\\
1953$\,-\,$011 & G 92--40 & DAP$^c$ & --442$^{1}$ & --699$^{1}$ & 11.39$^{2}$ & 7920 & 8.23 & 0.74 & 1.63 & 3.4 & 0.47 & 2.1 & 6\\
2007$\,-\,$219 & LTT 7983 & DA & 109$^{1}$ & --313$^{1}$ & 18.22$^{11}$ & 9887 & 8.14 & 0.69 & 0.76$^{*}$ & 3.0 & 0.63 & 1.4 & 12\\
2047$\,+\,$372 & G 210--36 & DA & 160$^{1}$ & 149$^{1}$ & 18.16$^{11}$ & 14630 & 8.13 & 0.69 & 0.26$^{*}$ & 3.0 & 0.63 & 0.89 & 13\\
2140$\,+\,$207 & LHS 3703 & DQ & --207$^{9}$ & --658$^{9}$ & 12.52$^{2}$ & 8200 & 7.84 & 0.49 & 0.82$^{*}$ & 1.5 & 3.56 & 4.4 & 3\\
2246$\,+\,$223 & G 67--23 & DA & 580$^{9}$ & 13$^{9}$ & 19.05$^{2}$ & 10330 & 8.57 & 0.97 & 1.56 & 5.1 & 0.17 & 1.7 & 6\\
2326$\,+\,$049 & G 29--38 & DAZ$^d$ & --360$^{9}$ & --302$^{9}$ & 13.62$^{2}$ & 11820 & 8.15 & 0.70 & 0.55 & 3.1 & 0.60 & 1.1 & 8\\
\hline
\end{tabular}
\begin{tabular}{p{0.95\textwidth}}
Columns: $\mu$ and $\theta$ are the R.A. and Dec components of the proper motion of the white dwarf, respectively, measured in milli arc seconds per year; $d$ is the distance to the white dwarf, measured in parsecs; $T_{\rm{eff}}$ is the effective temperature of the white dwarf, measured in Kelvin; log~$g$ is the log of the gravity of the white dwarf; $M_{\rm{WD}}$ is the mass of the white dwarf, measured in solar masses; $t_{\rm{WD}}$ is the cooling age of the white dwarf, measured in gigayears; $M_{\rm{MS}}$ is the mass of the main sequence progenitor, measured in solar masses, and is calculated using the IFMR of \citealt{dnb2006}; $t_{\rm{MS}}$ is the main sequence lifetime, measured in gigayears, and is calculated using Equation~\ref{mslifetime} \citep{w1992}; $t_{\rm{tot}}$ is the total age of the white dwarf, measured in gigayears.\\
R = References, which refer to the $T_{\rm{eff}}$, log~$g$, $M_{\rm{WD}}$ and $t_{\rm{WD}}$ columns. (1) \citet{sg2003}, (2) \citet*{vlh1995}, (3) \citet*{dbf2005}, (4) \citet{plk1997}, (5) \citet*{bsl1992}, (6) \citet*{blr2001}, (7) \citet*{fbb2007}, (8) \citet*{lbh2005}, (9) \citet*{bsn2002}, (10) \citet*{gbf2005}, (11) \citet*{hos2002}, (12) \citet{knc2001}, (13) \citet{gkk1998}, $^a$Magnetic field strength, $B=100$~MG; rotation period, $P=1.3$~days \citep{wf2000}, $^b$Magnetic field strength, $B=320$~MG \citep{wf2000}, $^c$Magnetic field strength, $B=70$~kG; rotation period, $P=1.4418$~days. WD~$1953-011$ is also photometrically variable at the $\sim2$\% level, an effect which is believed to be caused by a star spot \citep{bmm2005}, $^d$A dust disk is known to orbit this ZZ ceti white dwarf, $^{*}$The white dwarf cooling age was calculated using models from \citet*{fbb2001}.
\end{tabular}
\end{center}
\end{table*}

The cooling age of a white dwarf, $t_{\rm{WD}}$, can be calculated using 
evolutionary models. When the cooling age was unavailable in the literature, 
models from \citet*{fbb2001}, which use $T_{\rm{eff}}$ and log~$g$ values to 
calculate the cooling age, were used to estimate this value. The initial final 
mass relation (IFMR) determined by \citet{dnb2006}, based on the measurements 
of a small number of white dwarfs found in young open clusters, was used to 
determine the mass of the main sequence progenitor, $M_{\rm{MS}}$, from the 
mass of the white dwarf, $M_{\rm{WD}}$. This linear IFMR is given as 
\begin{equation} 
M_{\rm{WD}}=0.133\,M_{\rm{MS}}+0.289 
\label{ifmr} 
\end{equation} 
Recent observations of white dwarfs in older open clusters have placed 
constraints on the low mass end of the IFMR, suggesting that this equation 
is valid down to white dwarf masses of $0.54\,M_{\odot}$ \citep{khk2008}. The 
main sequence progenitor lifetime, $t_{\rm{MS}}$, is estimated using the 
equation 
\begin{equation}
t_{\rm{MS}}=10\,\left(\frac{M_{\rm{MS}}}{M_{\odot}}\right)^{-2.5}
\label{mslifetime}
\end{equation}
where $t_{\rm{MS}}$ is measured in Gyr \citep{w1992}.

\section{Observations}
\label{observations}
Observations of the 23 equatorial and northern hemisphere white dwarfs 
(Table~\ref{observed}) were acquired in the $J$ band primarily using Gemini 
North and \textit{NIRI} between 2003 and 2005, while a small number of 
observations of equatorial targets were acquired in 2002, using Gemini South 
and \textit{FLAMINGOS}. \textit{NIRI} consists of a $1024\times1024$~pixel 
ALADDIN--II array. When combined with the f/6 camera, \textit{NIRI} supplies a 
pixel scale of $0.117^{\prime\prime}\,\rm{pixel}^{-1}$ and a wide field of 
view of $120^{\prime\prime}\times120^{\prime\prime}$. \textit{FLAMINGOS} 
consists of a $2048\times2048$~pixel HAWAII--II array. When combined with the 
f/16 camera, \textit{FLAMINGOS} supplies a pixel scale of 
$0.078^{\prime\prime}\,\rm{pixel}^{-1}$ and a wide field of view of 
$160^{\prime\prime}\times160^{\prime\prime}$. \begin{table}
\begin{center}
\caption{Details of the observations of the 23 equatorial and northern hemisphere white dwarfs in the DODO survey\label{observed}}
\begin{tabular}{cccccc}
\hline
\hline
White Dwarf & Date & ET & FWHM & TW &\\ 
Number & Observed & [m] & [$^{\prime\prime}$] & &\\
\hline
0115$\,+\,$159 & 2003-08-17 & 20 & 0.65 & GN+N & \\
               & 2004-08-25 & 55 & 0.73 & GN+N & \\
0148$\,+\,$467 & 2003-08-09 & 54 & 0.57 & GN+N & 2\\
               & 2005-08-28 & 75 & 0.48 & GN+N & \\
0208$\,+\,$396 & 2004-08-25 & 55 & 0.57 & GN+N & \\
               & 2005-08-29 & 75 & 0.66 & GN+N & \\
0341$\,+\,$182 & 2004-12-21 & 55 & 0.62 & GN+N & \\
               & 2005-09-04 & 74 & 0.58 & GN+N & \\
0435$\,-\,$088 & 2004-12-24 & 39 & 0.55 & GN+N & \\
               & 2005-11-13 & 75 & 0.58 & GN+N & \\
0644$\,+\,$375 & 2003-11-03 & 52 & 0.59 & GN+N & 1\\
               & 2004-11-02 & 54 & 0.65 & GN+N & \\
0738$\,-\,$172 & 2004-12-24 & 53 & 0.56 & GN+N & \\
               & 2005-11-13 & 75 & 0.71 & GN+N & \\
0912$\,+\,$536 & 2003-03-22 & 54 & 0.63 & GN+N & 1\\
               & 2004-02-09 & 37 & 0.64 & GN+N & \\
               & 2005-01-28 & 27 & 0.56 & GN+N & \\
1055$\,-\,$072 & 2004-12-24 & 49 & 0.58 & GN+N & \\
               & 2005-11-16 & 63 & 0.77 & GN+N & \\
1121$\,+\,$216 & 2003-03-24 & 54 & 0.63 & GN+N & 1\\
               & 2004-02-04 & 52 & 0.77 & GN+N & 2\\
               & 2005-01-20 & 19 & 0.48 & GN+N & 1\\
1134$\,+\,$300 & 2003-03-22 & 40 & 0.73 & GN+N & 1\\
               & 2005-02-21 & 52 & 0.66 & GN+N & \\
1344$\,+\,$106 & 2003-03-22 & 54 & 0.62 & GN+N & 1\\
               & 2004-02-03 & 54 & 0.68 & GN+N & 2\\
1609$\,+\,$135 & 2003-06-10 & 53 & 0.44 & GN+N & 1\\
               & 2004-02-11 & 51 & 0.53 & GN+N & 2\\
1626$\,+\,$368 & 2003-06-10 & 46 & 0.43 & GN+N & 1\\
               & 2004-04-01 & 47 & 0.50 & GN+N & \\
1633$\,+\,$433 & 2003-06-09 & 53 & 0.61 & GN+N & 1\\
               & 2004-04-01 & 54 & 0.52 & GN+N & 3\\
1647$\,+\,$591 & 2003-05-17 & 54 & 0.58 & GN+N & \\
               & 2004-02-12 & 40 & 0.57 & GN+N & \\
               & 2005-02-20 & 54 & 0.81 & GN+N & \\
1900$\,+\,$705 & 2003-05-17 & 54 & 0.58 & GN+N & 2\\
               & 2004-04-06 & 54 & 0.54 & GN+N & \\
1953$\,-\,$011 & 2002-06-23 & 127.5 & 0.53 & GS+F & \\
               & 2003-08-10 & 51 & 0.52 & GN+N & 2\\
2007$\,-\,$219 & 2002-06-21 & 136.5 & 0.47 & GS+F & \\
               & 2003-08-15 & 53 & 0.53 & GN+N & \\
               & 2008-05-20 & 78 & 0.53 & GN+N & \\
2047$\,+\,$372 & 2002-06-16 & 51 & 0.53 & GN+N & \\
               & 2003-08-10 & 54 & 0.51 & GN+N & 2\\
               & 2004-06-06 & 54 & 0.64 & GN+N & \\
2140$\,+\,$207 & 2003-08-11 & 52 & 0.53 & GN+N & 2\\
               & 2004-06-07 & 47 & 0.80 & GN+N & \\
2246$\,+\,$223 & 2003-08-09 & 57 & 0.43 & GN+N & 2\\
               & 2004-06-07 & 53 & 0.70 & GN+N & \\
2326$\,+\,$049 & 2002-06-23 & 148 & 0.58 & GS+F & \\
               & 2003-08-09 & 54 & 0.49 & GN+N & 2\\
\hline
\end{tabular}
\begin{tabular}{p{0.45\textwidth}}
Columns: ET is the total exposure time, measured in minutes; FWHM is the full--width at half--maximum, calculated as the average FWHM of stars in the field using SExtractor, measured in arc seconds; TW is the telescope and instrument the data were taken with; GN+N indicates Gemini North and \textit{NIRI} were used; GS+F indicates Gemini South and \textit{FLAMINGOS} were used; Notes: (1) Moderate 60Hz signal, (2) Severe 60Hz signal, (3) Streak across the image due to a bright source just outside the field of view.
\end{tabular}
\end{center}
\end{table}

The science data were acquired using a dither pattern, which involved 
systematically offsetting the telescope, to allow the effective removal of the 
sky background. The total exposure time given in Table~\ref{observed} was 
achieved by obtaining $60$~second and $90$~second individual exposures per 
dither position for \textit{NIRI} and \textit{FLAMINGOS}, respectively. The 
number of coadds acquired for each individual exposure was adjusted to avoid 
saturating the white dwarf. Unfortunately, on occasion, the requested exposure 
time of $1$~hour was not always achieved. ``Lamps on'' dome flats were 
acquired by imaging a uniformly illuminated screen within the dome. ``Lamps 
off'' dome flats were acquired using the same method, except with no 
illumination of the screen. These dome flats were used for the calibration of 
the \textit{NIRI} science data. High and low twilight flats were acquired for 
the calibration of the \textit{FLAMINGOS} data. Short dark frames were also 
acquired to help with the identification of bad pixels.

\section{Data Reduction}
\label{datareduction}
All the data acquired for the DODO survey were reduced using the Image 
Reduction and Analysis Facility (IRAF; \citealt{t1986}) and the 
\textsc{gemini} package, versions 2.12.2a and 1.8, respectively. Raw 
\textit{NIRI} images are in the form of multi--extension fits (MEF) files with 
most of the header information in the Primary Header Unit (PHU), ``[0]'' 
extension and the raw image data in the second, ``[1]'' extension. Raw 
\textit{FLAMINGOS} images are in the form of single extension fits files. The 
\textsc{nprepare} and \textsc{fprepare} tasks in the \textsc{gemini} package 
were applied to all raw data acquired with \textit{NIRI} and 
\textit{FLAMINGOS}, respectively. These tasks add certain essential keywords 
to the header of each data file, allowing the subsequent data reduction tasks 
to be applied. In addition, the \textsc{fprepare} task converted all the 
\textit{FLAMINGOS} images into MEF files. This made it possible for both the 
\textit{NIRI} and the \textit{FLAMINGOS} data to be reduced in a homogeneous 
manner, using the \textsc{niri} tasks in the \textsc{gemini} package.

For the reduction of the science data, a sky frame was used instead of a dark 
frame as it gives a much better dark measurement, since it was acquired 
concurrently with the science data. The sky frame also removes any constant 
additions to the science data due to bias levels. The \textsc{nisky} task was 
used to create the sky frame by median combining each individual science image 
after masking the objects in each image. The \textsc{nireduce} task was used 
to subtract the sky frame from the individual science images. A flat field 
image was created by subtracting the median combined ``lamps off'' / low 
twilight flat image from each individual ``lamps on'' / high twilight flat 
images and then median combing the resultant images. This flat field image was 
divided by its mean pixel value to create a normalised flat field image. The 
\textsc{niflat} and \textsc{nireduce} tasks were used to create this 
normalised flat field image and to divide the individual science images by the 
normalised flat field image, respectively. For each individual science image, 
a sky background image was created by median combining the previous and 
subsequent $5$ and $4$ science images for the \textit{NIRI} and the 
\textit{FLAMINGOS} data, respectively, after masking the objects in each 
image. The \textsc{xmosaic} task in the \textsc{xdimsum} package was used to 
create and subtract each sky background image from each individual science 
image and then create an average combined final stacked image. This task also 
corrected the bad pixels in each sky subtracted image by using a mask created 
by the \textsc{niflat} task and linearly interpolating across bad columns in 
each image line. In addition, cosmic ray events are removed by replacing the 
value of the pixels, which are more than $5\sigma$ above the median of the 
pixel values in a surrounding box of $5\times5$ pixels and that are not part 
of an object, with the local median value. 

Saturated pixels (e.g., from the cores of bright stars) can lead to 
persistence effects in subsequent frames as residual charge remains in the 
pixels after the array has been reset. Persistence manifests itself as an 
apparent faint object at the location of the saturated object in the previous 
image. This faint object can remain for several frames depending on the level 
of saturation. For the ALADDIN--II \textit{NIRI} array, a highly saturated 
star will leave a faint object at the level of $\sim1\%$ in the subsequent 
exposure and $\sim0.2\%$ in the next one \citep{hji2003}. Persistence affects 
$\sim42\%$ of the science data in Table~\ref{observed}. To remove these 
persistence effects from the sky background image and from the final stacked 
image, a mask consisting of the cores of the bright stars in the field from 
the previous three science images was created for each individual science 
image and added to the object mask used in the \textsc{xmosaic} task.

Intermittent pattern noise substantially degraded a large amount of the 
\textit{NIRI} data throughout 2003 and has also affected some of the data from 
2004 and 2005. A diagonal herringbone pattern due to $60$~Hz interference 
\citep{hji2003} is seen in $\sim37\%$ of the \textit{NIRI} data 
(Table~\ref{observed}). This pattern is reflected symmetrically in each 
quadrant, due to the fact that the readout of the quadrants, from each corner 
of the array to the centre of the array along each row, is symmetrical. This 
pattern is not present in the lab and has been eliminated at times on the 
telescope, indicating that it arises from the telescope environment and not 
from the \textit{NIRI} electronics. It was not possible to remove this noise 
from the \textit{NIRI} data.

Another form of intermittent pattern noise found in the \textit{NIRI} data is 
due to $50$~Hz interference, which creates a horizontal pattern. The constant 
variations in the rows and columns of each image due to this pattern were 
removed by collapsing each quadrant along rows and columns and then 
subtracting the median value from each row and each column separately for each 
of the quadrants.

Another form of intermittent pattern noise found in the \textit{NIRI} data, 
which manifests itself as a combined quadrant pattern plus a vertical pattern, 
was also removed in the same manner. The quadrant pattern occurs due to 
mismatched bias levels between the quadrants and the vertical pattern has a 
periodicity of $8$ pixels, which corresponds to the $8$ amplifier channels 
reading out each quadrant.

\section{Data Analysis}
\label{dataanalysis}
The cleaned final stacked images created using \textsc{xmosaic} were 
astrometrically calibrated to within $\sim1-2^{\prime\prime}$ using Two Micron 
All Sky Survey (2MASS; \citealt{scs2006}) objects in the field of each white 
dwarf. The IRAF tasks \textsc{ccfind} and \textsc{ccmap} in the 
\textsc{images} package were used to calculate the transformation required to 
match the positions of the objects in the final stacked image with the 
positions of the objects in the 2MASS catalogue. Objects near the edge of the 
final stacked image were removed from the transformation calculations. The 
\textsc{ccsetwcs} task was then used to apply this transformation to the final 
stacked image. The astrometrically calibrated final stacked images were then 
photometrically calibrated using aperture photometry. Objects in the final 
stacked image, with instrumental magnitudes $m_{\rm{i}}$, were matched with 
objects in the 2MASS catalogue. The instrumental magnitude of the matched 
object could then be associated with their apparent magnitudes, $m$, using the 
linear formula $m=m_{\rm{i}}+zp$. The zeropoint, $zp$, is therefore equal to 
the y intercept of the line of best fit to the points in the plot of 
$m_{\rm{i}}$ against $m$. Only photometric 2MASS stars, with a $J$ band 
photometric quality flag equal to ``A'' (given to objects with a SNR~$>10$ and 
a photometric uncertainty, $\sigma<0.109$), were used to determine the $zp$. 
In addition, objects near the edge of the final stacked image or that were 
saturated were excluded from this step.

\subsection{Point Source Detection}
\label{pointsource}
The SExtractor program \citep{ba1996} was used to detect all objects in the 
final stacked images with a signal to noise ratio (SNR) $>3$. SExtractor 
determines the position of the centre of an object by computing the first 
order moments of the isophotal profile of the object, which is adequate for 
detecting point sources in all the final stacked images in the DODO survey. A 
``weight map'', created by the \textsc{xmosaic} task, was used to normalise 
the noise over the final stacked image. This avoided the detection of spurious 
sources around the edges of the final stacked image, where only a few of the 
individual science images contribute. The first pass of SExtractor used 
apertures with diameters ranging from $1$ to $20$ pixels to determine the 
aperture size that delivered the highest SNR. Only objects with an internal 
flag equal to $0$, indicating that no problems were found with the detection, 
were used in the determination of the optimum aperture size. In addition, the 
ellipticity of the objects, given as $1-B/A$, where $A$ and $B$ are the 
semi--major and semi--minor axes of the object, respectively, was chosen to be 
$<0.2$. This excluded objects with high ellipticities, such as background 
galaxies, from the determination of the optimum aperture size. Also, only 
those objects within the central region of the final stacked image that have a 
full--width at half--maximum (FWHM) $<1^{\prime\prime}$ were used. This 
further assisted in removing extended objects. The resulting ideal aperture 
size was then used to detect all objects in the final stacked images.

\subsection{Measurement of Proper Motions}
\label{propermotions}
\begin{center}
\begin{figure*}
\mbox{\includegraphics[bb=0 0 757 563,clip,angle=0,scale=0.329]{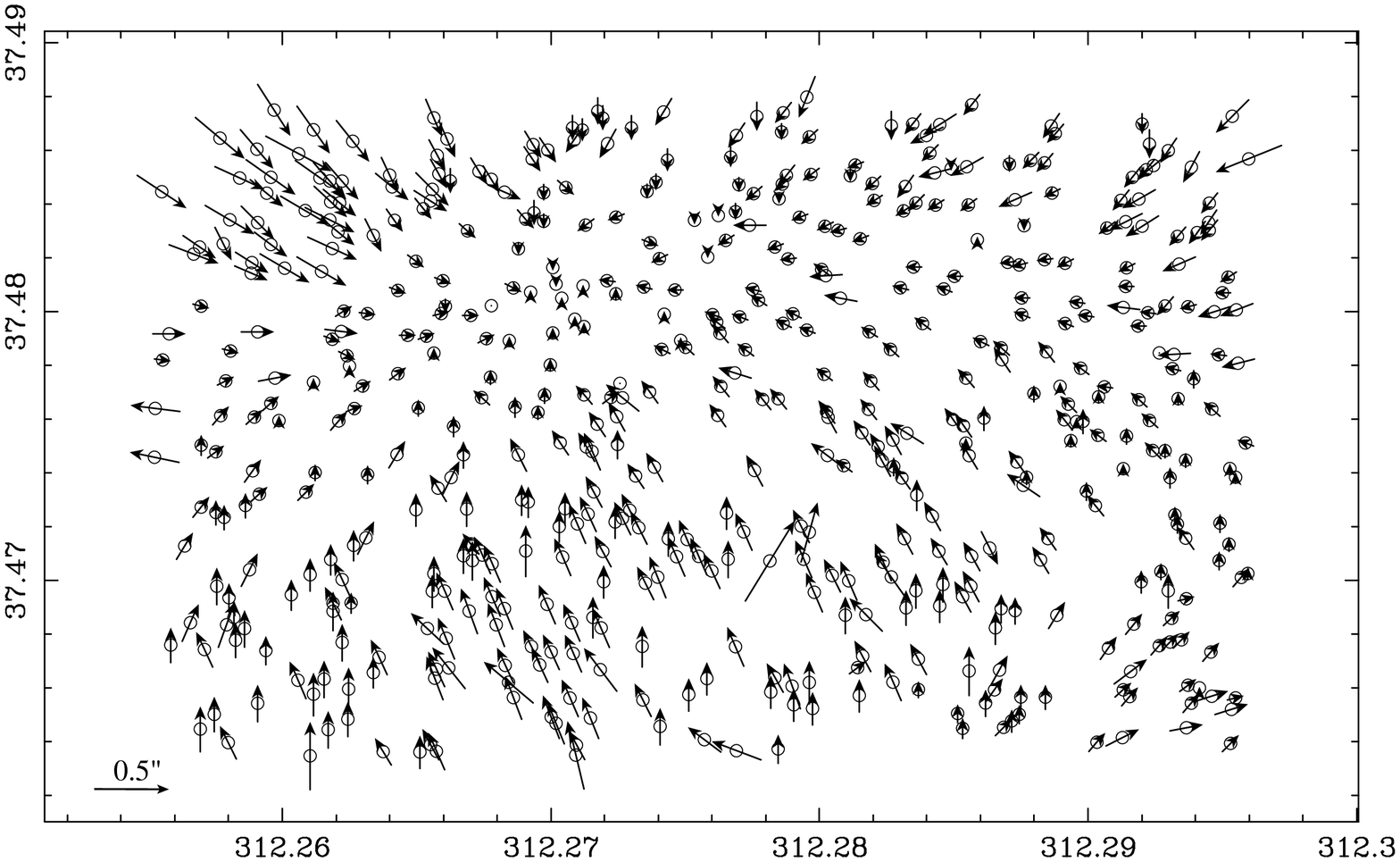}}
\mbox{\includegraphics[bb=0 0 757 563,clip,angle=0,scale=0.329]{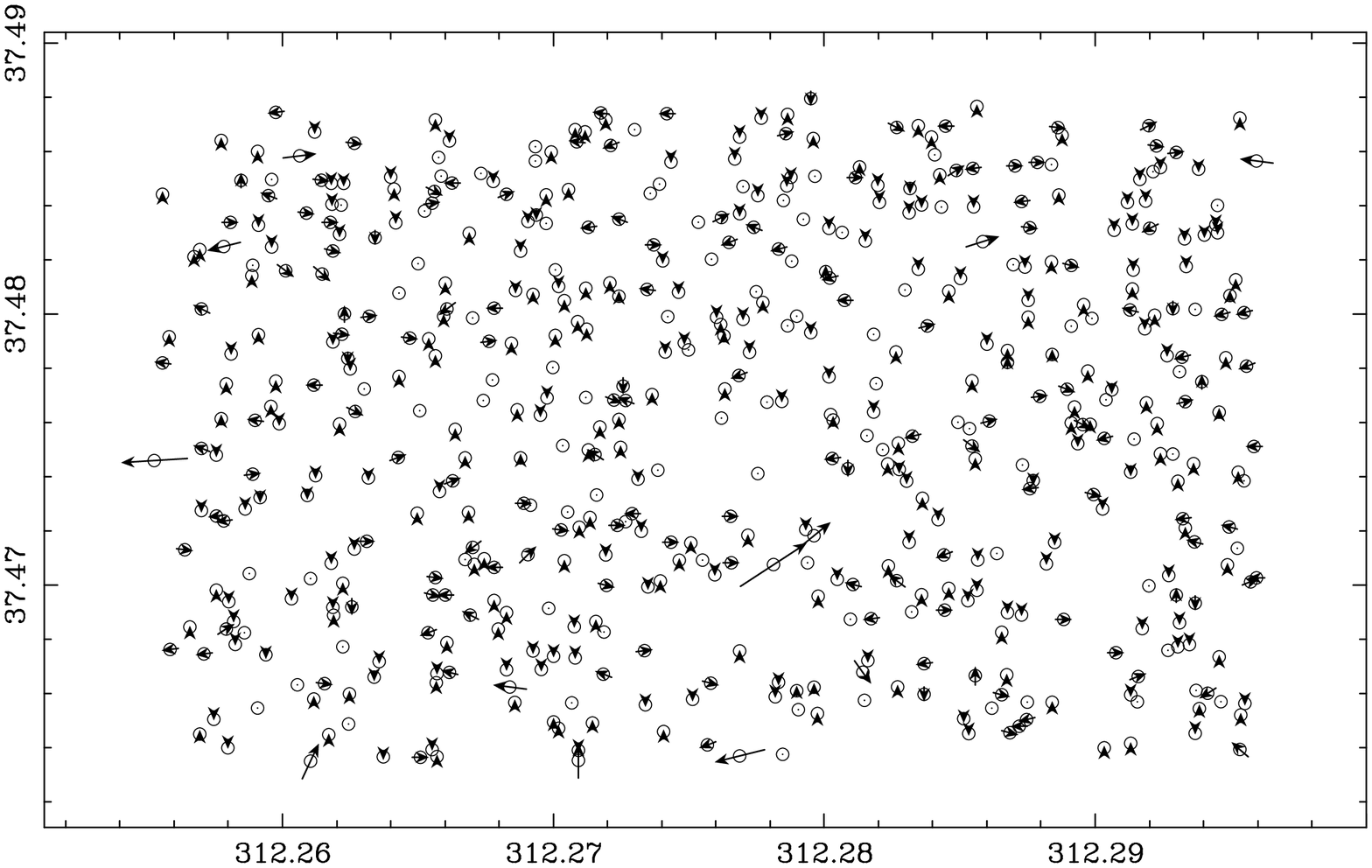}}
\caption{Distortion effects before (left) and after (right) the distortion correction was applied to the images of WD~$2047+372$. The x-- and y--axes are the R.A. and Dec, respectively, measured in degrees. The arrows in the image indicate the direction and magnitude, multiplied by 20, of the motion of each object in the field between the first epoch and second epoch images.}
\label{distortion}
\end{figure*}
\end{center}
The motion of the objects in the field of each white dwarf between the first 
epoch and second (or third) epoch images was calculated. Since the white dwarf 
is rarely positioned on the same pixels in each epoch, spurious distortion 
effects can be seen, which are caused by optical aberrations. As a result, the 
motion of objects between the two epoch images is seen to be a function of 
field position (Figure~\ref{distortion}). When a large number of background 
reference stars are present, the two epoch images can be well matched and the 
distortions can be effectively removed. However, as the number of background 
reference stars decreases, the two epoch images can not be accurately matched 
and this is often the limiting factor for astrometric accuracy. The 
\textsc{geomap} and \textsc{geoxytran} tasks in the \textsc{images} package 
were used to correct for these distortion effects. Objects from the first 
epoch image were matched to the closest object present in the second (or 
third) epoch image only if 1) their magnitudes are within $1$~mag, 2) the 
SExtractor internal flag~$\leq3$, which indicates a good detection, 3) the 
ellipticity of the object is $<0.5$. This excluded objects with very high 
ellipticities, such as background galaxies, from the matching procedure. In 
some cases, the closest object was too far away to be a true match, so a 
clipping factor was introduced to remove these mismatches.

\subsection{Limits and Errors}
\label{limitsanderrors}
The completeness limit for each final stacked image was estimated by 
determining the magnitude at which $90\%$ and $50\%$ of inserted artificial 
stars were recovered from each image. The \textsc{starlist} task was used to 
create a list of $200$ randomly positioned artificial stars at a magnitude of 
$J=19.0$~mag. The \textsc{mkobjects} task was used to insert the artificial 
stars into the final stacked image. SExtractor was then used to detect all 
objects in the image, including the artificial stars. The calculated 
magnitudes of the artificial stars were checked to ensure they were equal to 
$J=19.0$~mag. Using the same artificial star list, the \textsc{mkobjects} and 
SExtractor steps were repeated for magnitudes between 
$19.1\leq{J}\leq24.0$~mag in $0.1$ magnitude steps. The entire process was 
then repeated a further $50$ times, equivalent to a total of $10,000$ inserted 
artificial stars for each $0.1$ magnitude bin. Plots of the percentage of 
artificial stars recovered against the apparent $J$ magnitudes of the 
artificial stars were created. The number of artificial stars recovered was 
often much less than $100\%$ at the brighter $J$ magnitudes as some stars were 
lost within the point spread function (PSF) of other real objects or 
artificial stars. The motion of an object can be calculated only when the 
object is detected in both epochs. Assuming that the probability of detecting 
an object in the first epoch image, $P_1$, is independent from the probability 
of detecting an object in the second epoch image, $P_2$, the probability of 
detecting an object in both epochs is $P_{1} \times P_{2}$. Therefore, by 
multiplying the individual completeness limits for each epoch, a combined 
completeness limit for both epoch images can be determined. This assumption is 
valid for objects near the completeness limit. However, it is not valid for 
the bright objects not detected due to the fact that they are within the PSF 
of other real objects or artificial stars. Therefore, the completeness limits 
at the brighter $J$ magnitudes are underestimated. 

The ``COND'' evolutionary models for cool brown dwarfs and extrasolar planets 
\citep{bcb2003}, along with the magnitudes at which $90\%$ and $50\%$ of 
artificial stars were recovered, were used to estimate the minimum mass of a 
companion which could be detected in both epoch images. The models predict the 
absolute magnitudes of substellar objects depending upon their age. Isabelle 
Baraffe kindly supplied these models for the total ages determined for all the 
white dwarfs in the DODO survey. The total age is equal to the sum of the main 
sequence progenitor lifetime and white dwarf cooling age, both of which depend 
upon evolutionary models. While the cooling age errors are small and well 
constrained \citep{fbb2001}, and the scatter in the empirical IFMR is 
significantly reducing as more and higher quality observations are made of 
white dwarfs in open clusters \citep{cdn2008}, the main sequence progenitor 
lifetimes rely on models which are difficult to calibrate (e.g., 
\citealt{cig2008}). Therefore, to take these uncertainties into account, a 
conservative error of $\pm25\%$ is applied to the total age of each white 
dwarf (note that the white dwarf cooling age is the dominant timescale for 
most of the targets in the DODO survey, as shown in Table~\ref{parameters}). 
However, at ages $>1$~Gyr, the ``COND'' evolutionary models indicate that the 
absolute magnitudes of substellar objects are relatively insensitive to 
changes in their age, implying that even with a $\pm25\%$ error, the resulting 
error on the mass of a companion is small (Table~\ref{limits}).

The detection of a companion with a mass equal to the minimum mass determined 
using the completeness limit will only be possible if the companion is outside 
the extent of the PSF of the white dwarf. In addition, it is expected that the 
orbital radius of any companions that avoid direct contact with the red giant 
envelope will expand, which would increase the projected physical separation 
between the companion and the white dwarf. The majority of the DODO survey 
observations were acquired in good seeing conditions, so the minimum projected 
physical separation at which a companion could be found around each white 
dwarf was taken to be $3^{\prime\prime}$. Beyond this distance, the 
contribution of the flux from the white dwarf was assumed to be minimal. A 
more careful treatment of the PSF of the white dwarf could allow companions to 
be uncovered within $3^{\prime\prime}$ and will be dealt with in a future 
publication. The maximum projected physical separation at which a companion 
could be found around each white dwarf is limited by the field of view covered 
by both epochs. The completeness limit is only valid in the central region of 
each final stacked image, where all the individual images contribute. The 
useable field of view decreases further when the two epoch images are matched 
as the white dwarf is rarely positioned on the same pixels in each epoch 
image. The minimum and maximum projected physical separations at which a 
companion could be found around the main sequence progenitor can also be 
estimated, since the orbital radius of any companions around the main sequence 
progenitor will expand by a factor of $M_{\rm{MS}}/M_{\rm{WD}}$ during stellar 
evolution.

\section{Results}
\label{results}
Figures~\ref{WD0115_plots}~--~\ref{WD2326_plots} show the results for each of 
the 23 equatorial and northern hemisphere white dwarfs from the DODO survey. 
Using the combined completeness limit for each white dwarf, an estimate of the 
minimum mass of a companion which could be detected in both epoch images is 
calculated. The range of projected physical separations at which a companion 
of this mass could be found around each white dwarf and the corresponding 
range of projected physical separations around the main sequence progenitors 
is determined. These results are summarised in Table~\ref{limits}. Comments 
are made only on interesting objects.

\subsection{WD~{\boldmath$0644+375$}} 
\label{WD0644}
The mass of WD~$0644+375$ used throughout this paper was determined by 
assuming that the core of this white dwarf is made partly of strange matter 
\citep{mso2006}. This unusual core composition was suggested as a way to 
explain the inconsistency between the radius determined from the parallax of 
WD~$0644+375$, obtained from Hipparcos data, and the radius predicted using a 
mass--radius relation, which assumes the core of the white dwarf is composed 
primarily of carbon \citep{psh1998}. The mass of the white dwarf was 
originally determined by \citet{bsl1992} to be $0.66\,M_{\odot}$, but this 
predicted a radius that was significantly larger than predicted using the 
parallax. Therefore, the slightly lower mass of $0.54\,M_{\odot}$ determined 
by \citet{fbb2007} is used here to determine the total age of the white dwarf, 
since this mass provides a radius consistent with observations. Note that a 
companion with a mass of $5\pm1\,M_{\rm{Jup}}$ could have been detected if the 
larger white dwarf mass was used to determine the total age of the white dwarf.

\subsection{WD~{\boldmath$0738-172$}} 
\label{WD0738}
WD~$0738-172$ is a member of a known common proper motion binary. The 
secondary star of this binary system is an M6 main sequence star 
\citep{mjh2006} with an orbital radius of $\sim262$~AU \citep{pha1994}. The 
main sequence secondary does not appear in the proper motion diagram as it was 
saturated in the 2005 second epoch image, making it unavailable for proper 
motion measurements. The overall decrease in the completeness limit, compared 
to the other white dwarfs, of the images acquired of WD~$0738-172$ is due to 
the higher proportion of artificial stars inserted within the PSF of the 
bright secondary.

\subsection{WD~{\boldmath$1626+368$}} 
\label{WD1626}
Recent MIR observations of the helium atmosphere DZ white dwarf WD~$1626+368$ 
show no evidence of a dust disk \citep{mkr2007}. However, the abundance of 
carbon relative to iron in the atmosphere of WD~$1626+368$ is 10 times below 
the solar abundance, similar to the carbon deficient asteroids in the Solar 
System. Therefore, external pollution from such asteroids naturally explains 
the abundances of the metals in the atmosphere of this white dwarf 
\citep{j2006}. The possible presence of asteroids in orbit around 
WD~$1626+368$ represents an increased probability of the existence of an old 
planetary system. However, the motion of this white dwarf between the first 
epoch and second epoch images is large enough to confidently state that there 
are no common proper motion companions to WD~$1626+368$ within the limits 
given in Table~\ref{limits}. 

\subsection{WD~{\boldmath$1633+433$}} 
\label{WD1633}
\begin{figure} 
\begin{center} 
\mbox{\includegraphics[angle=270,scale=0.316]{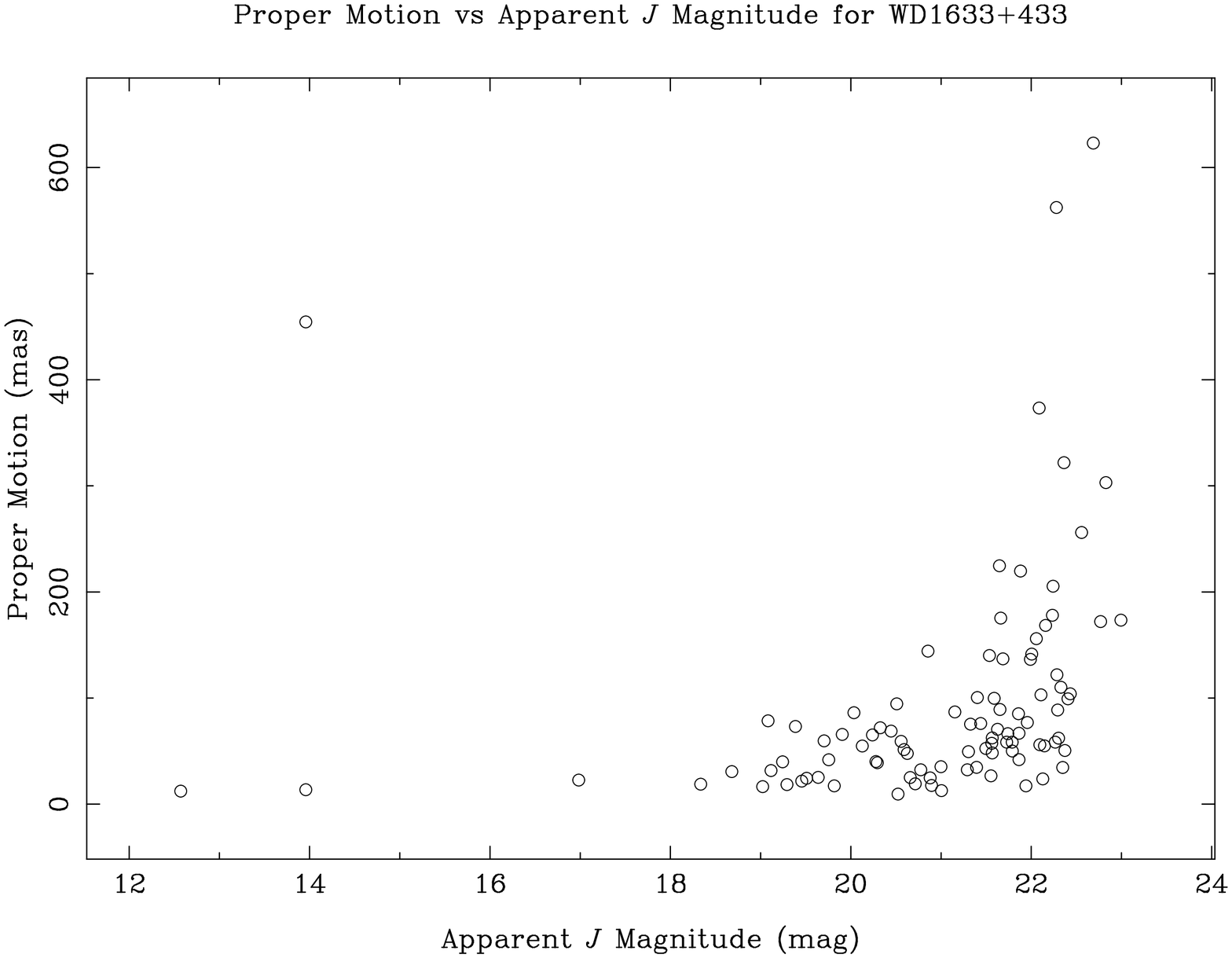}} 
\caption{The motion of the objects between the first epoch and second epoch images of WD~$1633+433$. The white dwarf has a magnitude of $J\sim14$~mag.} 
\label{WD1633_mag} 
\end{center} 
\end{figure}

Although no dust disk has been found in orbit around the DAZ white dwarf 
WD~$1633+433$, the presence of metals in its atmosphere may indicate the 
existence of an old planetary system. The 2003 first epoch image of 
WD~$1633+433$ was degraded by $60$~Hz interference (Table~\ref{observed}), 
which has decreased the completeness limit of this image. In addition, the 
2004 second epoch image of WD~$1633+433$ was degraded by a large streak across 
the image, due to a source just outside the field of view. However, this 
streak is not present in the first epoch image. It is likely that these 
effects have introduced the large scatter in the motions of objects with 
magnitudes $J>21$~mag between the 2003 first epoch and 2004 second epoch 
images (Figure~\ref{WD1633_mag}), particularly in the region of the streak. 
This suggests that the error on the motion of these faint objects is 
comparable to the motion of WD~$1633+433$ (Table~\ref{parameters}). As a 
result, multiple objects appear to have motions similar to the motion of the 
white dwarf (the two objects with motions closest to the motion of 
WD~$1633+433$ lie on the streak). Real common proper motion companions to 
WD~$1633+433$ cannot be distinguished from non--moving background objects. 
Therefore, a third

\begin{center}
\begin{figure*}
\mbox{\includegraphics[bb=45 121 576 616,clip,angle=270,scale=0.370]{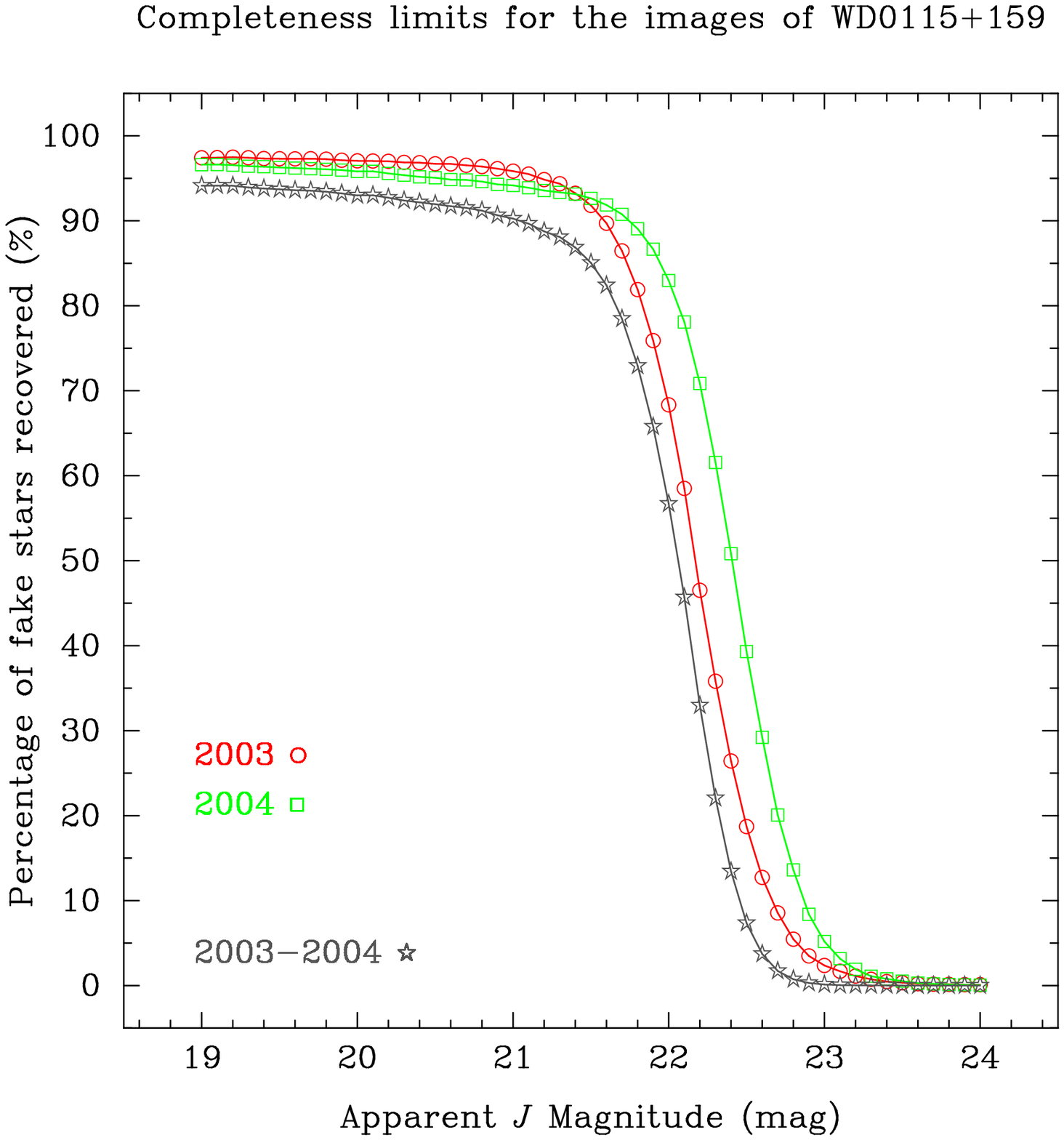}}
\mbox{\includegraphics[bb=45 130 579 616,clip,angle=270,scale=0.370]{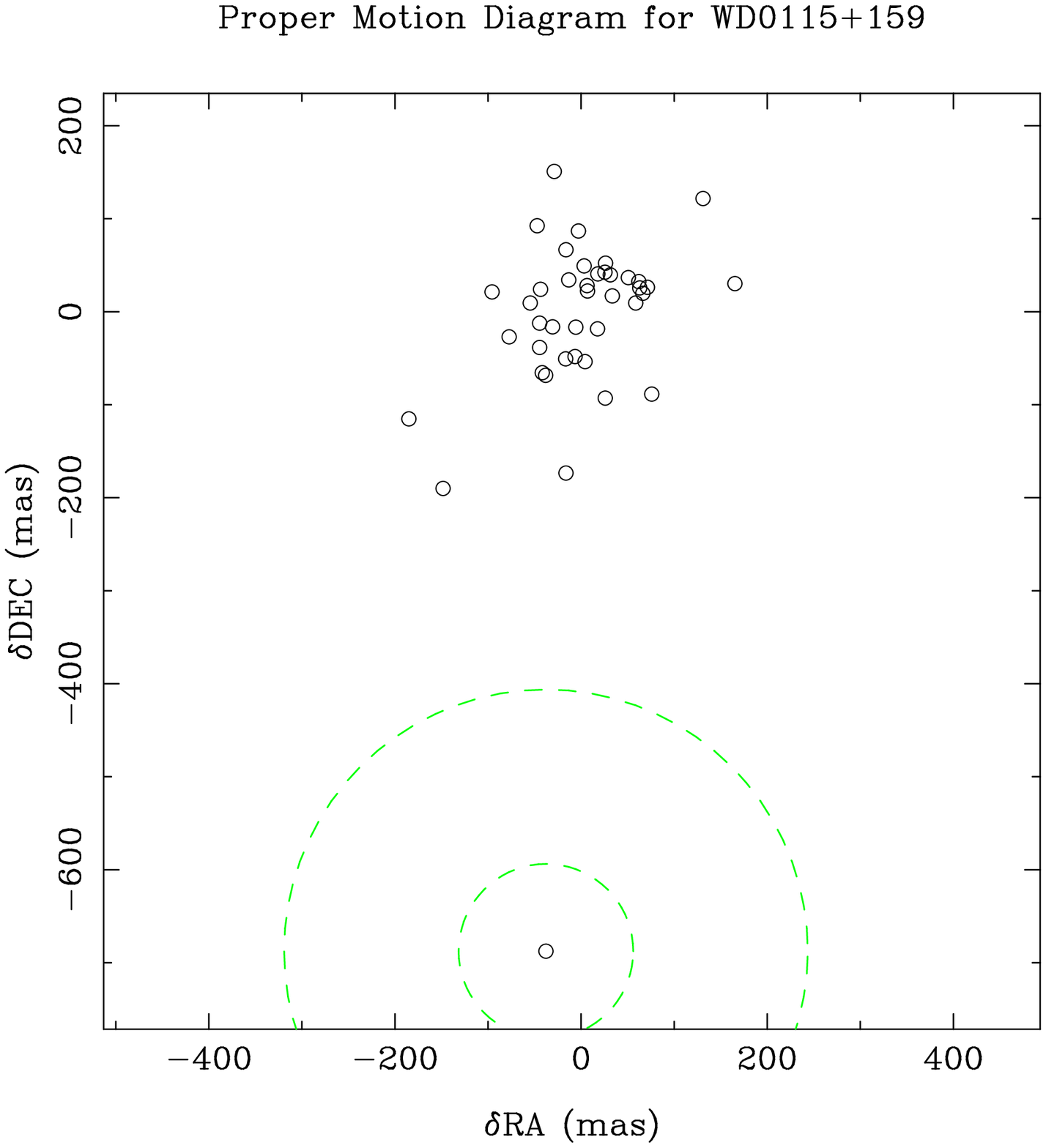}}
\caption{The completeness limit (left) shows the percentage of artificial 
stars recovered by SExtractor from the 2003 first epoch and 2004 second epoch 
images acquired of WD~$0115+159$, against the apparent $J$ magnitude of the 
artificial stars. The proper motion diagram (right) shows the motion of all 
objects in the field of WD~$0115+159$ between the first epoch and second epoch 
images. The dashed green circles represent the $1\sigma$ and $3\sigma$ scatter 
of the distribution of the motions of all objects excluding the white dwarf, 
centred on the white dwarf, to help determine possible common proper motion 
companions to the white dwarf.}
\label{WD0115_plots}
\end{figure*}

\begin{figure*}
\mbox{\includegraphics[bb=45 121 576 616,clip,angle=270,scale=0.370]{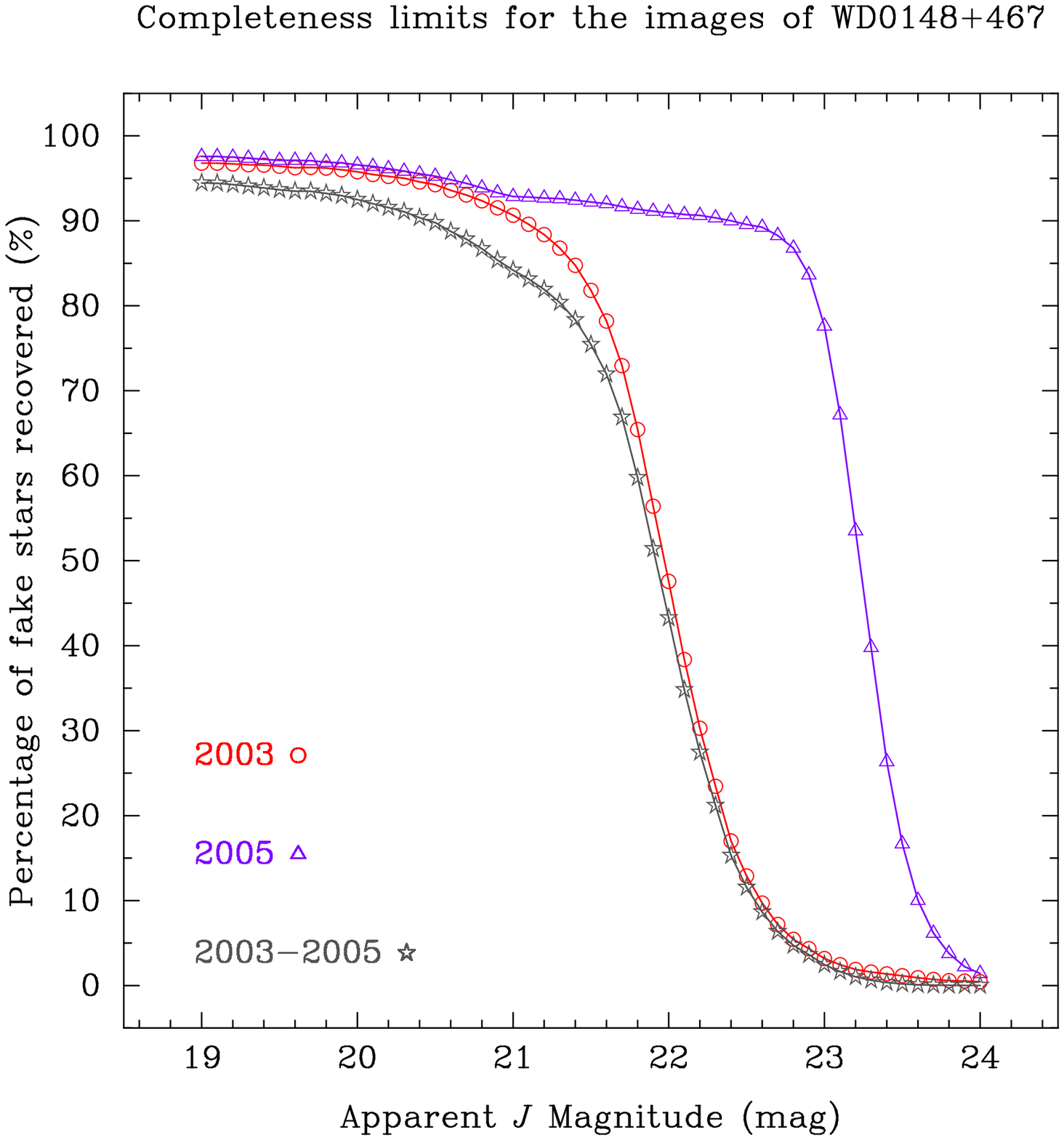}}
\mbox{\includegraphics[bb=45 130 579 616,clip,angle=270,scale=0.370]{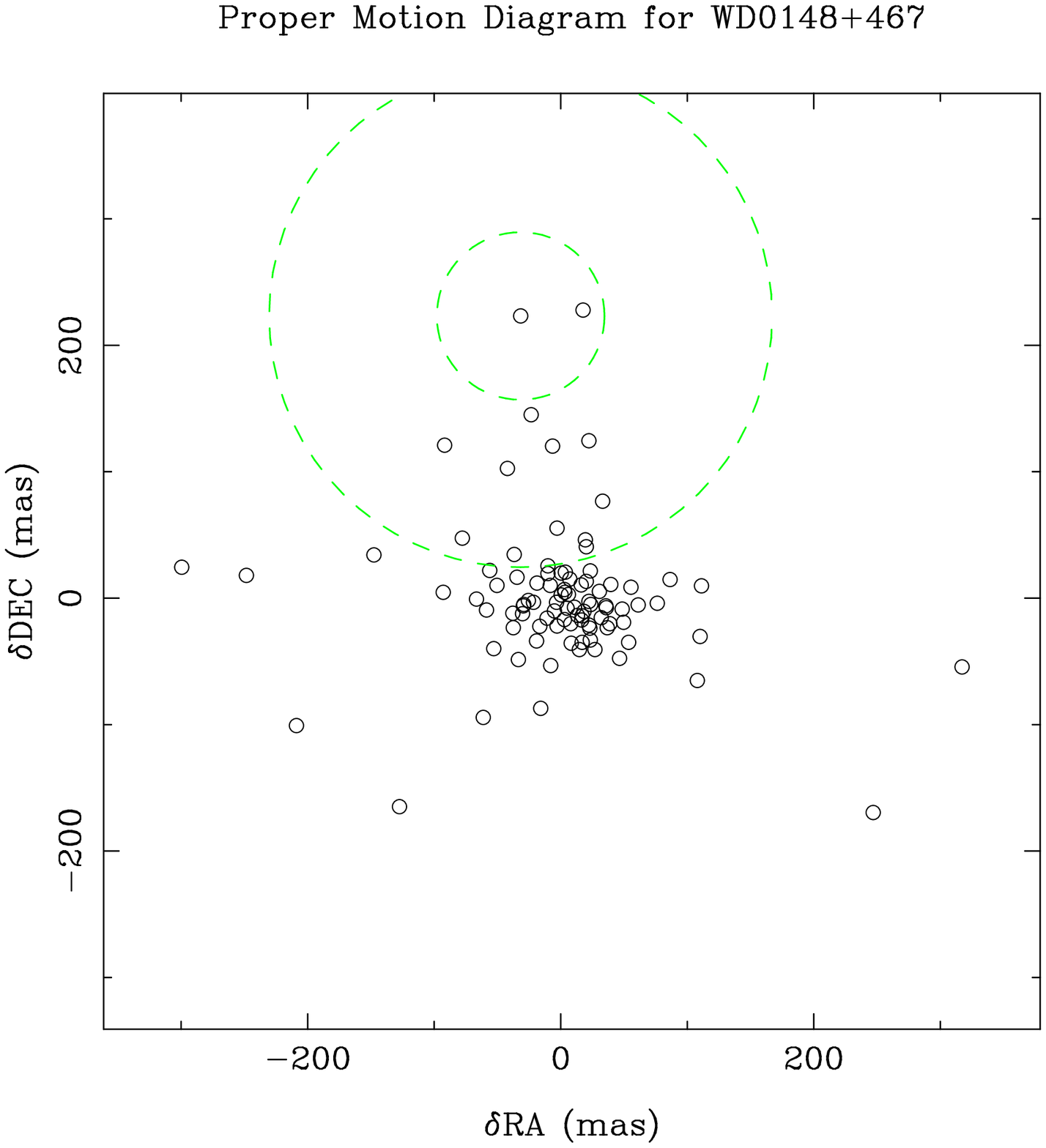}}
\caption{The completeness limit (left) and the proper motion diagram (right) for WD~$0148+467$.}
\label{WD0148_plots}
\end{figure*}
\clearpage

\begin{figure*}
\mbox{\includegraphics[bb=45 121 576 616,clip,angle=270,scale=0.370]{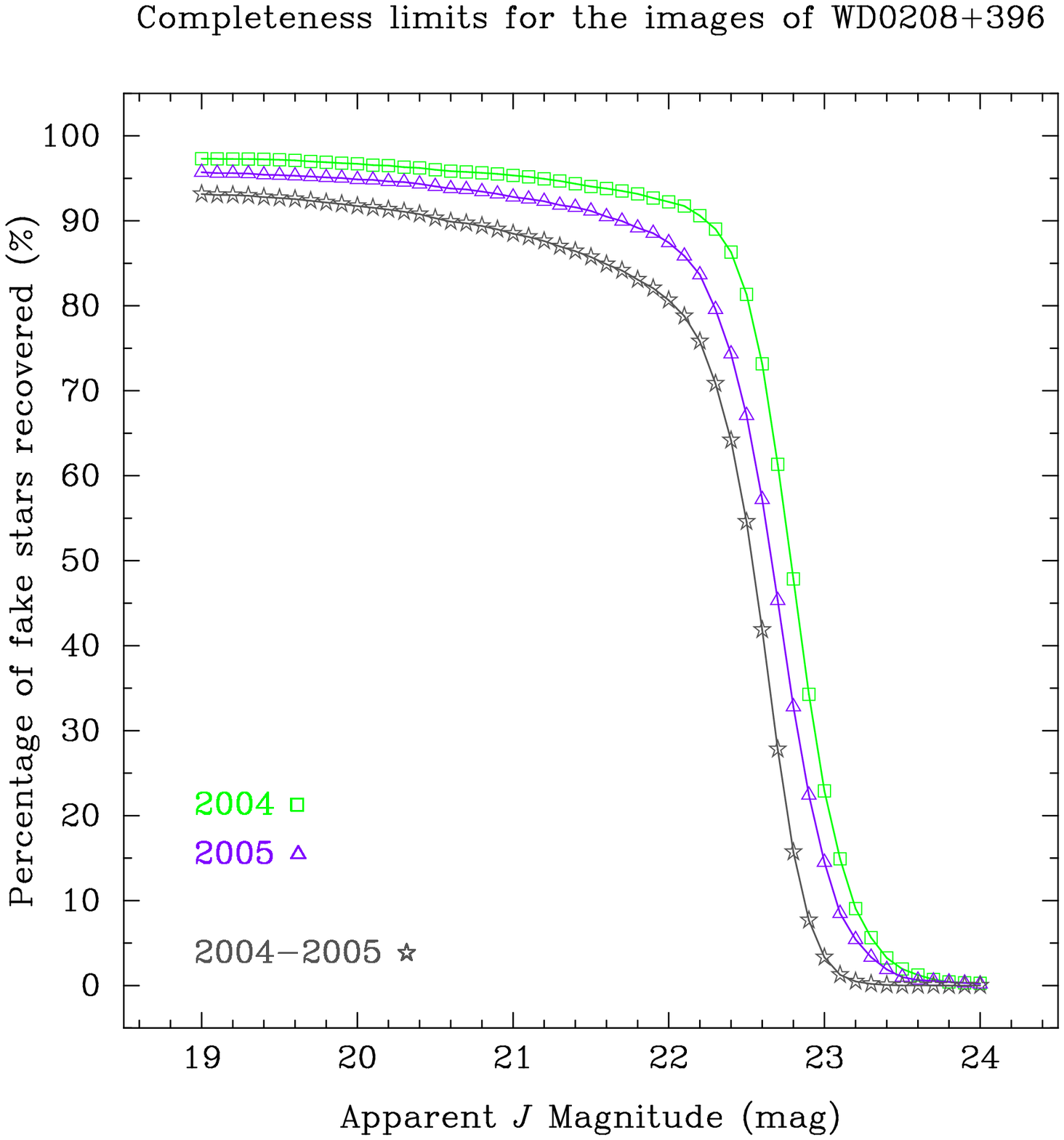}}
\mbox{\includegraphics[bb=45 130 579 616,clip,angle=270,scale=0.370]{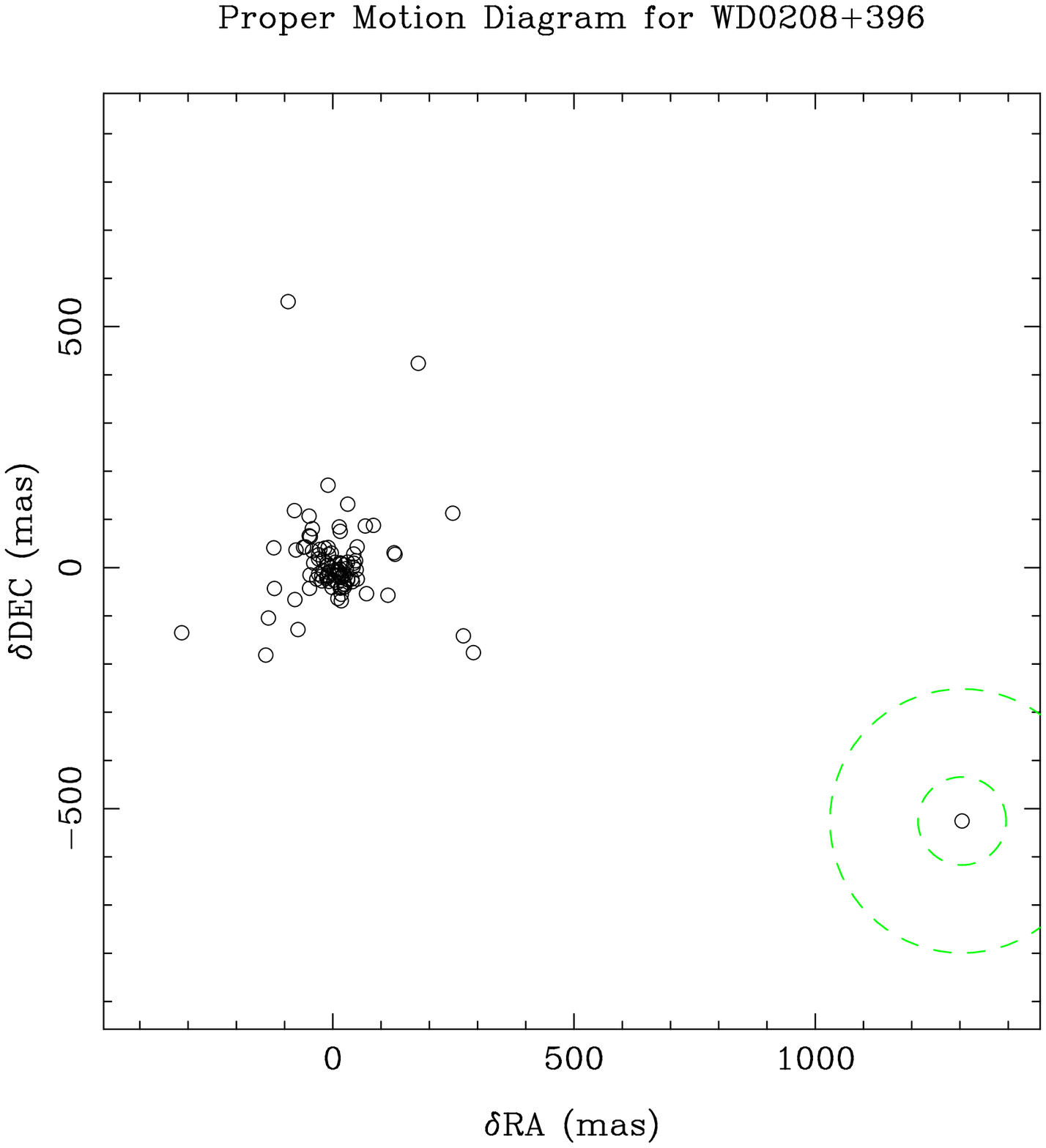}}
\caption{The completeness limit (left) and the proper motion diagram (right) for WD~$0208+396$.}
\label{WD0208_plots}
\end{figure*}

\begin{figure*}
\mbox{\includegraphics[bb=45 121 576 616,clip,angle=270,scale=0.370]{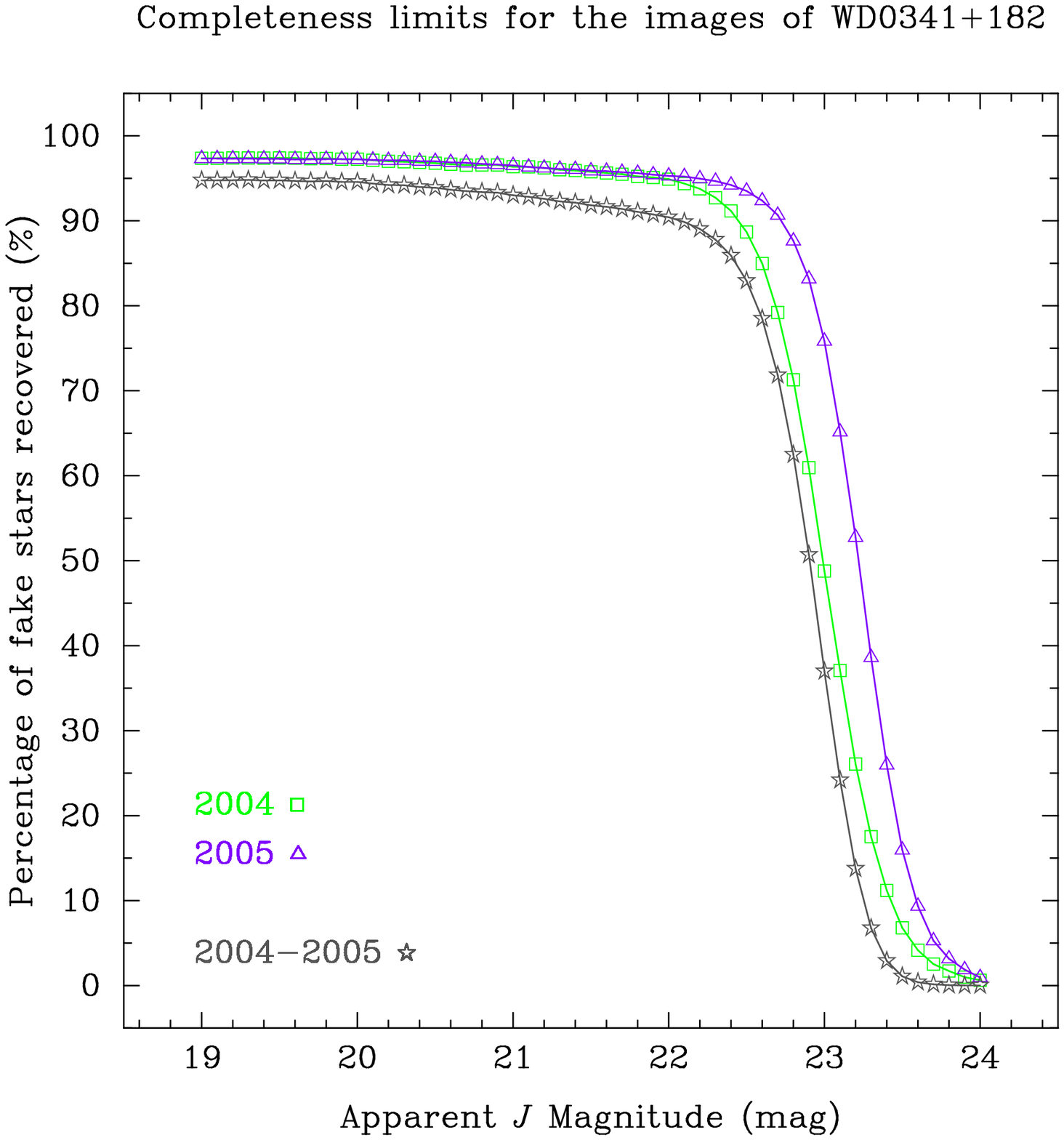}}
\mbox{\includegraphics[bb=45 130 579 616,clip,angle=270,scale=0.370]{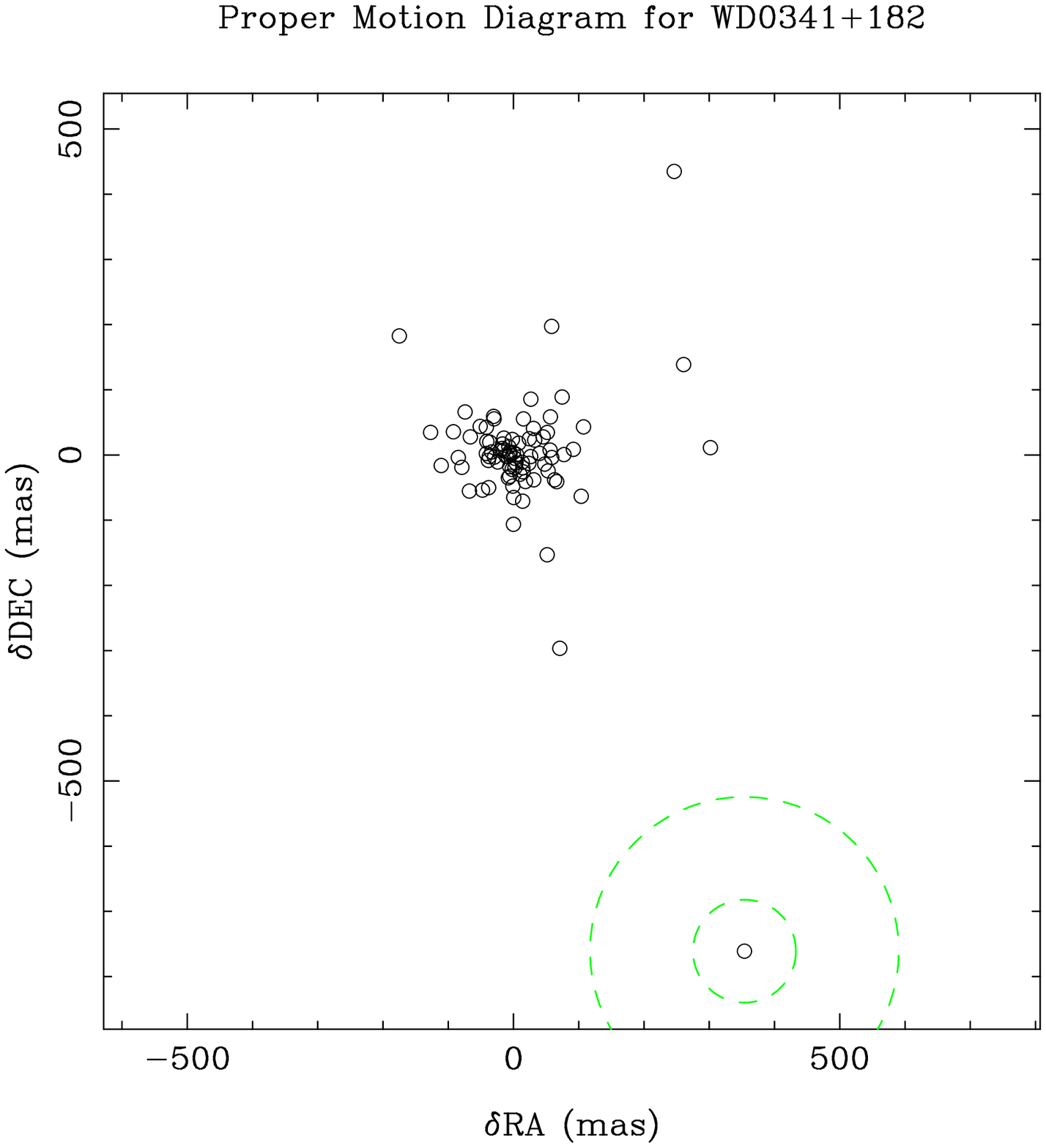}}
\caption{The completeness limit (left) and the proper motion diagram (right) for WD~$0341+182$.}
\label{WD0341_plots}
\end{figure*}

\begin{figure*}
\mbox{\includegraphics[bb=45 121 576 616,clip,angle=270,scale=0.370]{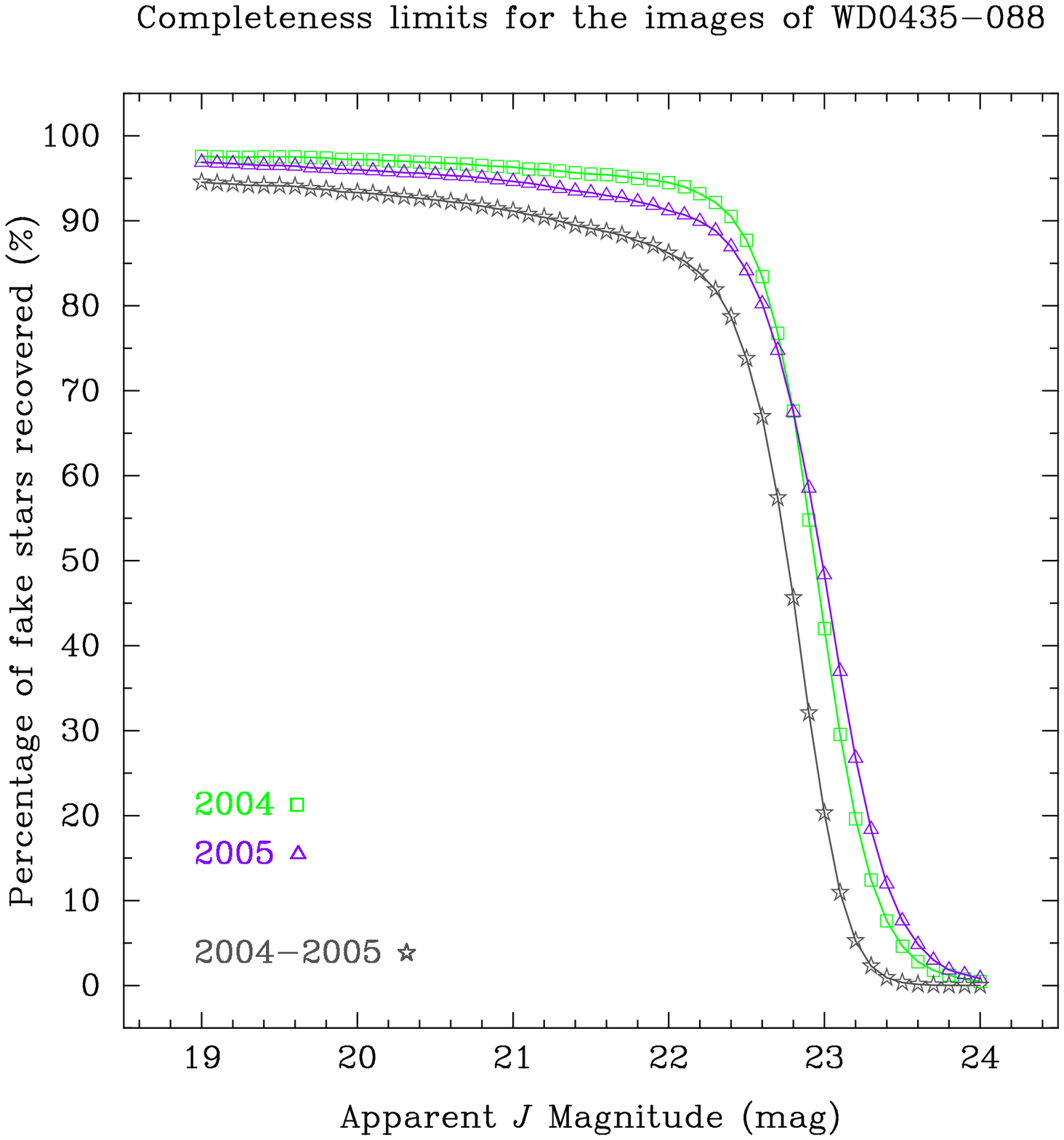}}
\mbox{\includegraphics[bb=45 130 579 616,clip,angle=270,scale=0.370]{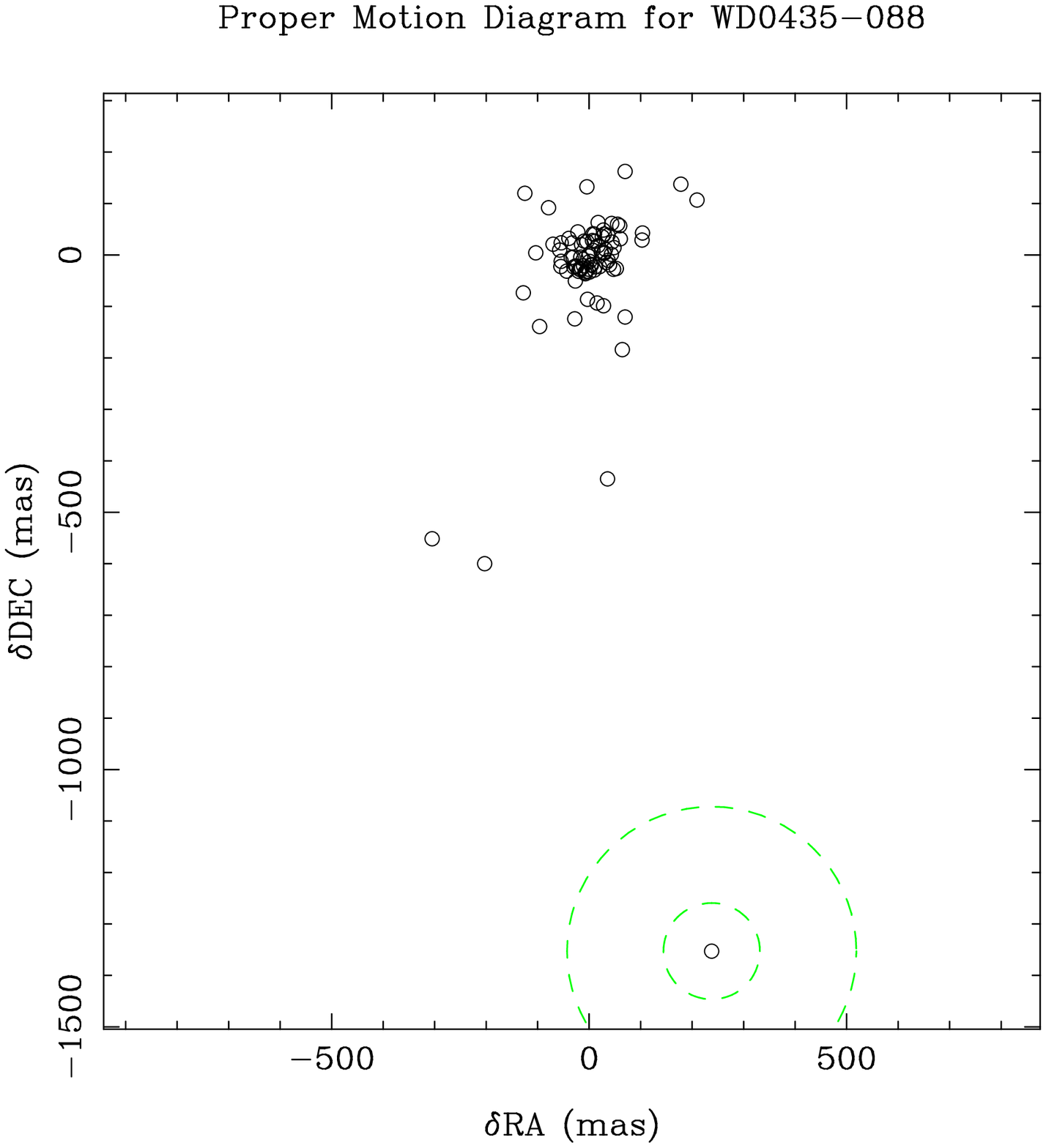}}
\caption{The completeness limit (left) and the proper motion diagram (right) for WD~$0435-088$.}
\label{WD0435_plots}
\end{figure*}
\clearpage

\begin{figure*}
\mbox{\includegraphics[bb=45 121 576 616,clip,angle=270,scale=0.370]{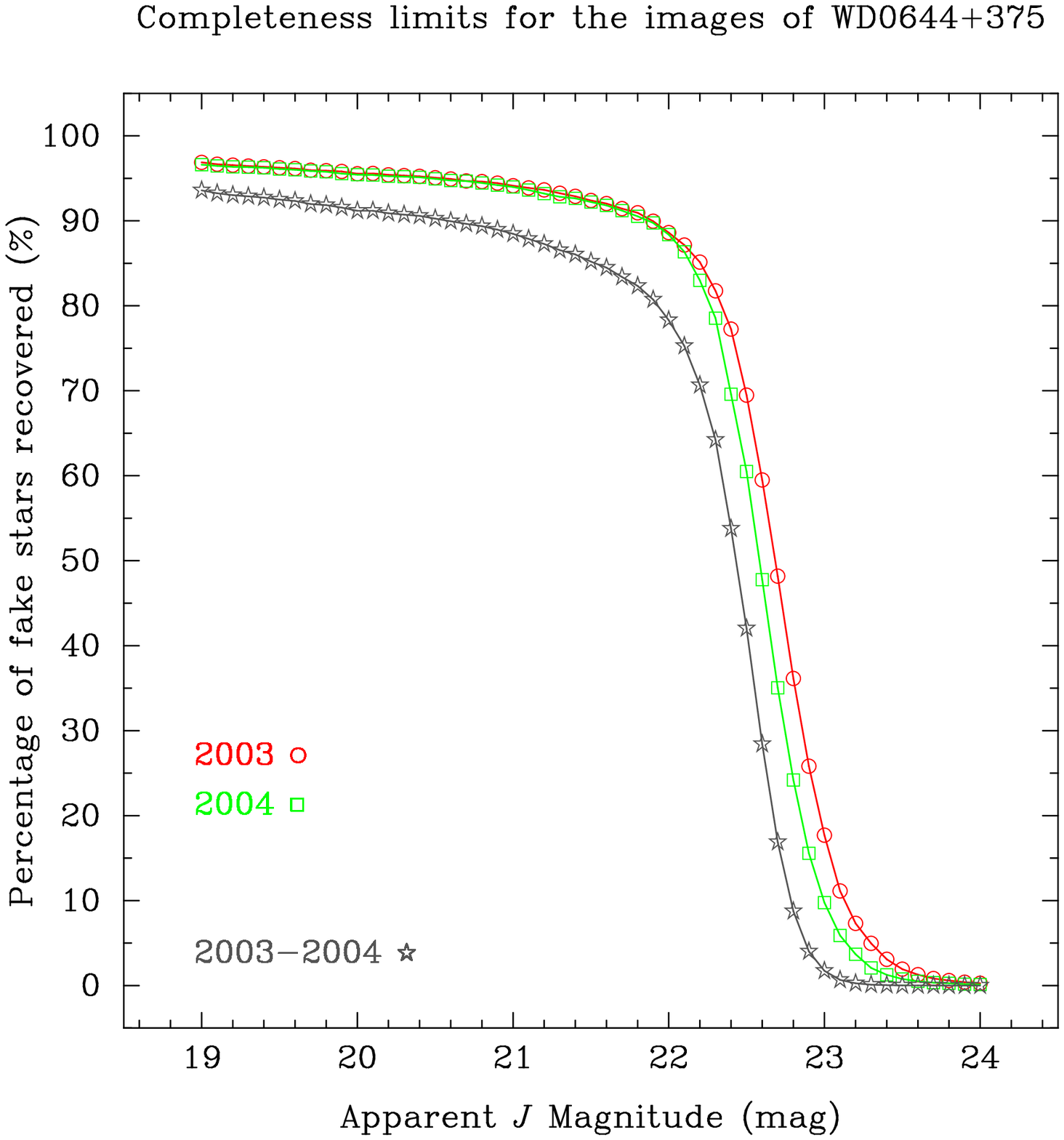}}
\mbox{\includegraphics[bb=45 130 579 616,clip,angle=270,scale=0.370]{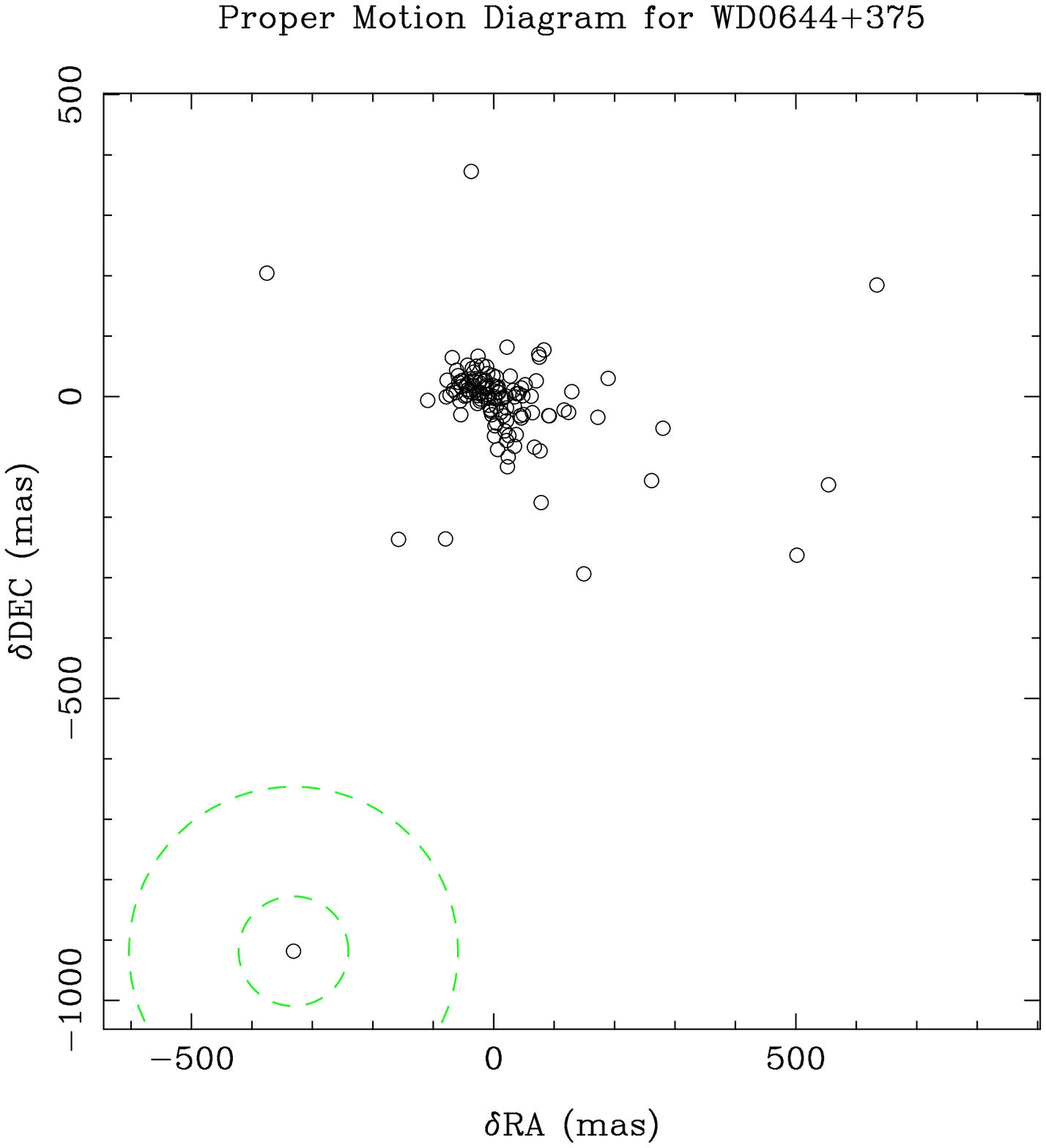}}
\caption{The completeness limit (left) and the proper motion diagram (right) for WD~$0644+375$.}
\label{WD0644_plots}
\end{figure*}

\begin{figure*}
\mbox{\includegraphics[bb=45 121 576 616,clip,angle=270,scale=0.370]{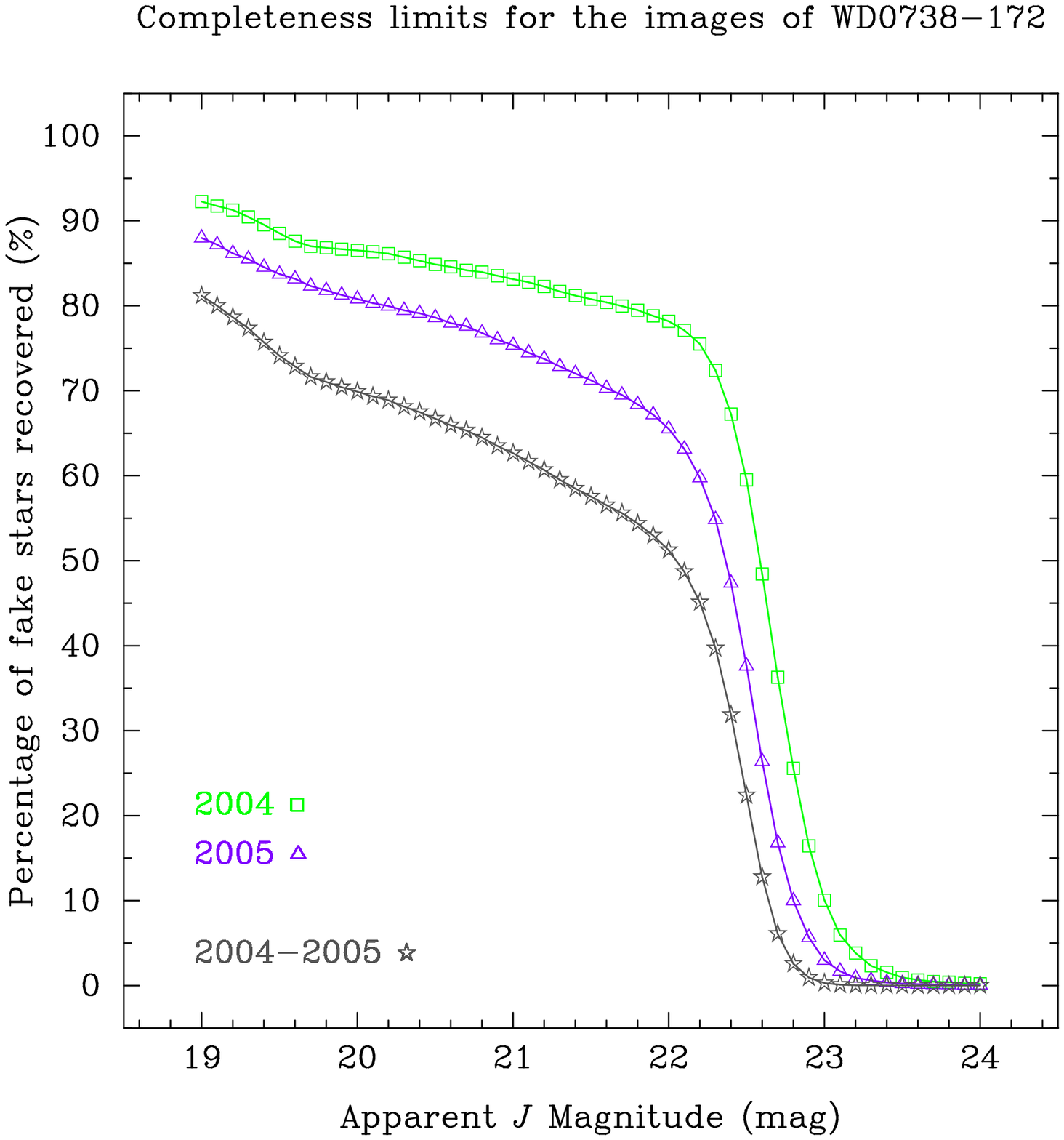}}
\mbox{\includegraphics[bb=45 130 579 616,clip,angle=270,scale=0.370]{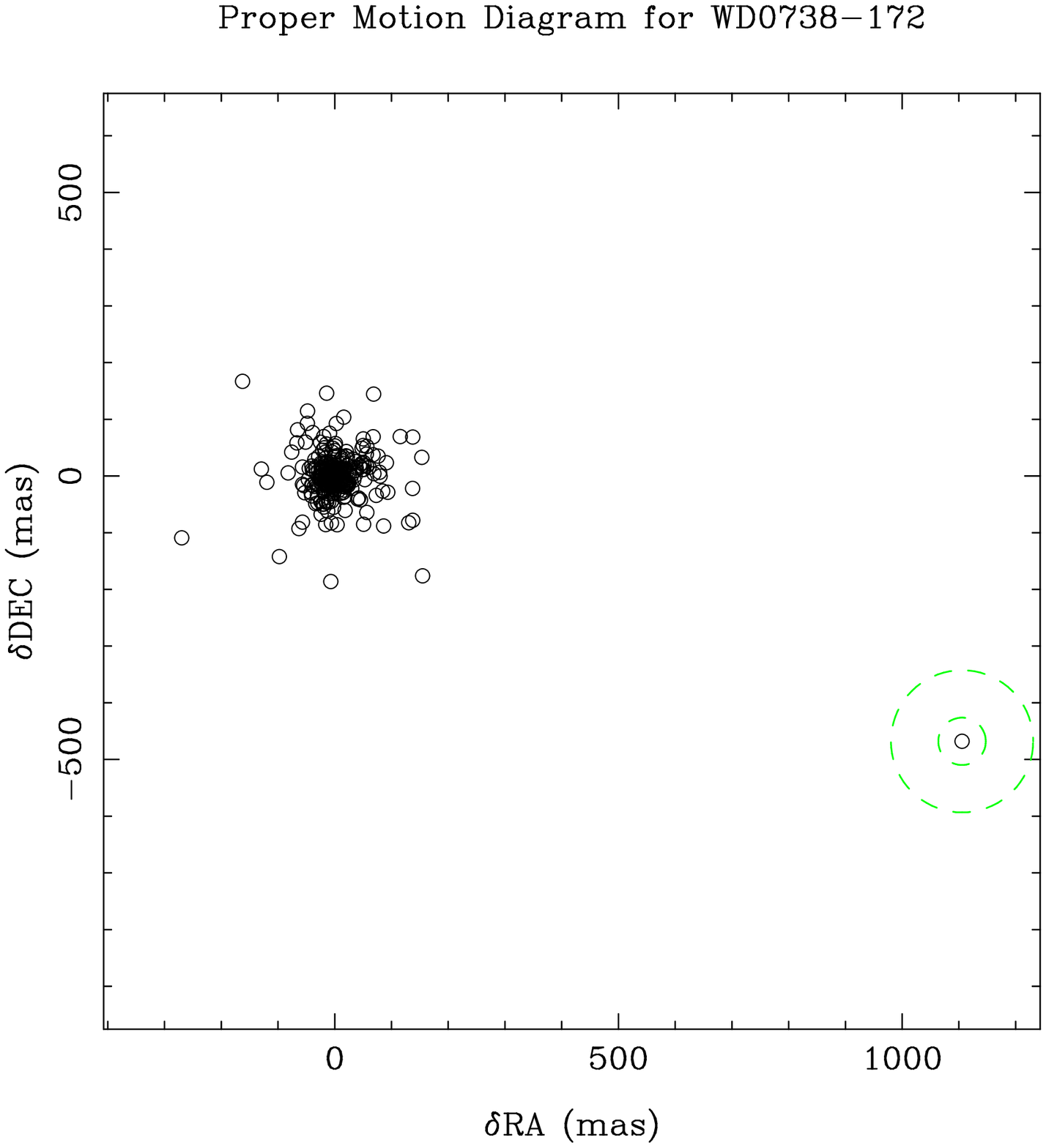}}
\caption{The completeness limit (left) and the proper motion diagram (right) for WD~$0738-172$.}
\label{WD0738_plots}
\end{figure*}

\begin{figure*}
\mbox{\includegraphics[bb=45 121 576 616,clip,angle=270,scale=0.370]{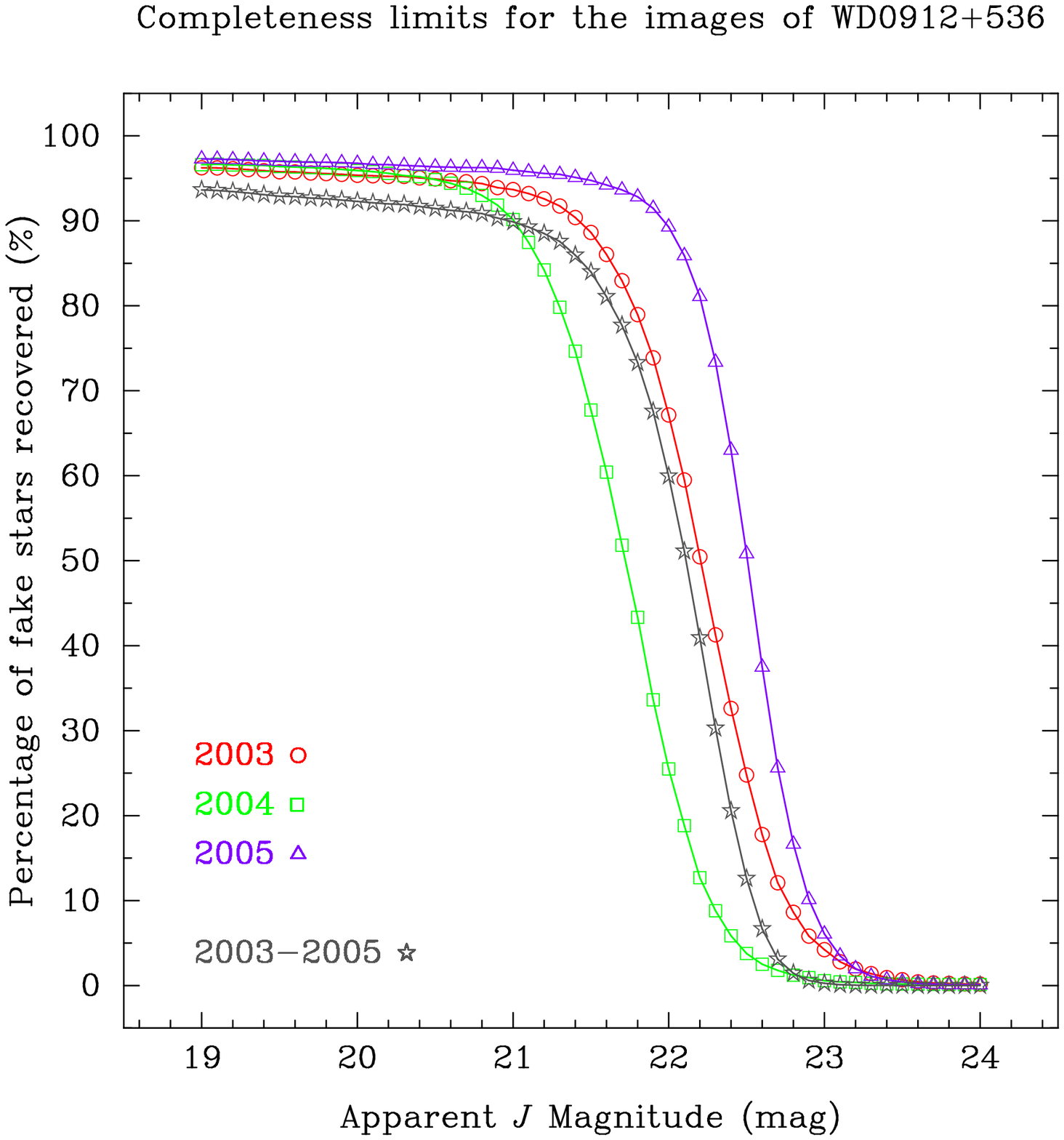}}
\mbox{\includegraphics[bb=45 130 579 616,clip,angle=270,scale=0.370]{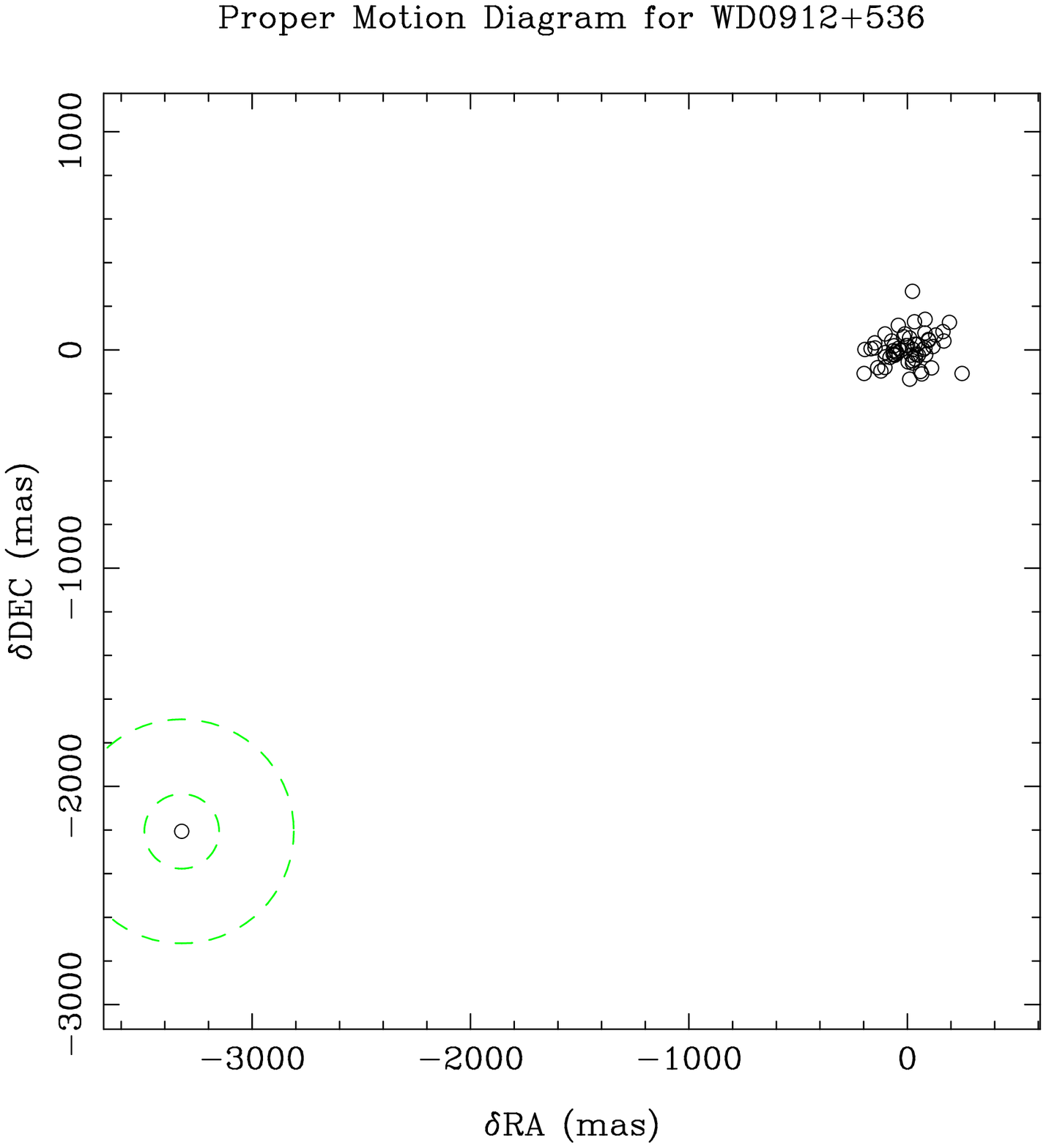}}
\caption{The completeness limit (left) and the proper motion diagram (right) for WD~$0912+536$.}
\label{WD0912_plots}
\end{figure*}
\clearpage

\begin{figure*}
\mbox{\includegraphics[bb=45 121 576 616,clip,angle=270,scale=0.370]{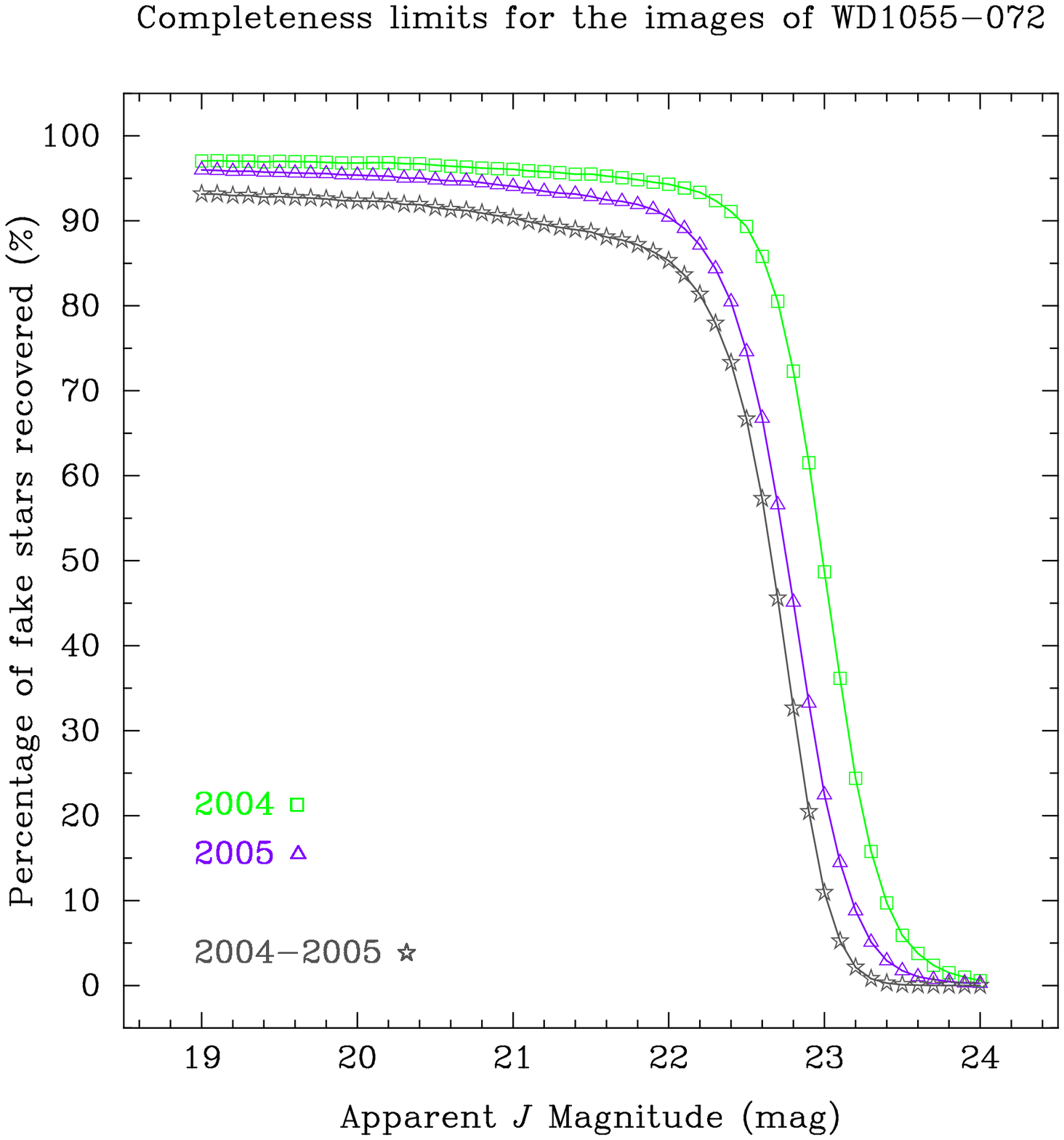}}
\mbox{\includegraphics[bb=45 130 579 616,clip,angle=270,scale=0.370]{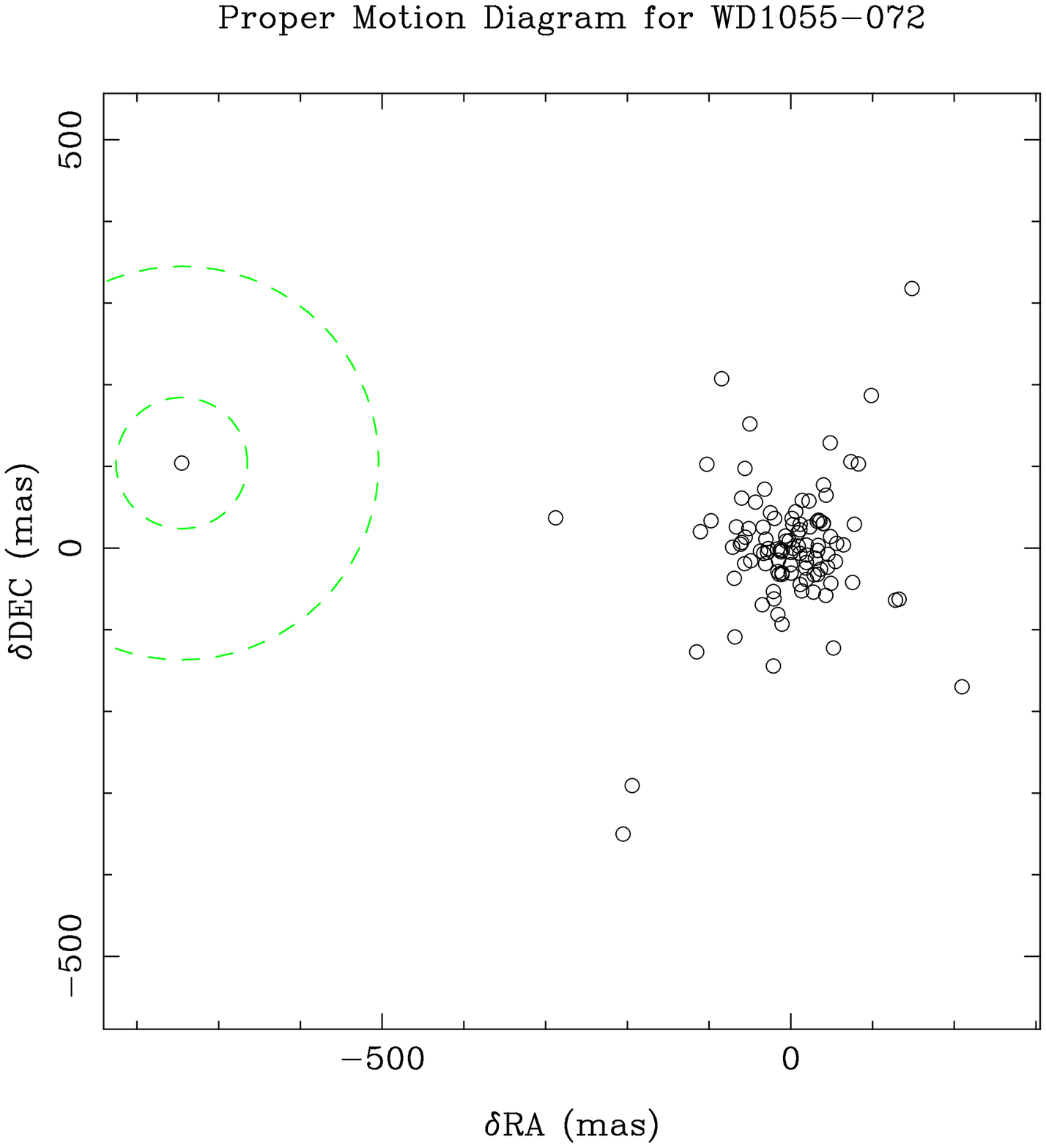}}
\caption{The completeness limit (left) and the proper motion diagram (right) for WD~$1055-072$.}
\label{WD1055_plots}
\end{figure*}

\begin{figure*}
\mbox{\includegraphics[bb=45 121 576 616,clip,angle=270,scale=0.370]{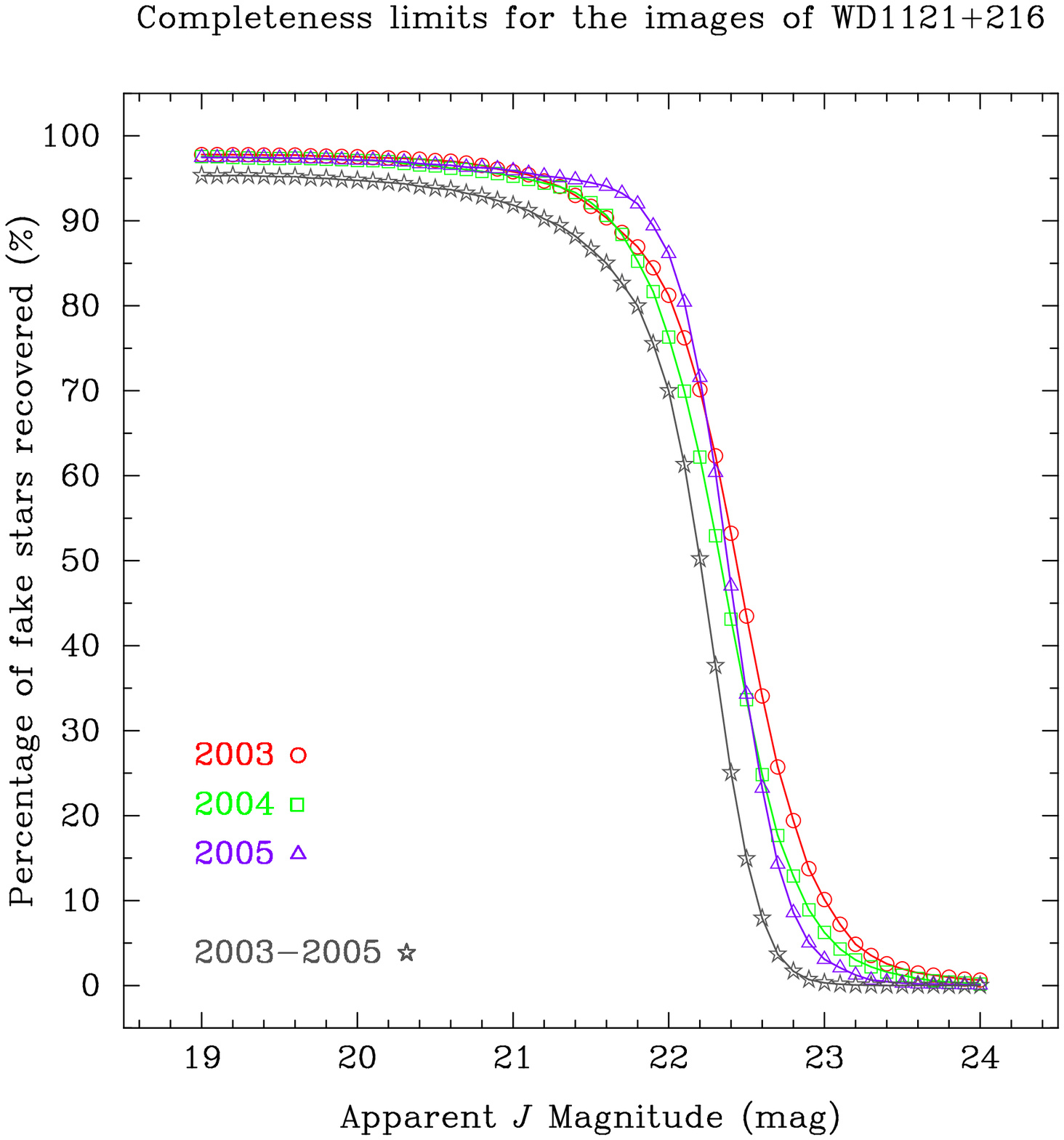}}
\mbox{\includegraphics[bb=45 130 579 616,clip,angle=270,scale=0.370]{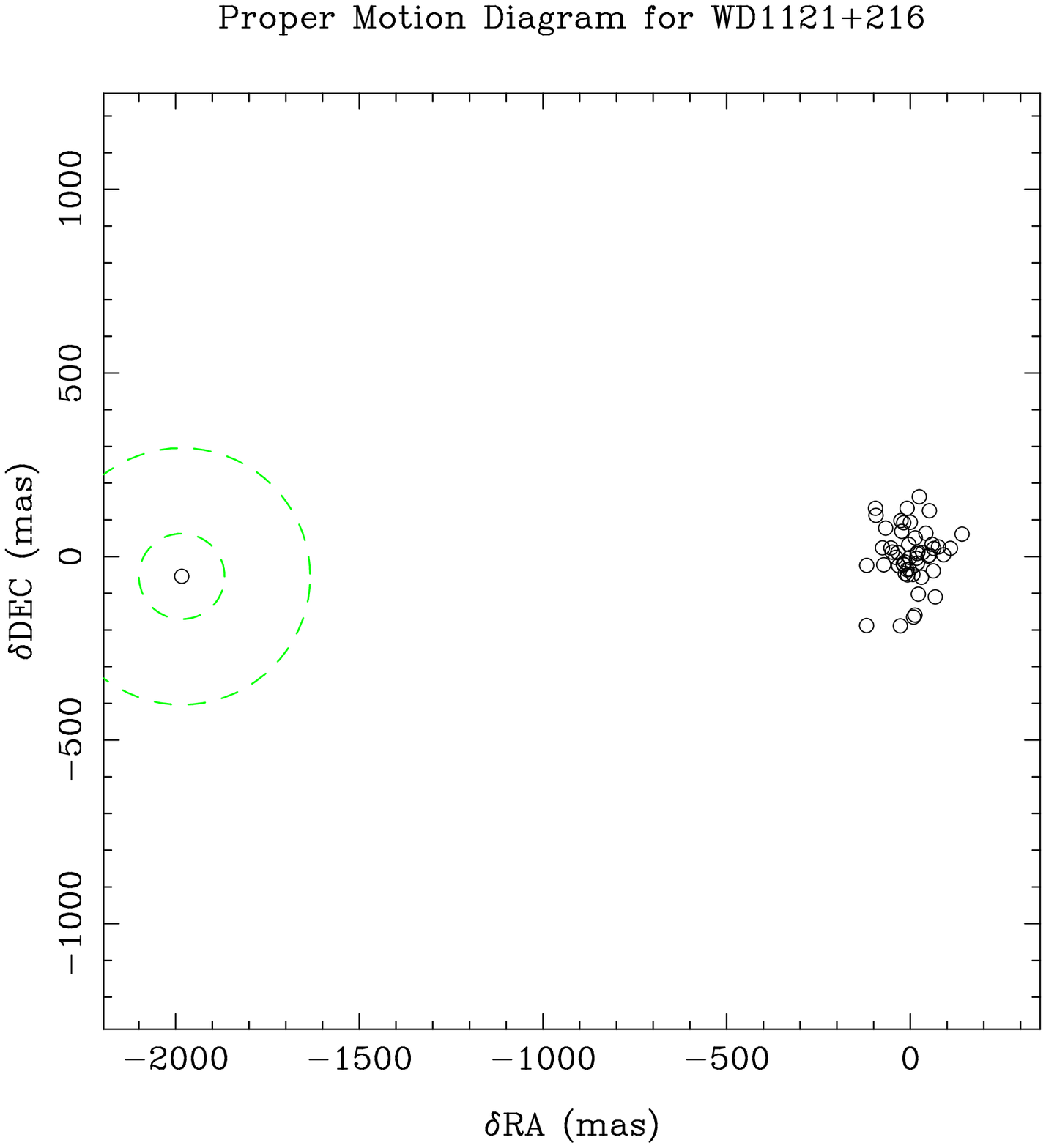}}
\caption{The completeness limit (left) and the proper motion diagram (right) for WD~$1121+216$.}
\label{WD1121_plots}
\end{figure*}

\begin{figure*}
\mbox{\includegraphics[bb=45 121 576 616,clip,angle=270,scale=0.370]{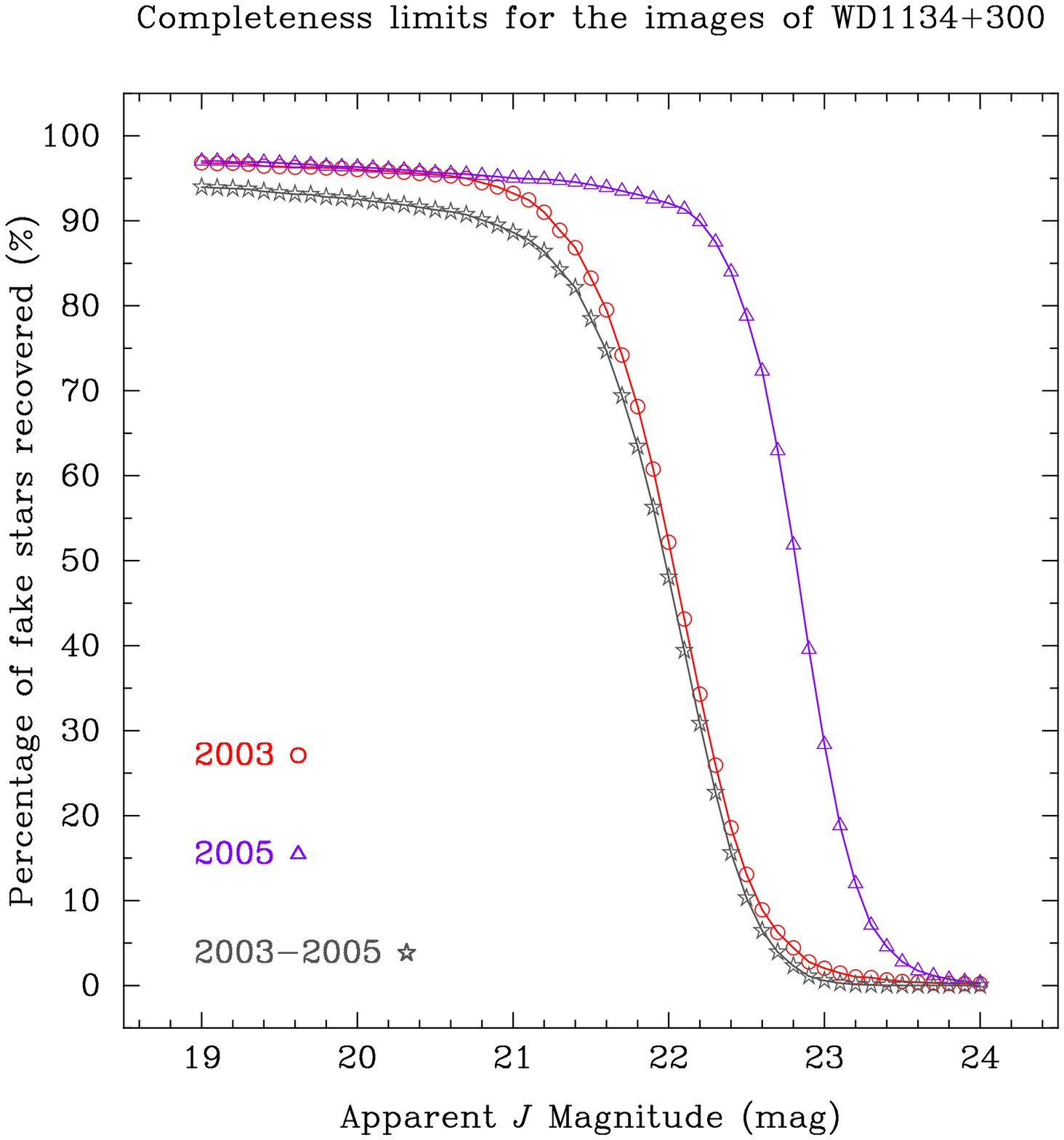}}
\mbox{\includegraphics[bb=45 130 579 616,clip,angle=270,scale=0.370]{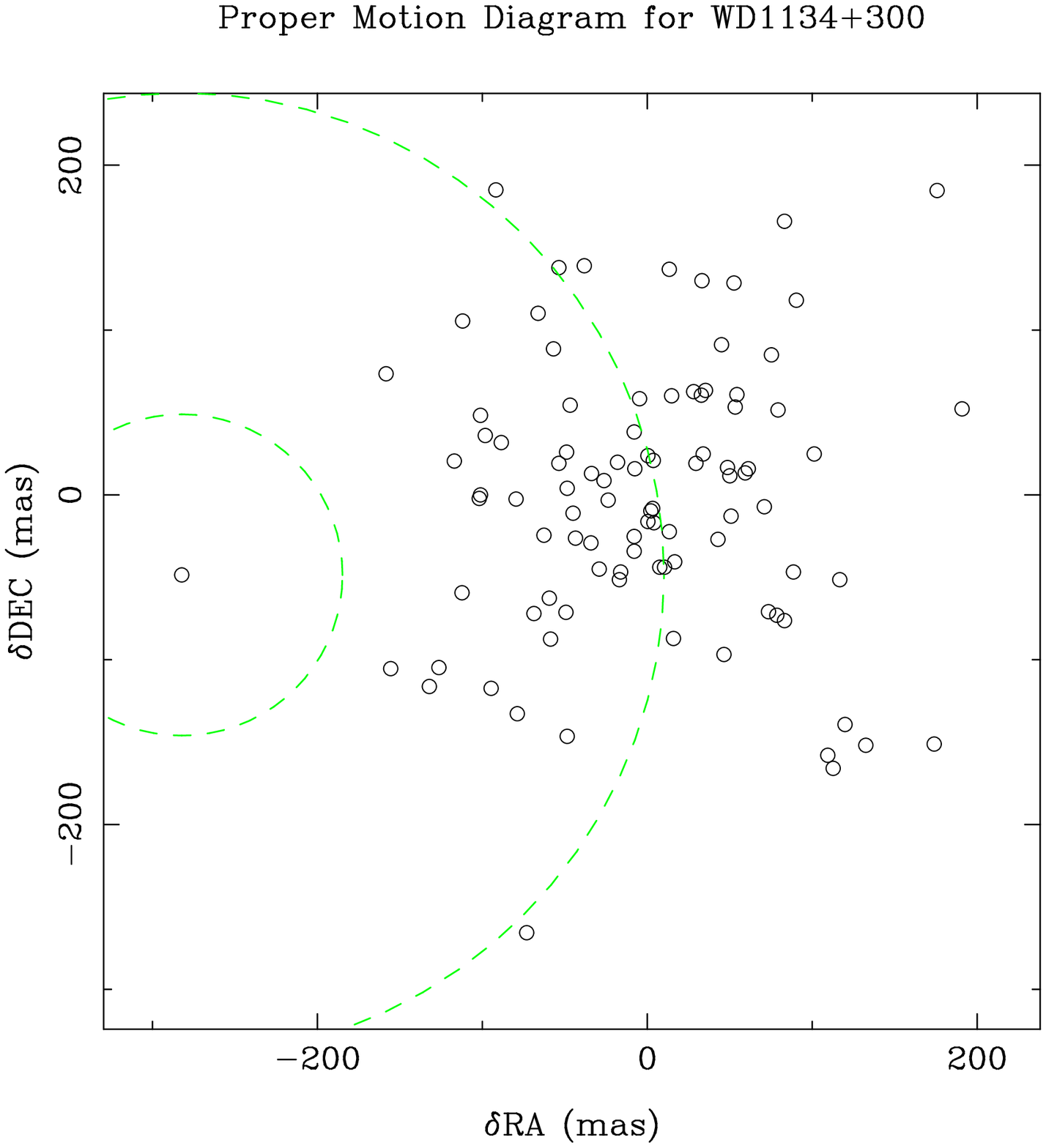}}
\caption{The completeness limit (left) and the proper motion diagram (right) for WD~$1134+300$.}
\label{WD1134_plots}
\end{figure*}
\clearpage

\begin{figure*}
\mbox{\includegraphics[bb=45 121 576 616,clip,angle=270,scale=0.370]{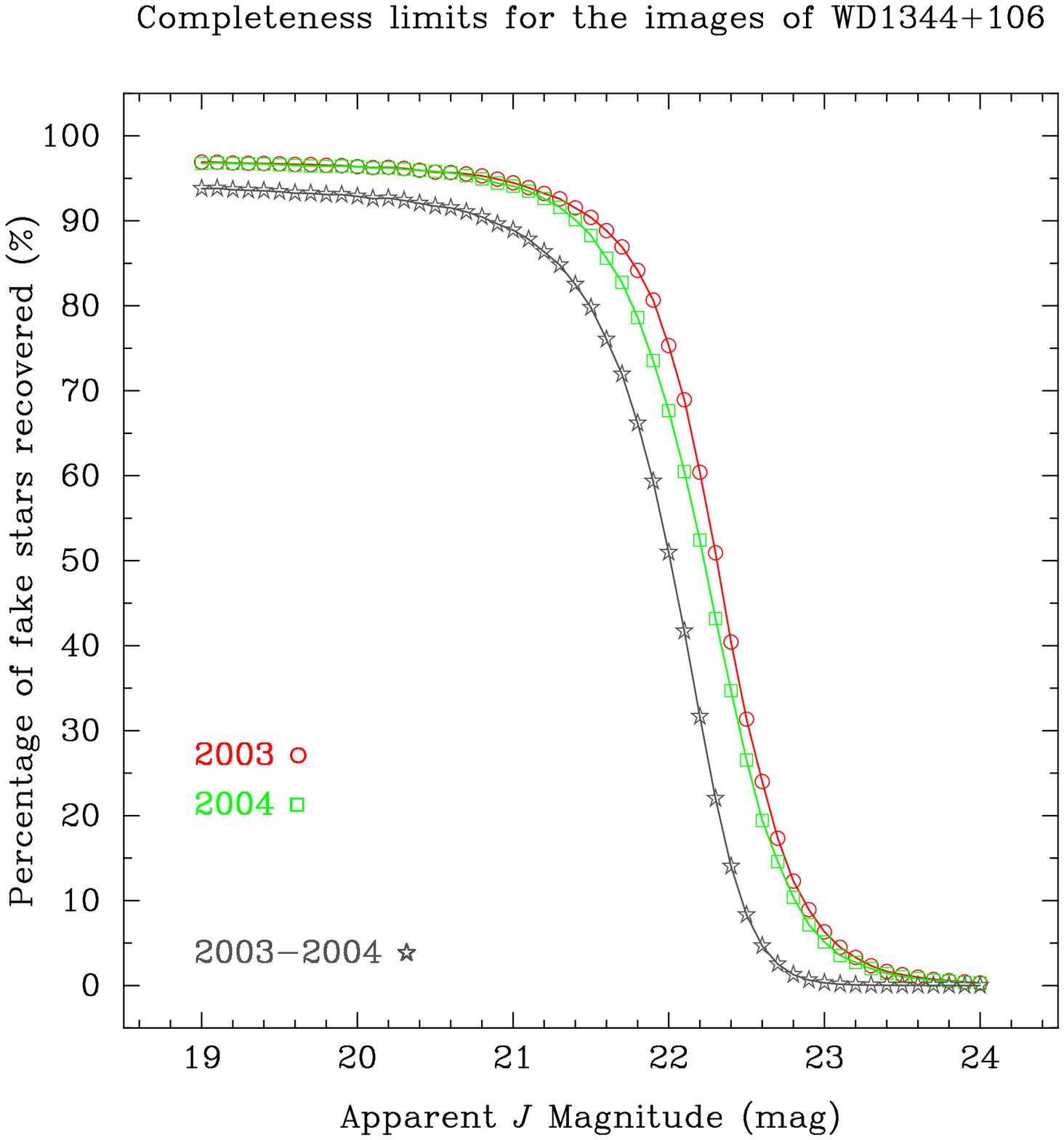}}
\mbox{\includegraphics[bb=45 130 579 616,clip,angle=270,scale=0.370]{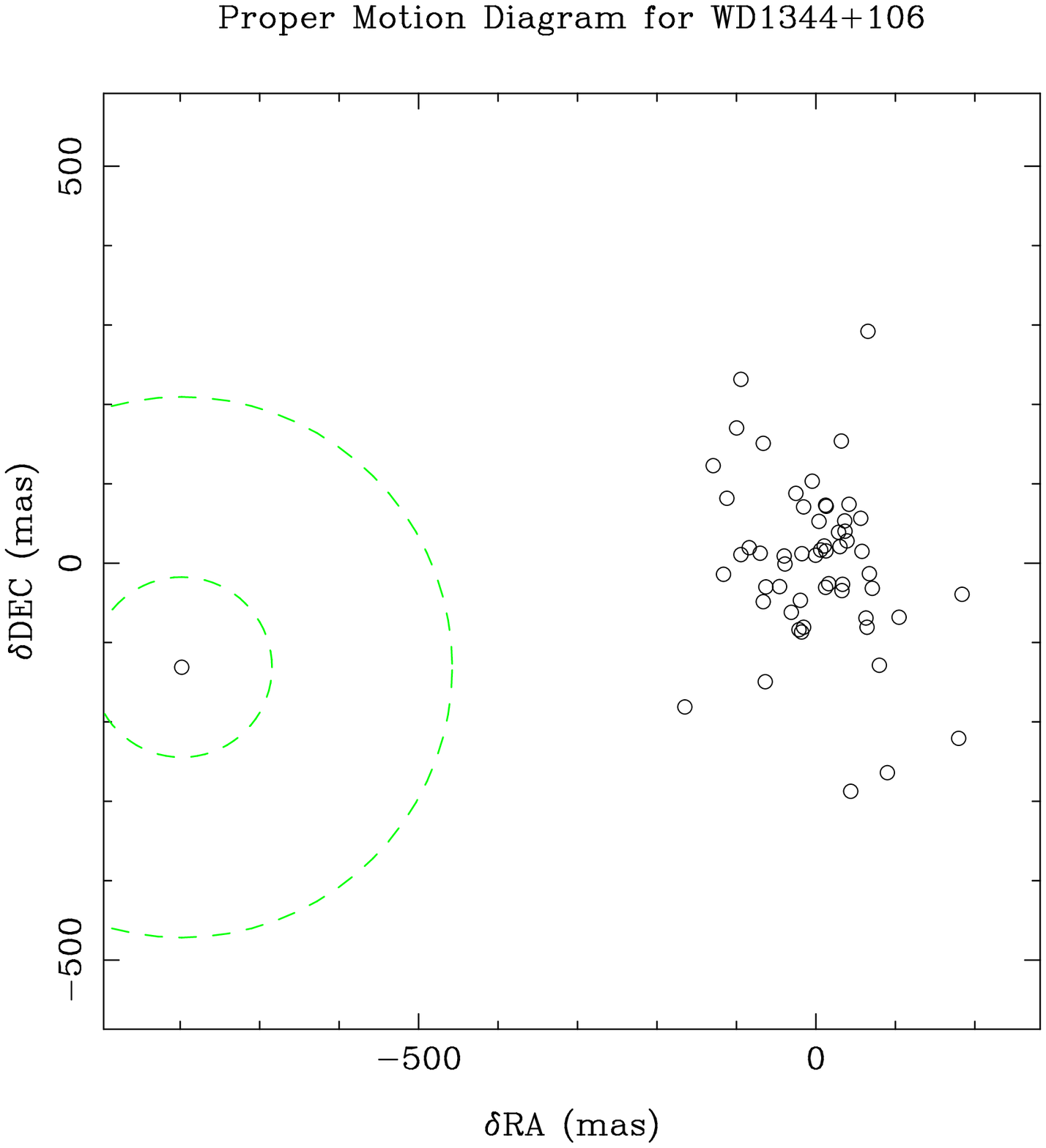}}
\caption{The completeness limit (left) and the proper motion diagram (right) for WD~$1344+106$.}
\label{WD1344_plots}
\end{figure*}

\begin{figure*}
\mbox{\includegraphics[bb=45 121 576 616,clip,angle=270,scale=0.370]{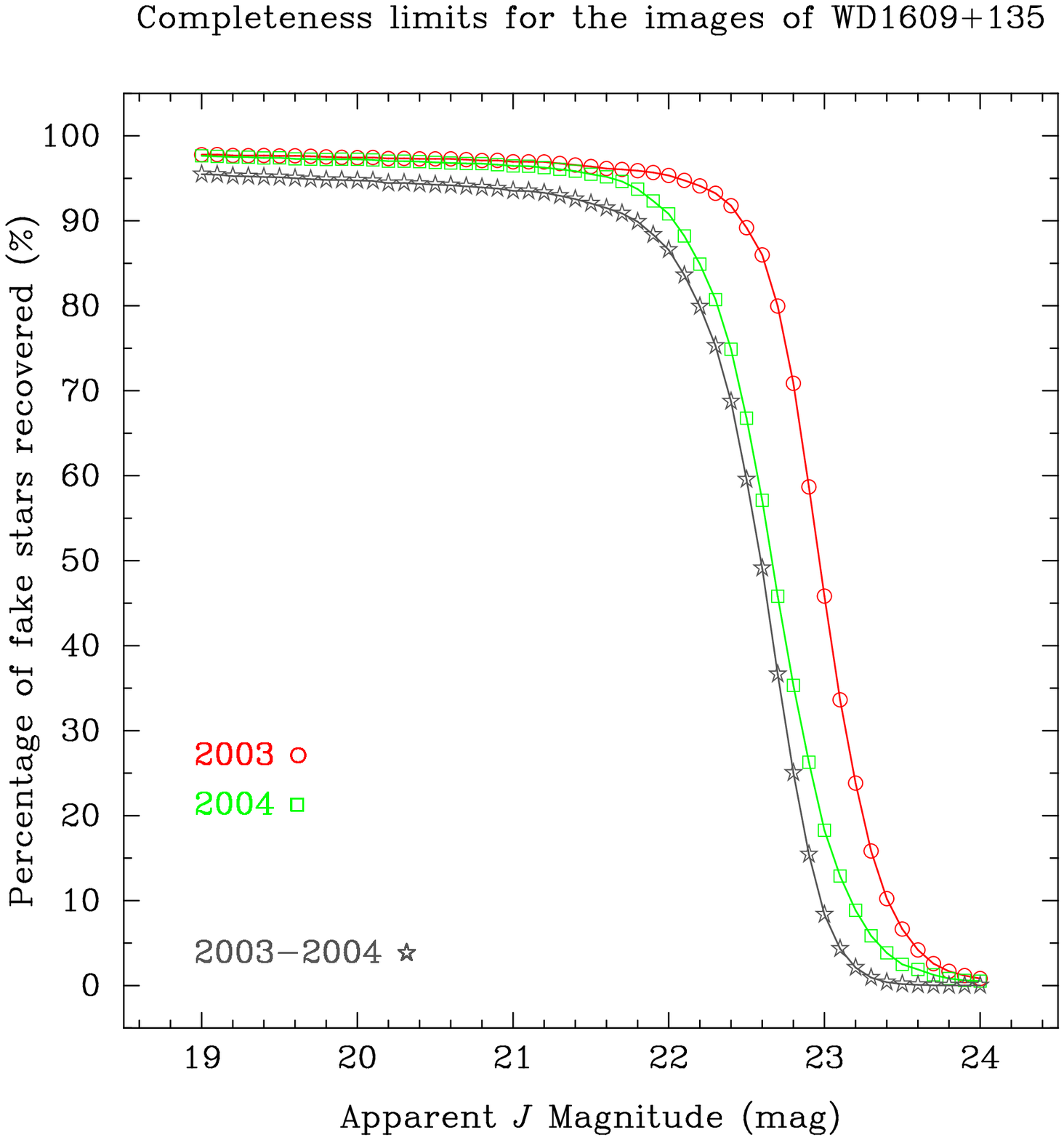}}
\mbox{\includegraphics[bb=45 130 579 616,clip,angle=270,scale=0.370]{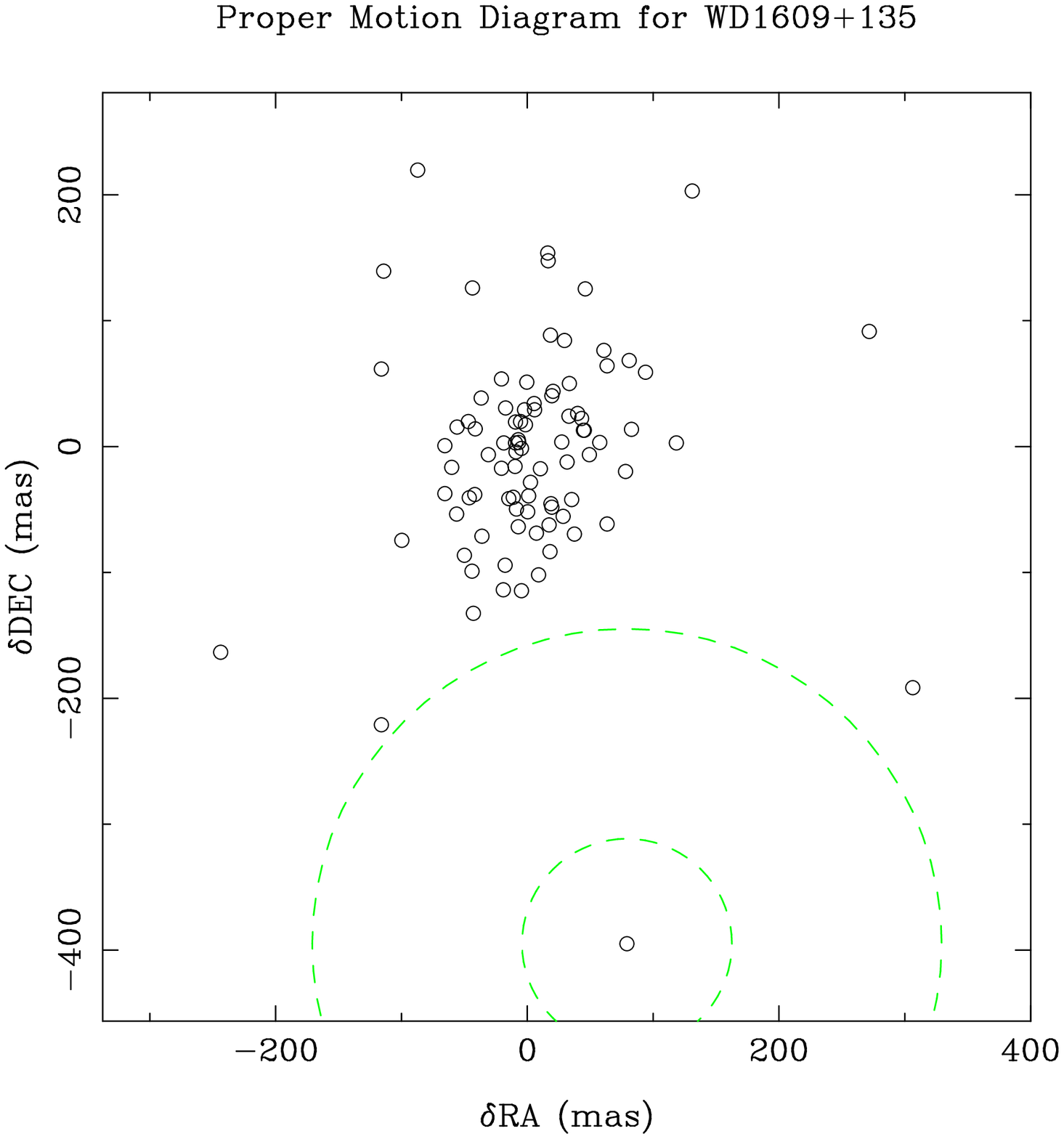}}
\caption{The completeness limit (left) and the proper motion diagram (right) for WD~$1609+135$.}
\label{WD1609_plots}
\end{figure*}

\begin{figure*}
\mbox{\includegraphics[bb=45 121 576 616,clip,angle=270,scale=0.370]{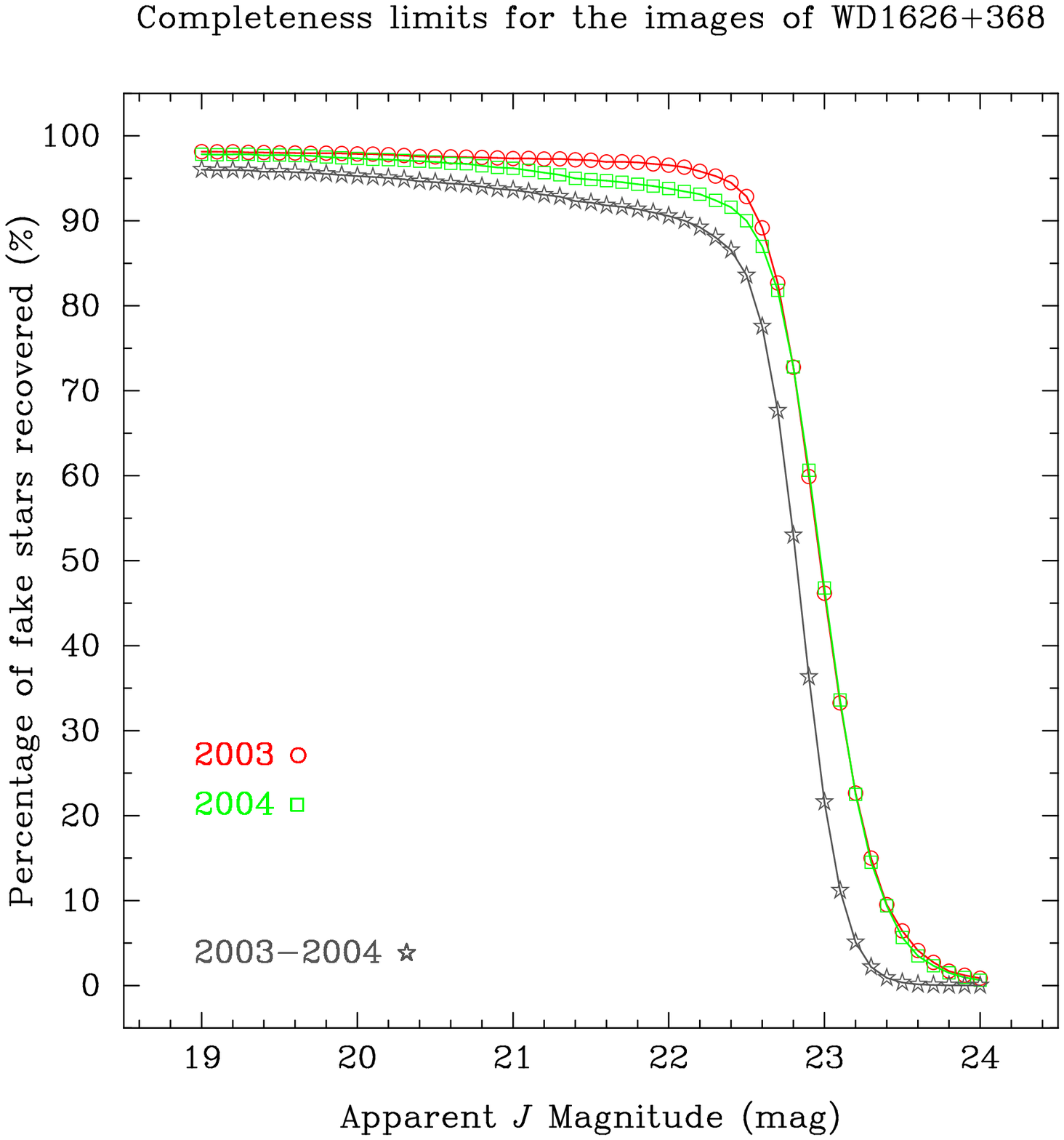}}
\mbox{\includegraphics[bb=45 130 579 616,clip,angle=270,scale=0.370]{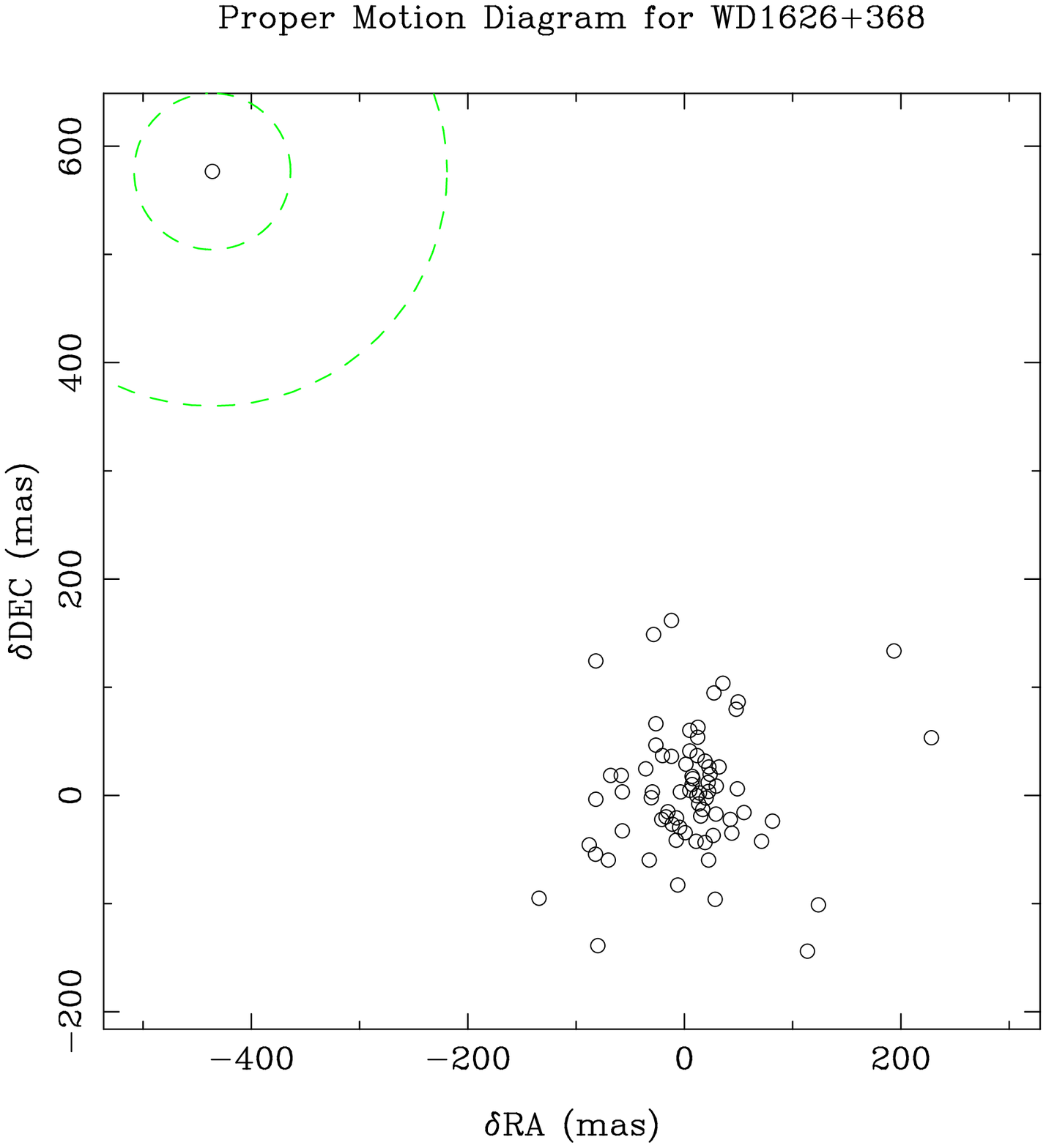}}
\caption{The completeness limit (left) and the proper motion diagram (right) for WD~$1626+368$.}
\label{WD1626_plots}
\end{figure*}
\clearpage

\begin{figure*}
\mbox{\includegraphics[bb=45 121 576 616,clip,angle=270,scale=0.370]{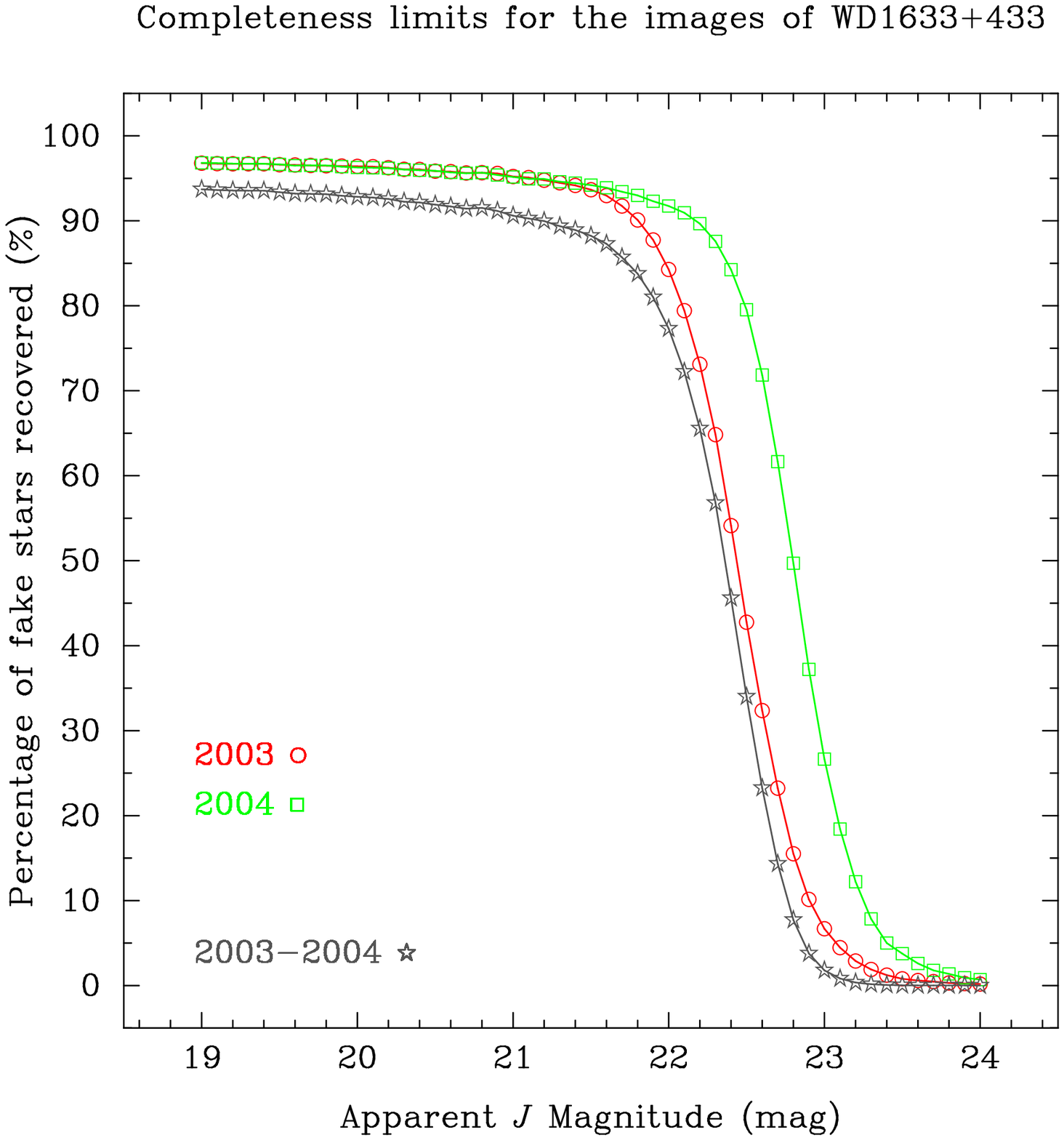}}
\mbox{\includegraphics[bb=45 130 579 616,clip,angle=270,scale=0.370]{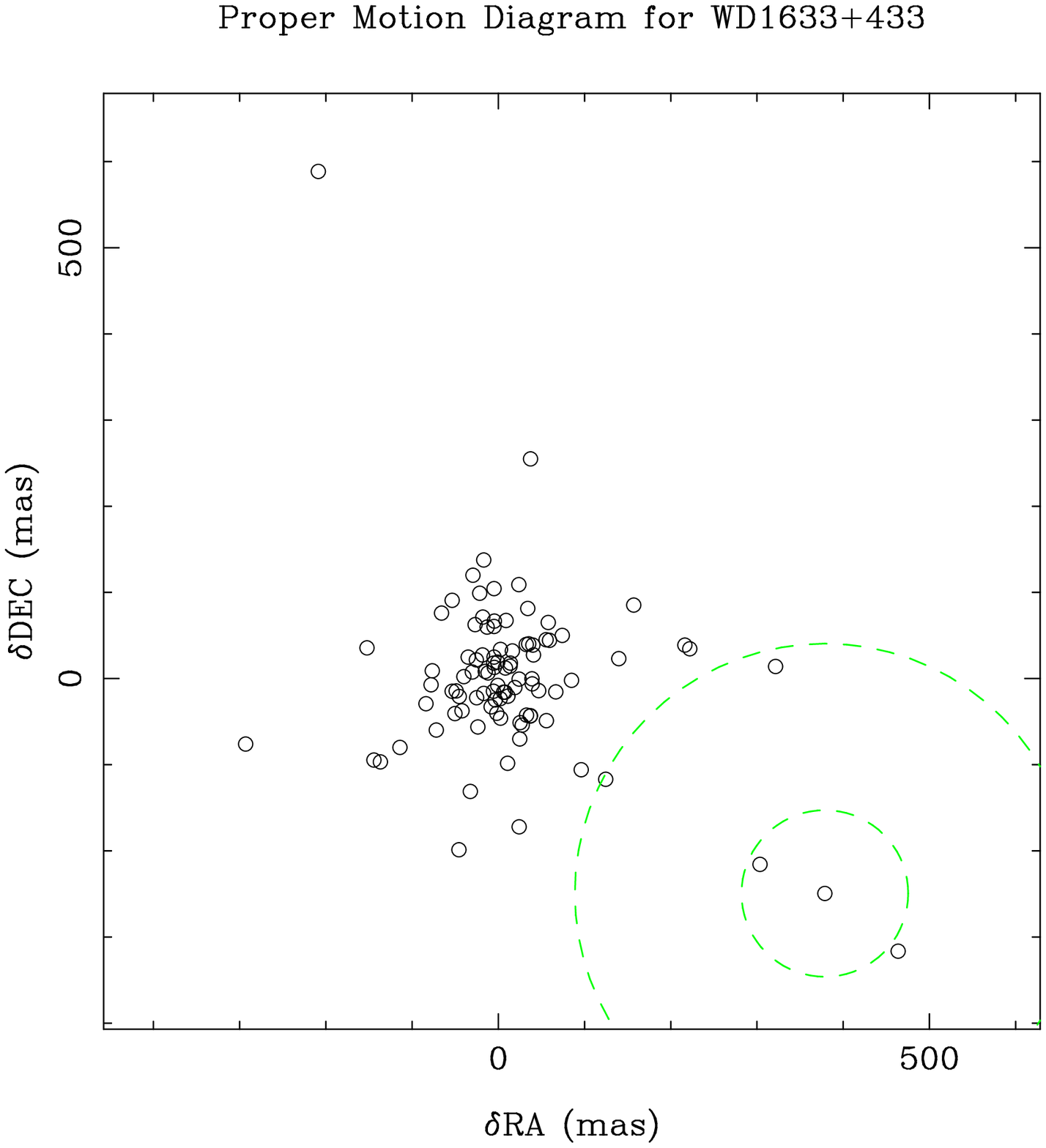}}
\caption{The completeness limit (left) and the proper motion diagram (right) for WD~$1633+433$.}
\label{WD1633_plots}
\end{figure*}

\begin{figure*}
\mbox{\includegraphics[bb=45 121 576 616,clip,angle=270,scale=0.370]{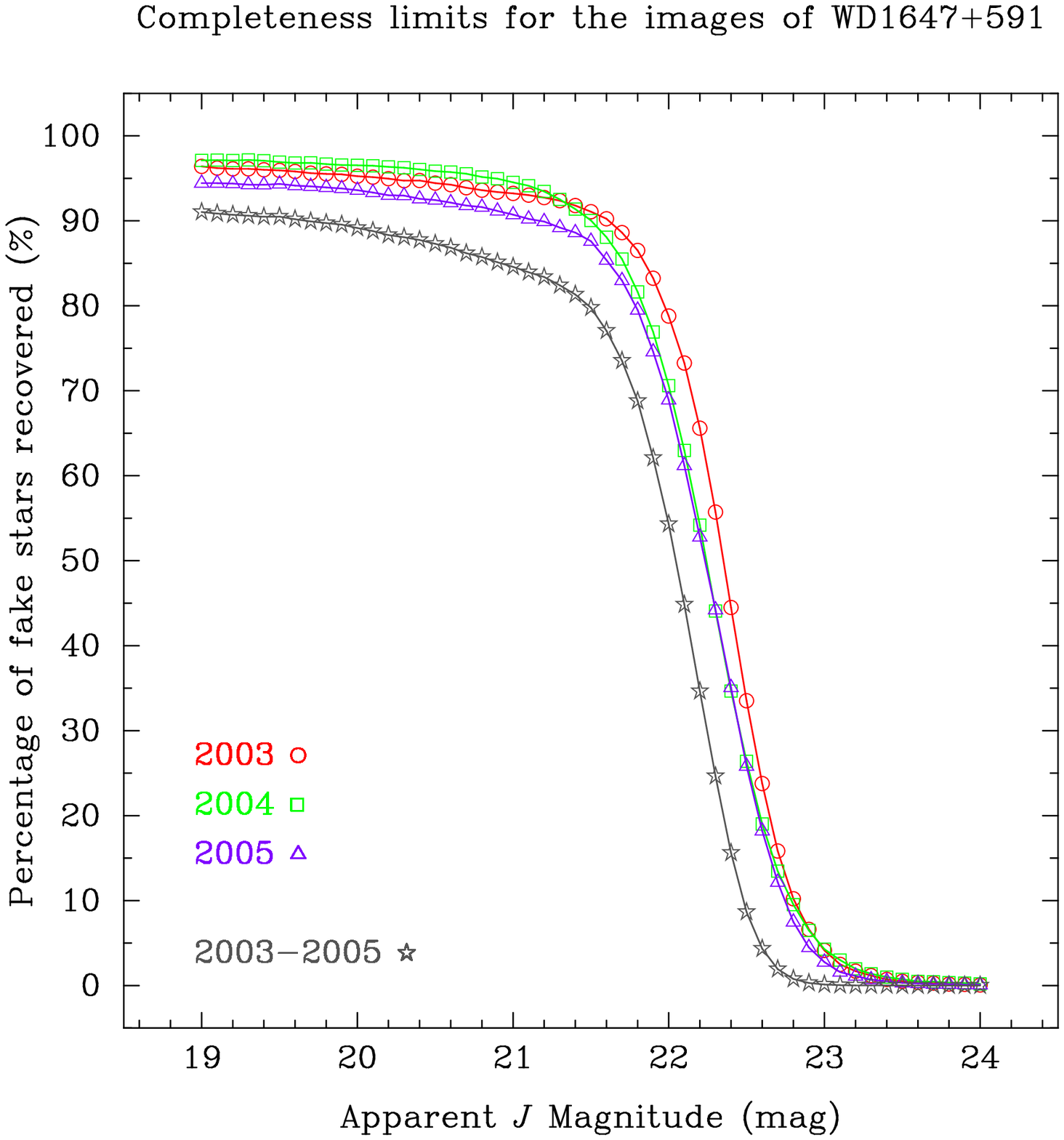}}
\mbox{\includegraphics[bb=45 130 579 616,clip,angle=270,scale=0.370]{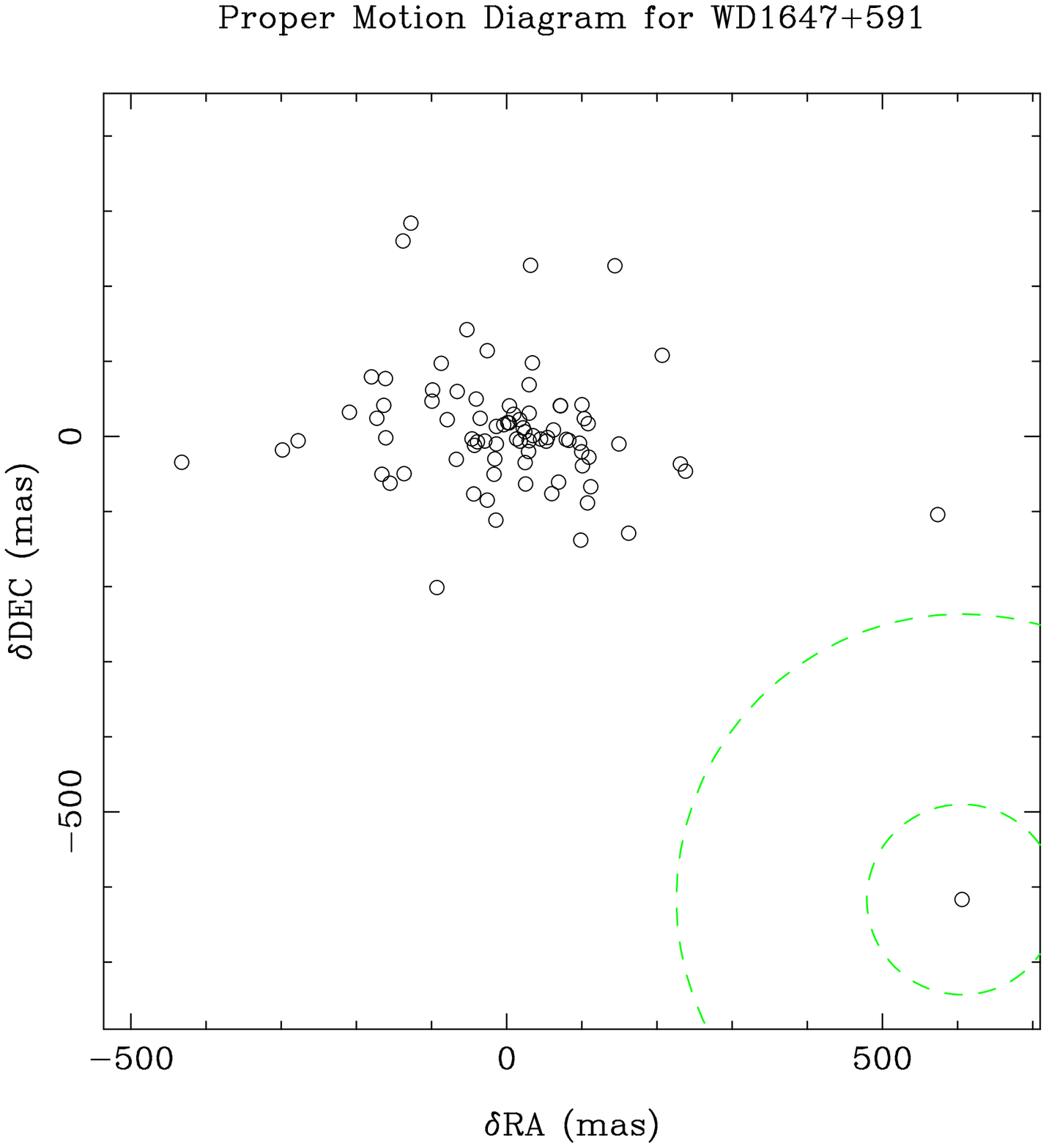}}
\caption{The completeness limit (left) and the proper motion diagram (right) for WD~$1647+591$.}
\label{WD1647_plots}
\end{figure*}

\begin{figure*}
\mbox{\includegraphics[bb=45 121 576 616,clip,angle=270,scale=0.370]{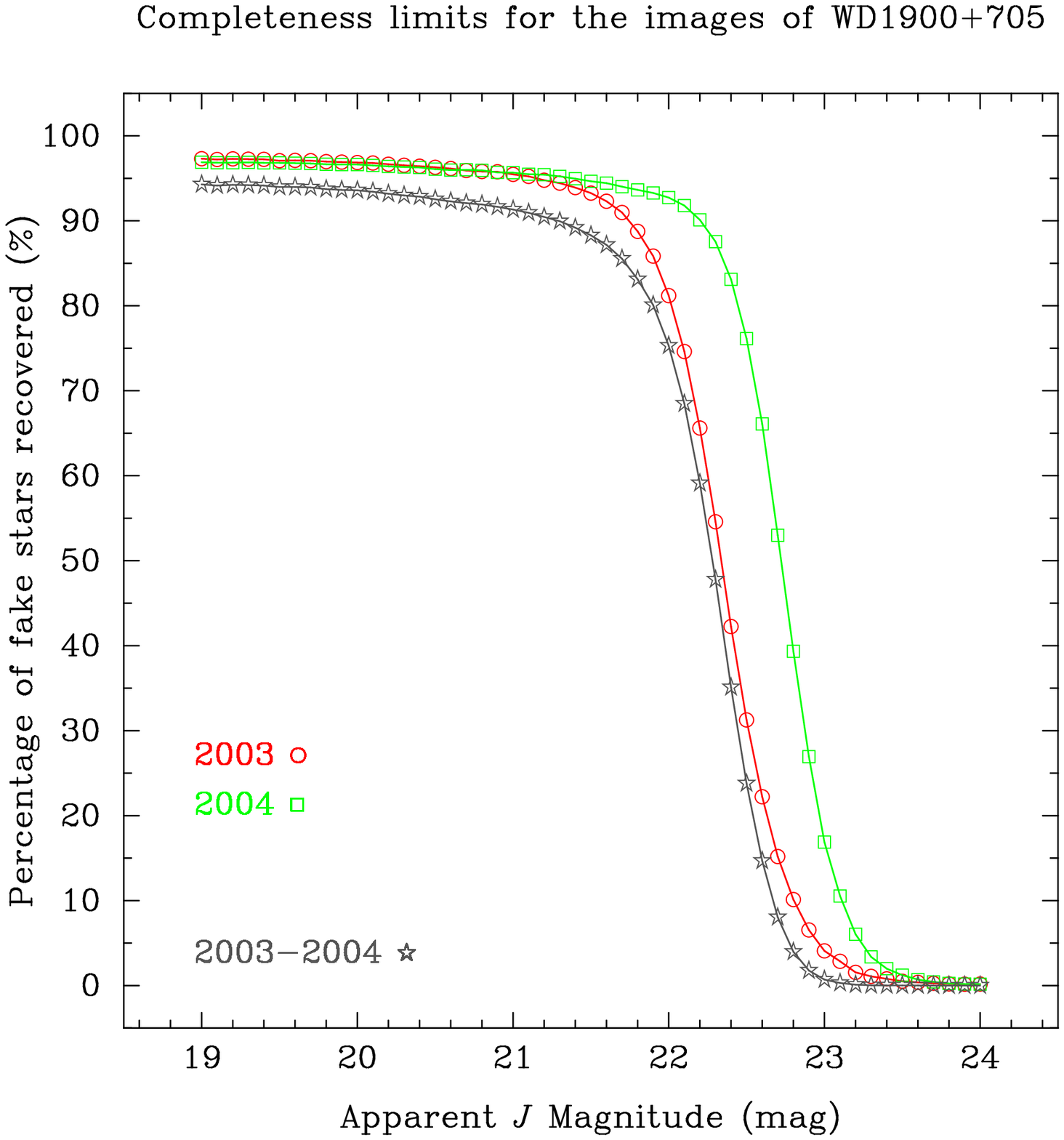}}
\mbox{\includegraphics[bb=45 130 579 616,clip,angle=270,scale=0.370]{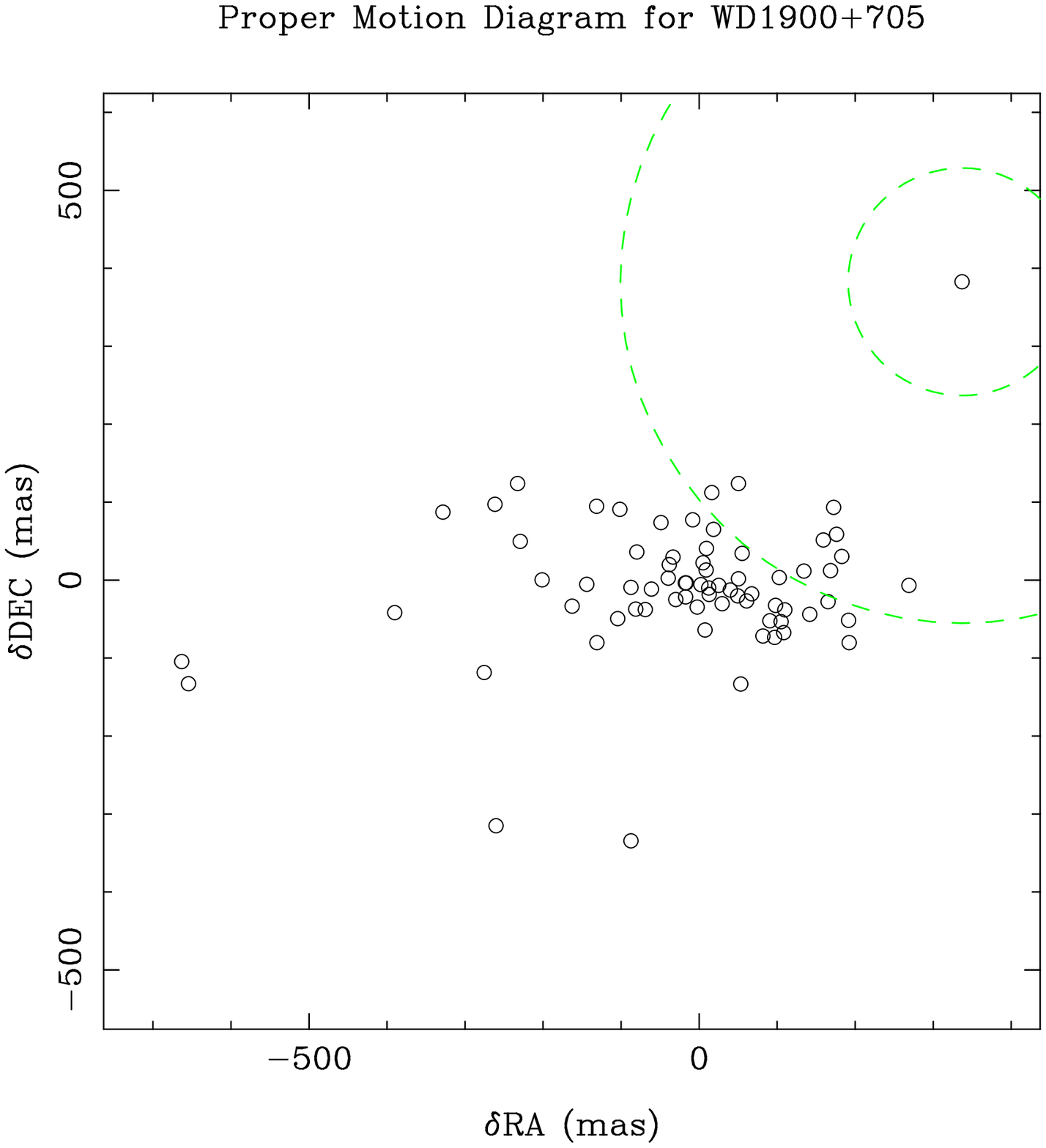}}
\caption{The completeness limit (left) and the proper motion diagram (right) for WD~$1900+705$.}
\label{WD1900_plots}
\end{figure*}
\clearpage

\begin{figure*}
\mbox{\includegraphics[bb=45 121 576 616,clip,angle=270,scale=0.370]{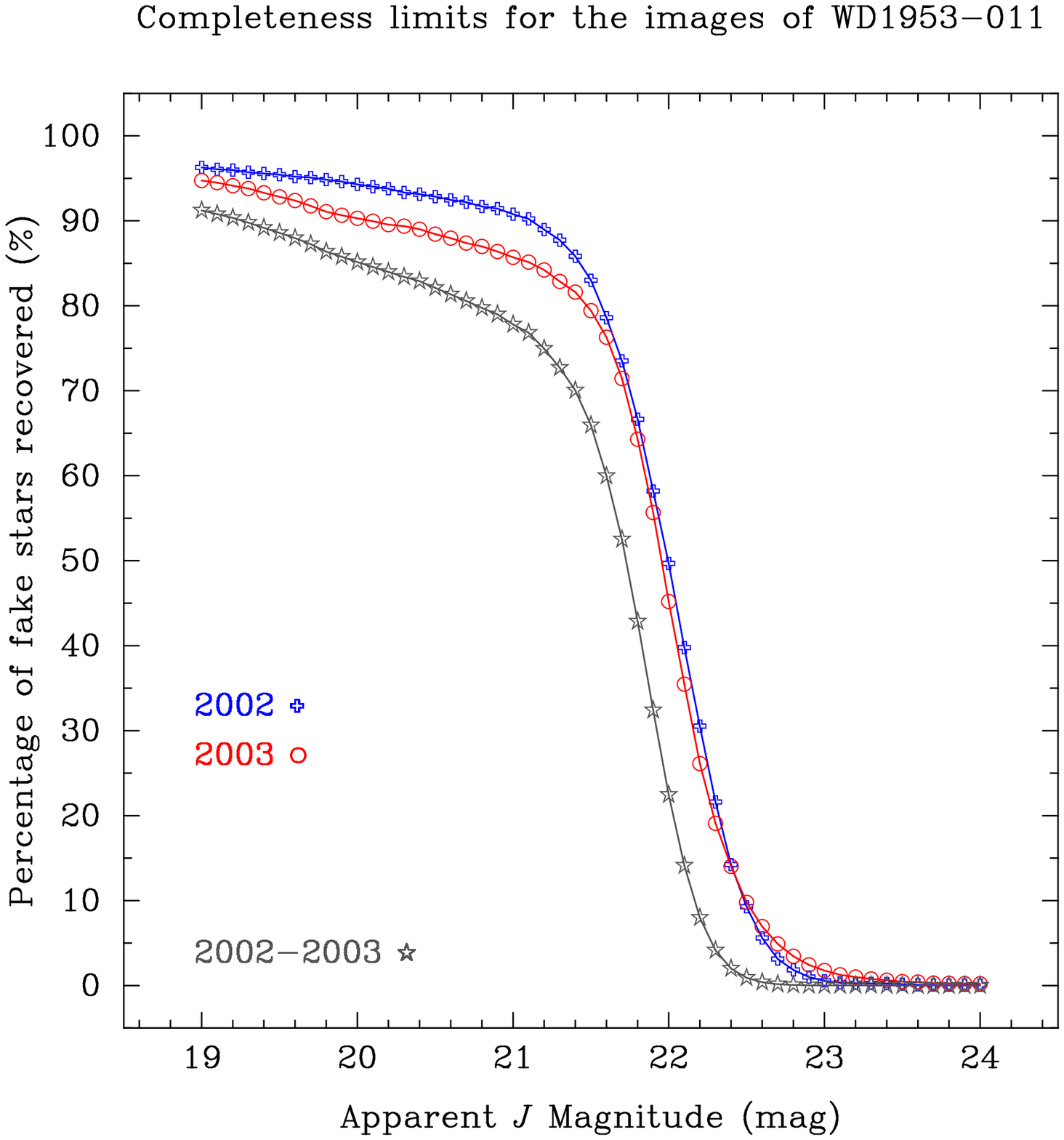}}
\mbox{\includegraphics[bb=45 130 579 616,clip,angle=270,scale=0.370]{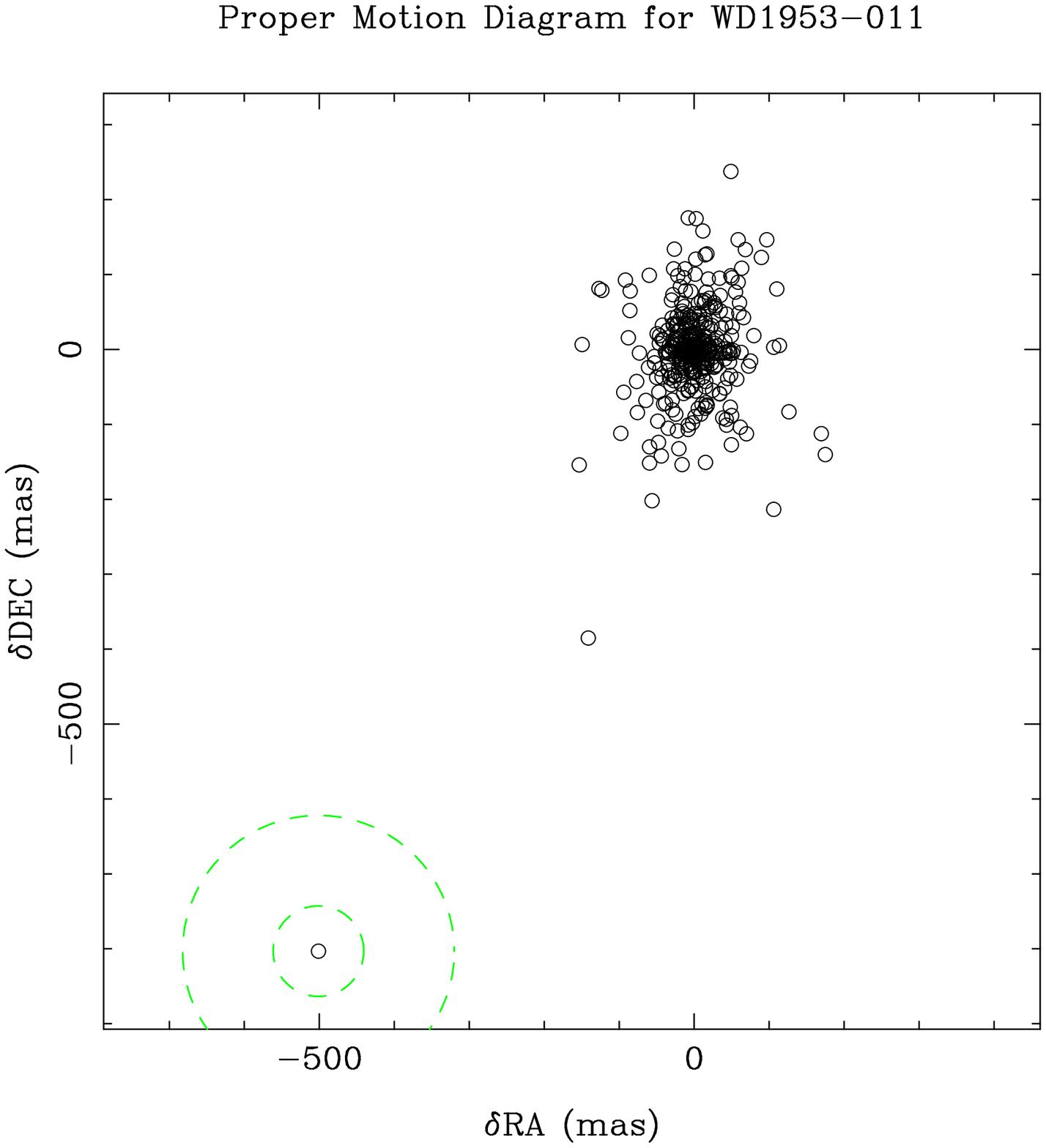}}
\caption{The completeness limit (left) and the proper motion diagram (right) for WD~$1953-011$.}
\label{WD1953_plots}
\end{figure*}

\begin{figure*}
\mbox{\includegraphics[bb=45 121 576 616,clip,angle=270,scale=0.370]{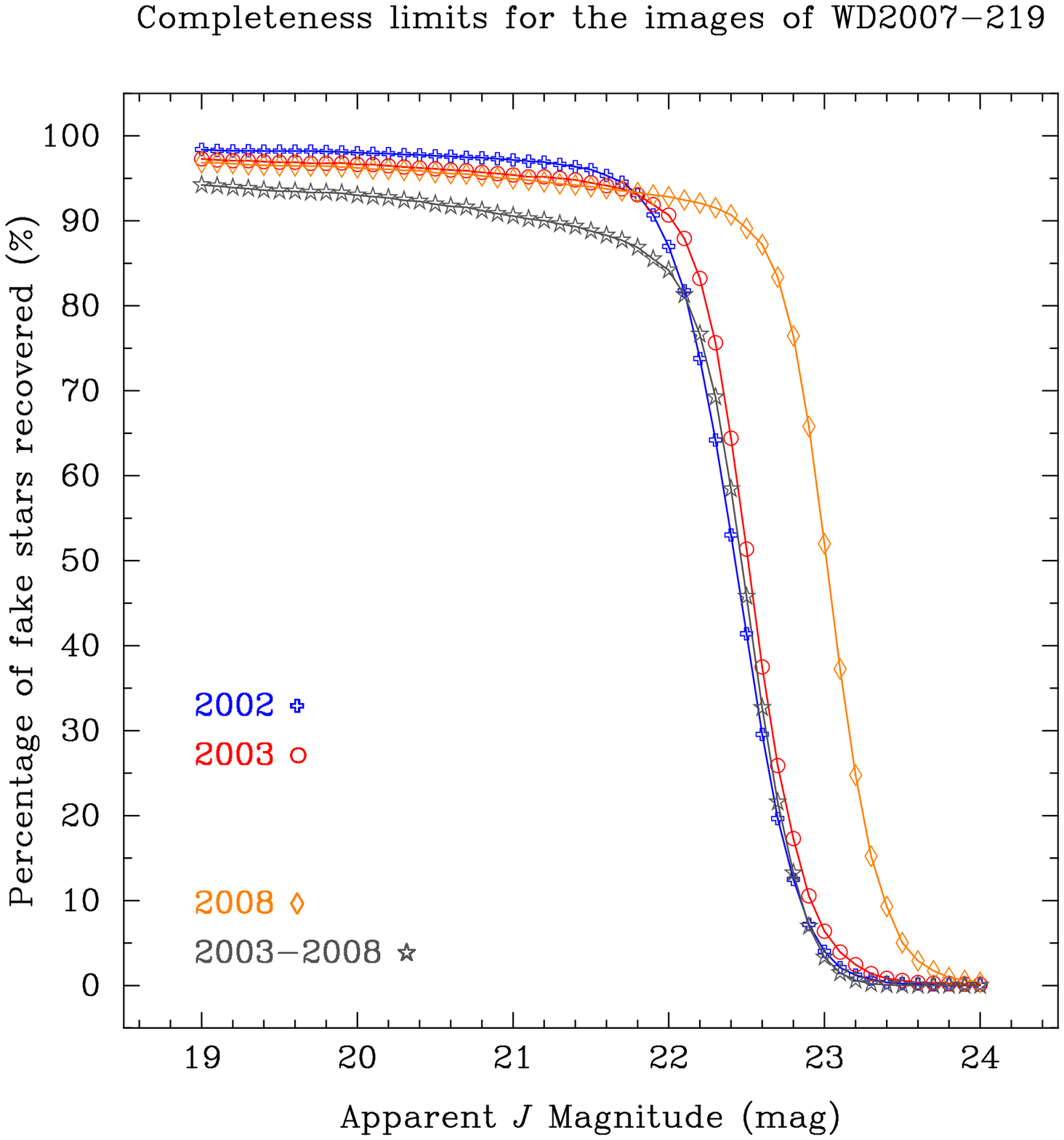}}
\mbox{\includegraphics[bb=45 130 579 616,clip,angle=270,scale=0.370]{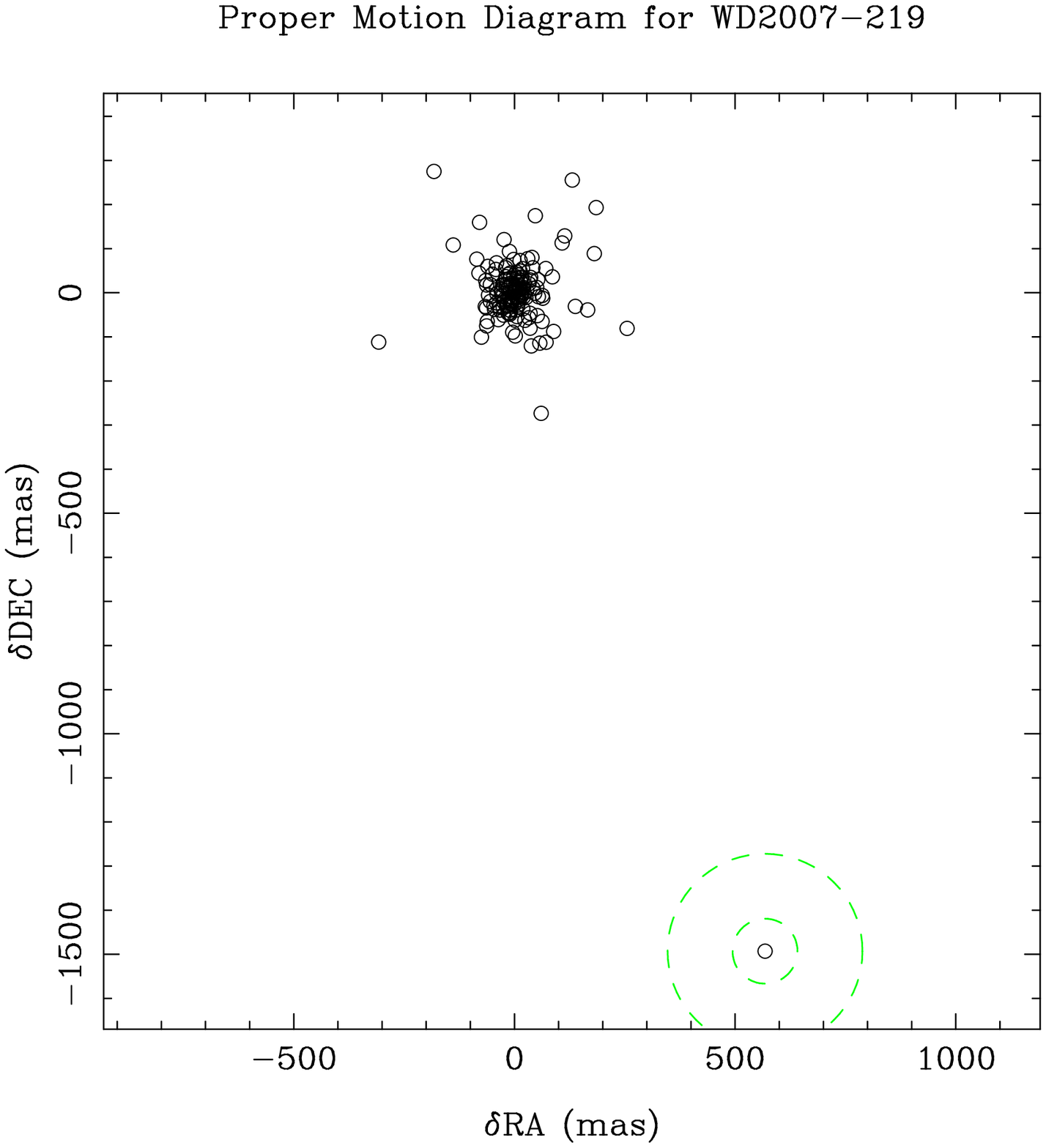}}
\caption{The completeness limit (left) and the proper motion diagram (right) for WD~$2007-219$.}
\label{WD2007_plots}
\end{figure*}

\begin{figure*}
\mbox{\includegraphics[bb=45 121 576 616,clip,angle=270,scale=0.370]{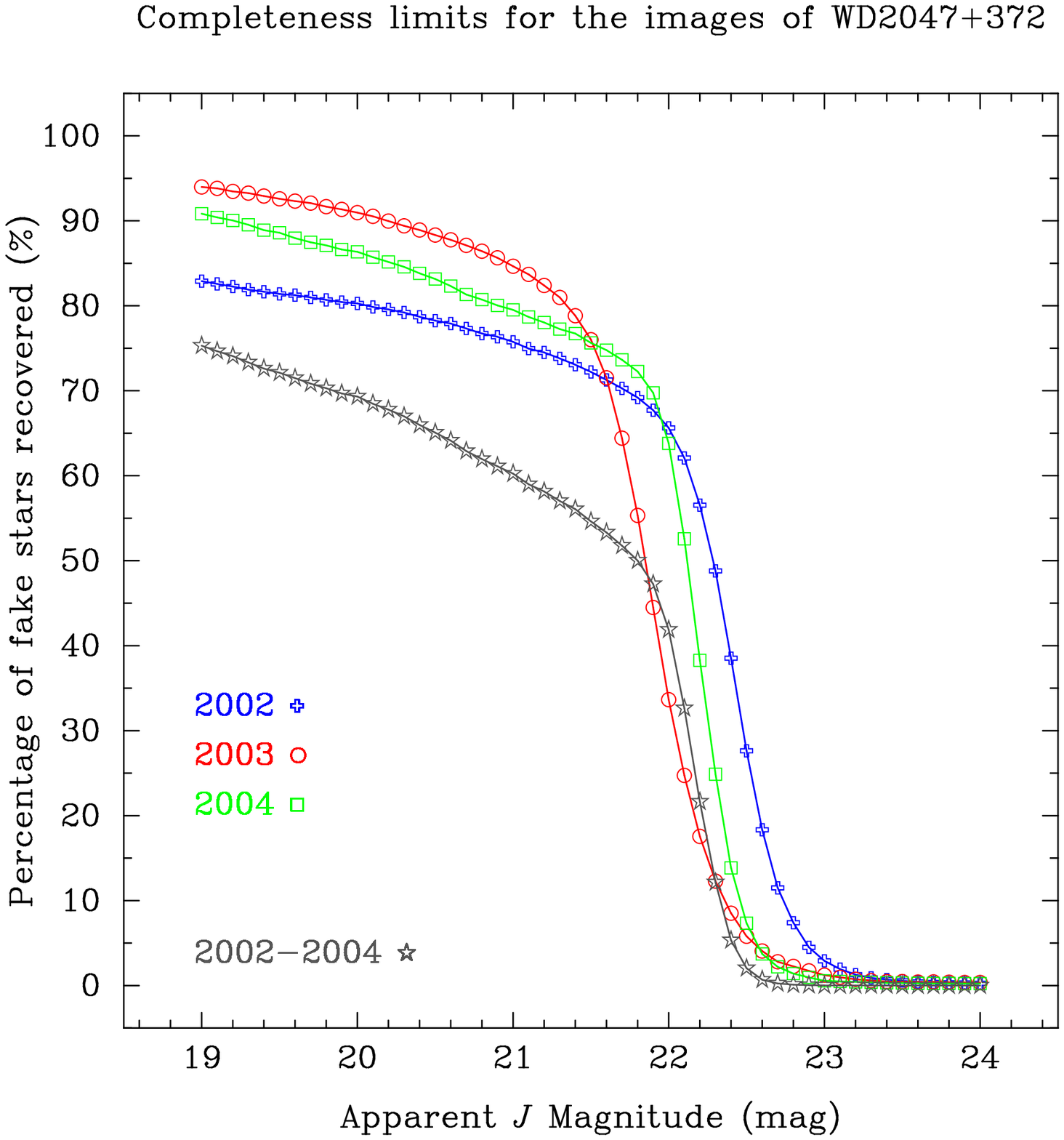}}
\mbox{\includegraphics[bb=45 130 579 616,clip,angle=270,scale=0.370]{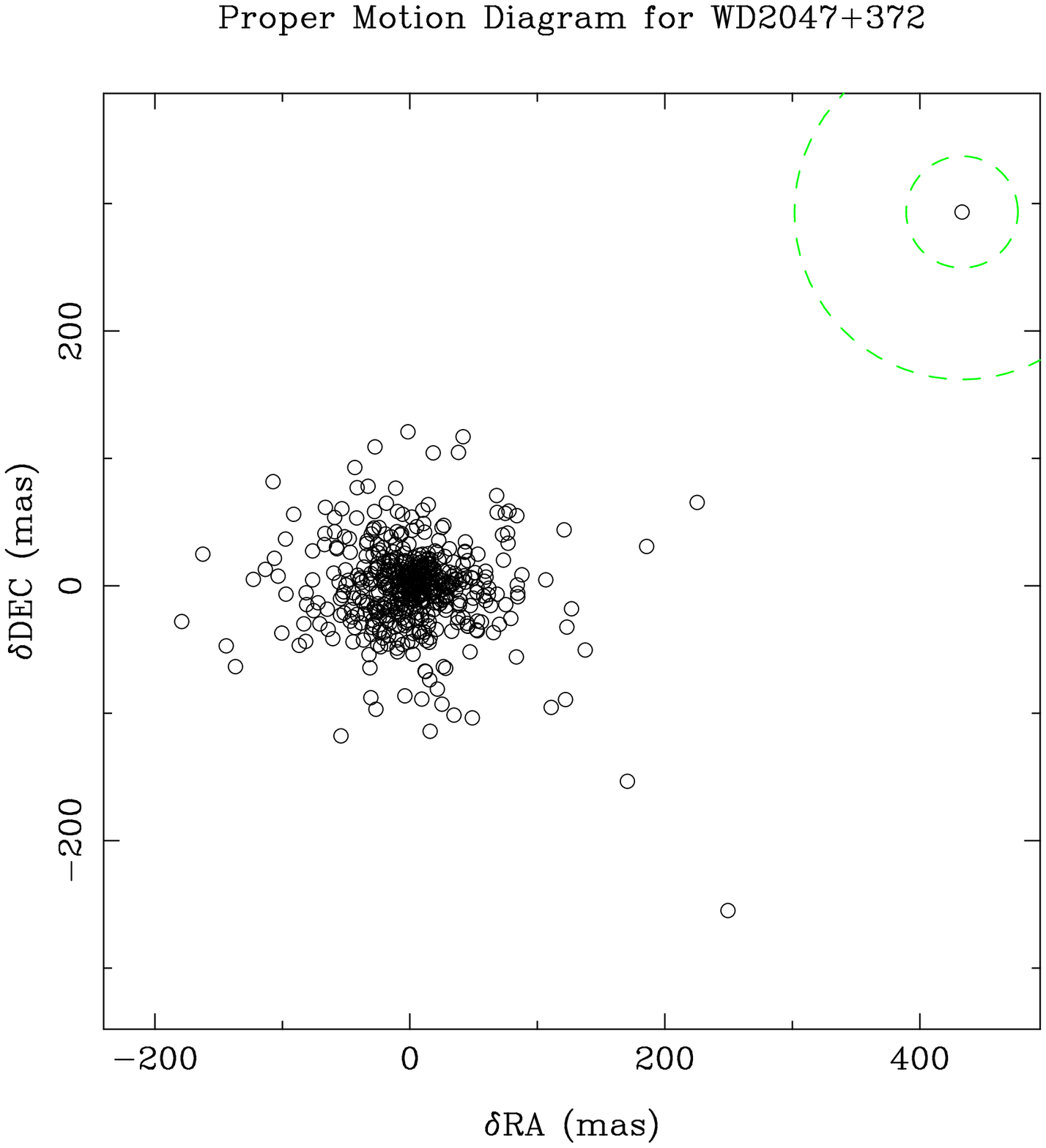}}
\caption{The completeness limit (left) and the proper motion diagram (right) for WD~$2047+372$.}
\label{WD2047_plots}
\end{figure*}
\clearpage

\begin{figure*}
\mbox{\includegraphics[bb=45 121 576 616,clip,angle=270,scale=0.370]{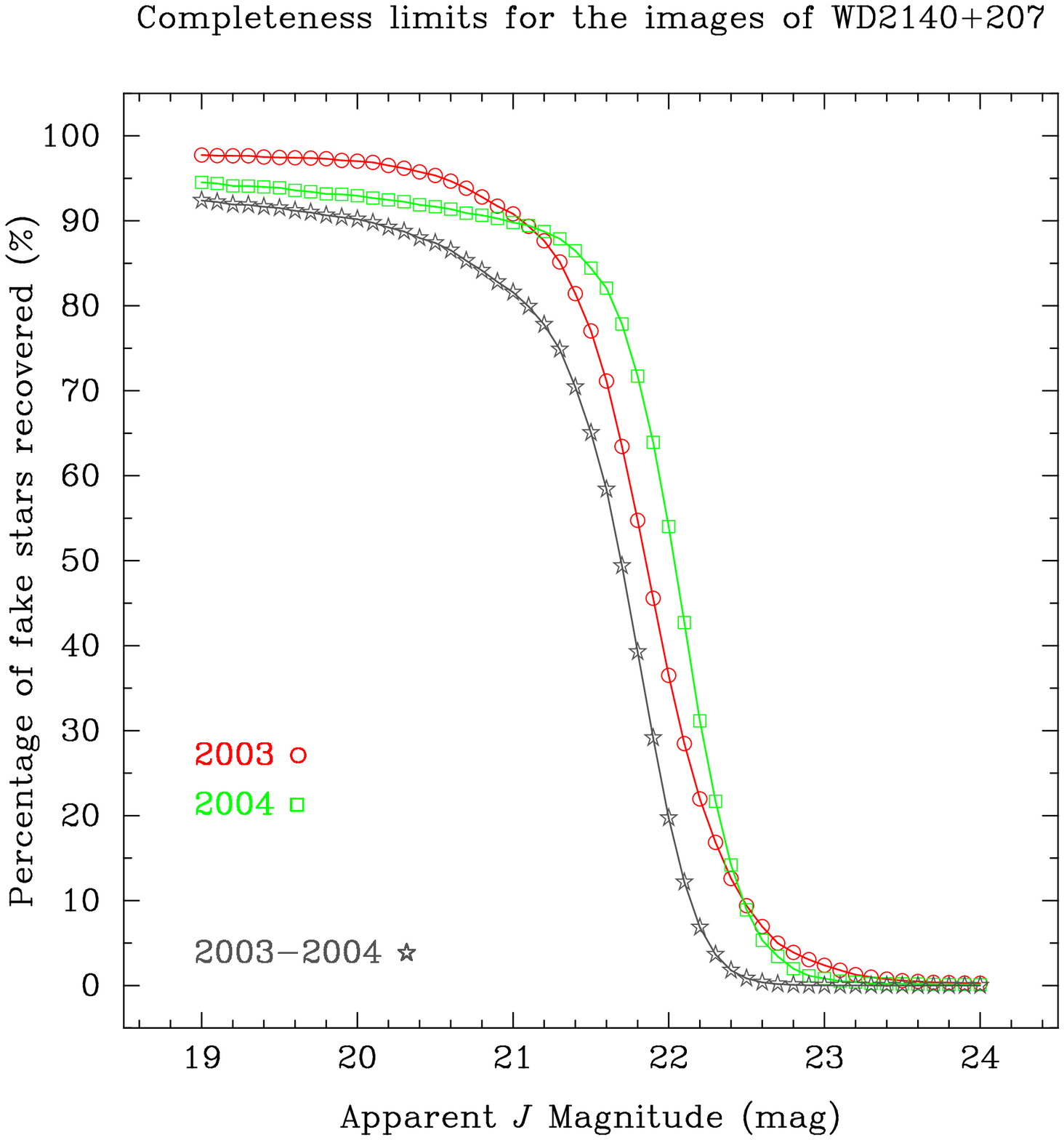}}
\mbox{\includegraphics[bb=45 130 579 616,clip,angle=270,scale=0.370]{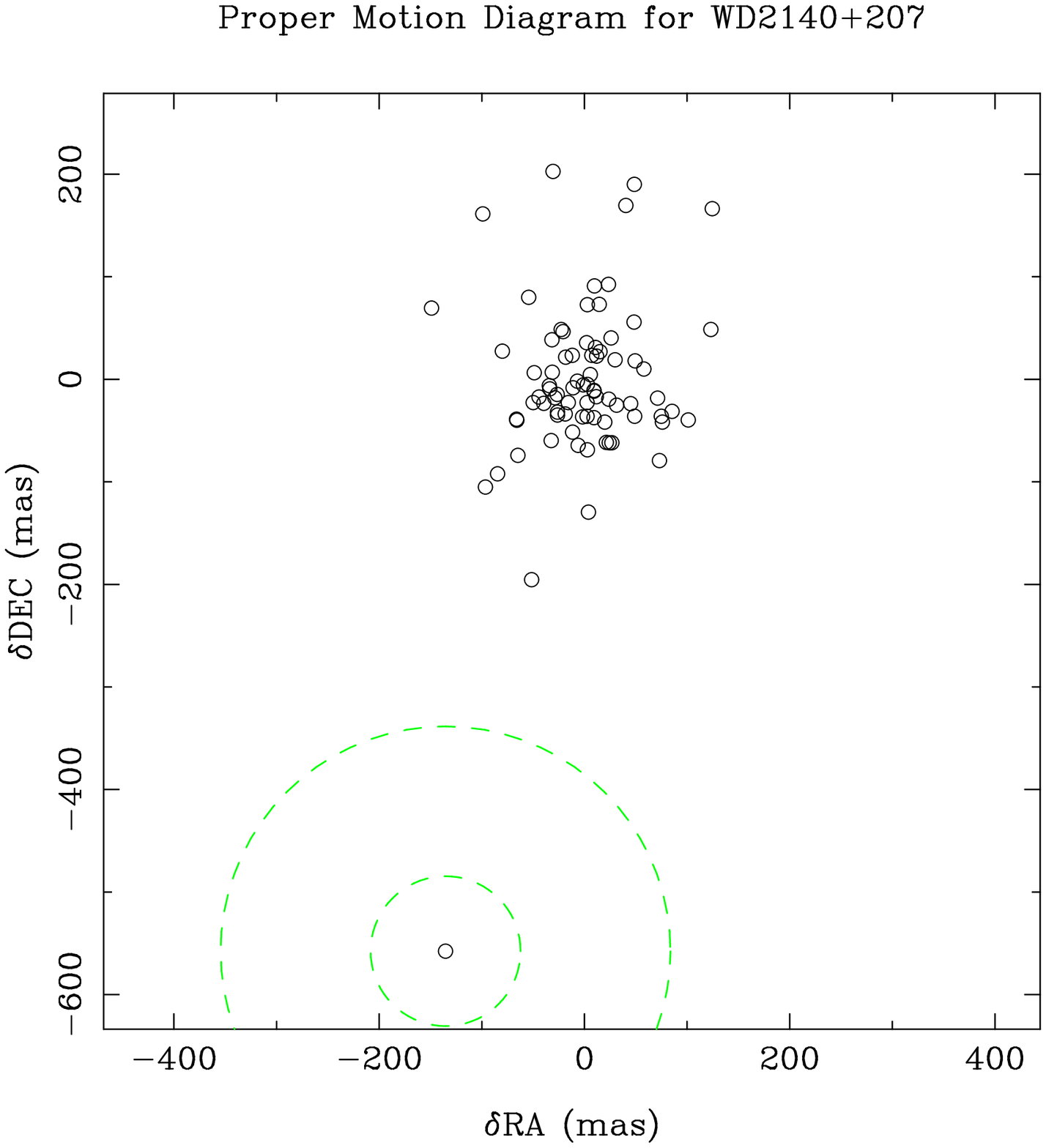}}
\caption{The completeness limit (left) and the proper motion diagram (right) for WD~$2140+207$.}
\label{WD2140_plots}
\end{figure*}

\begin{figure*}
\mbox{\includegraphics[bb=45 121 576 616,clip,angle=270,scale=0.370]{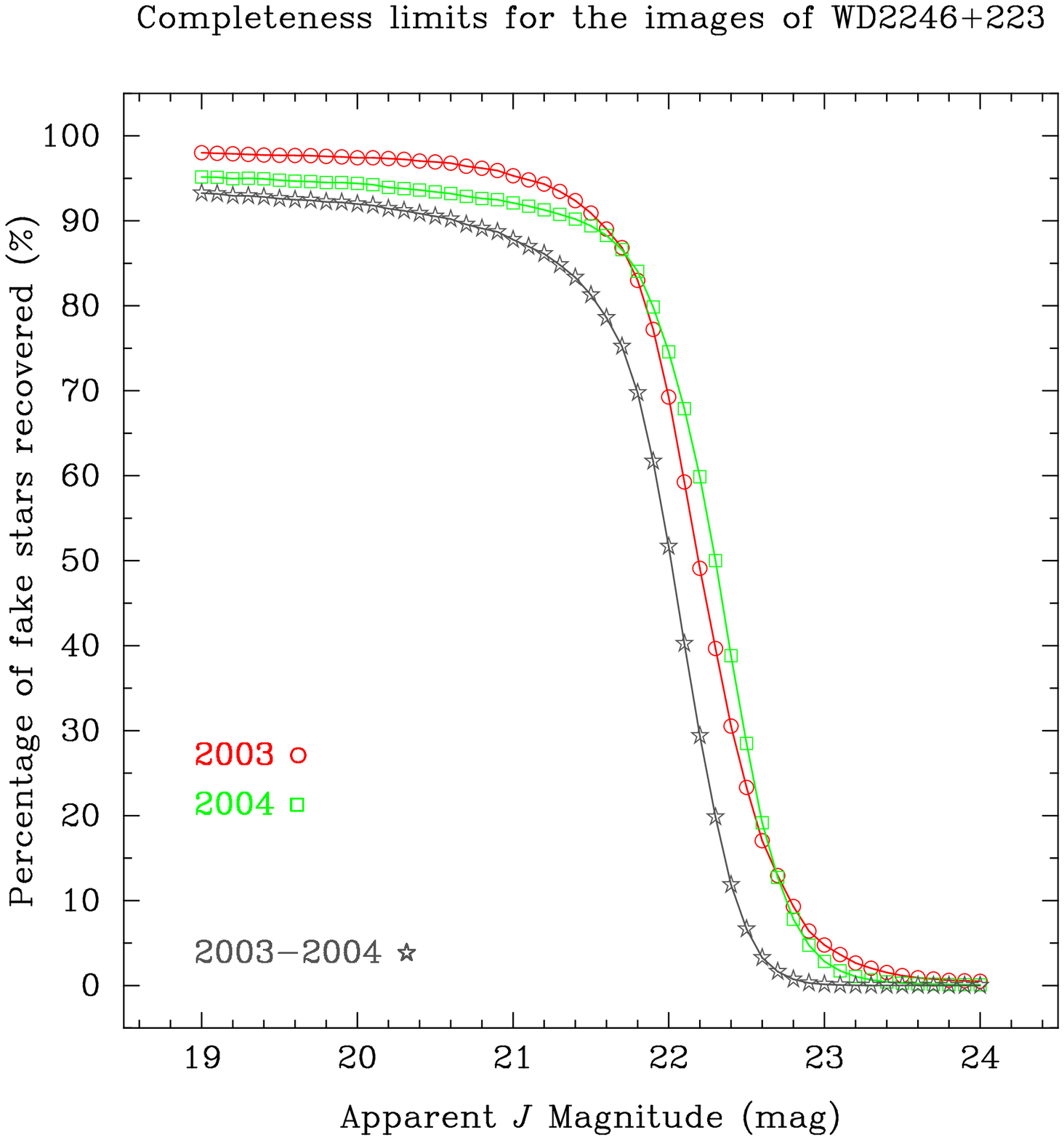}}
\mbox{\includegraphics[bb=45 130 579 616,clip,angle=270,scale=0.370]{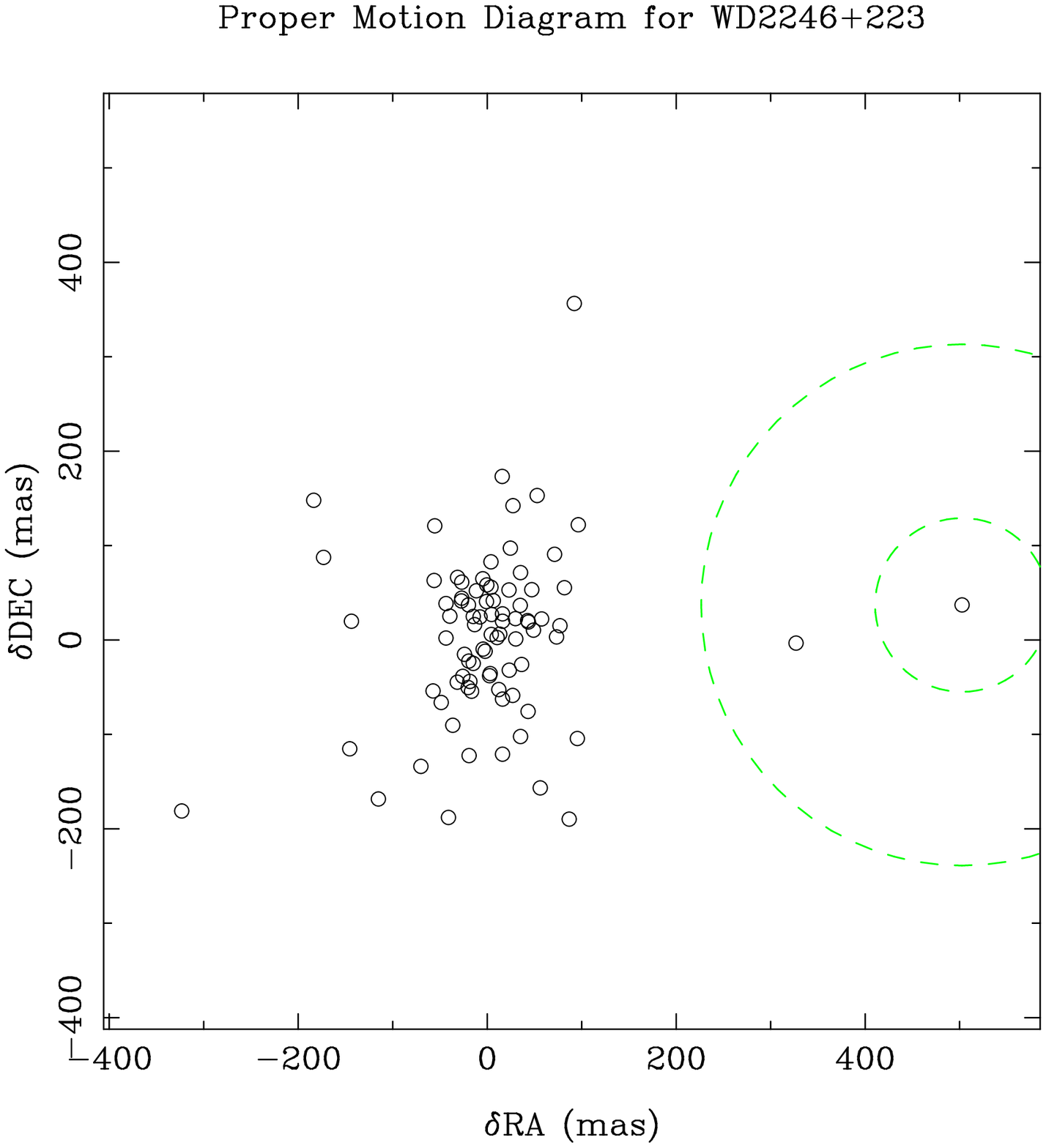}}
\caption{The completeness limit (left) and the proper motion diagram (right) for WD~$2246+223$.}
\label{WD2246_plots}
\end{figure*}

\begin{figure*}
\mbox{\includegraphics[bb=45 121 576 616,clip,angle=270,scale=0.370]{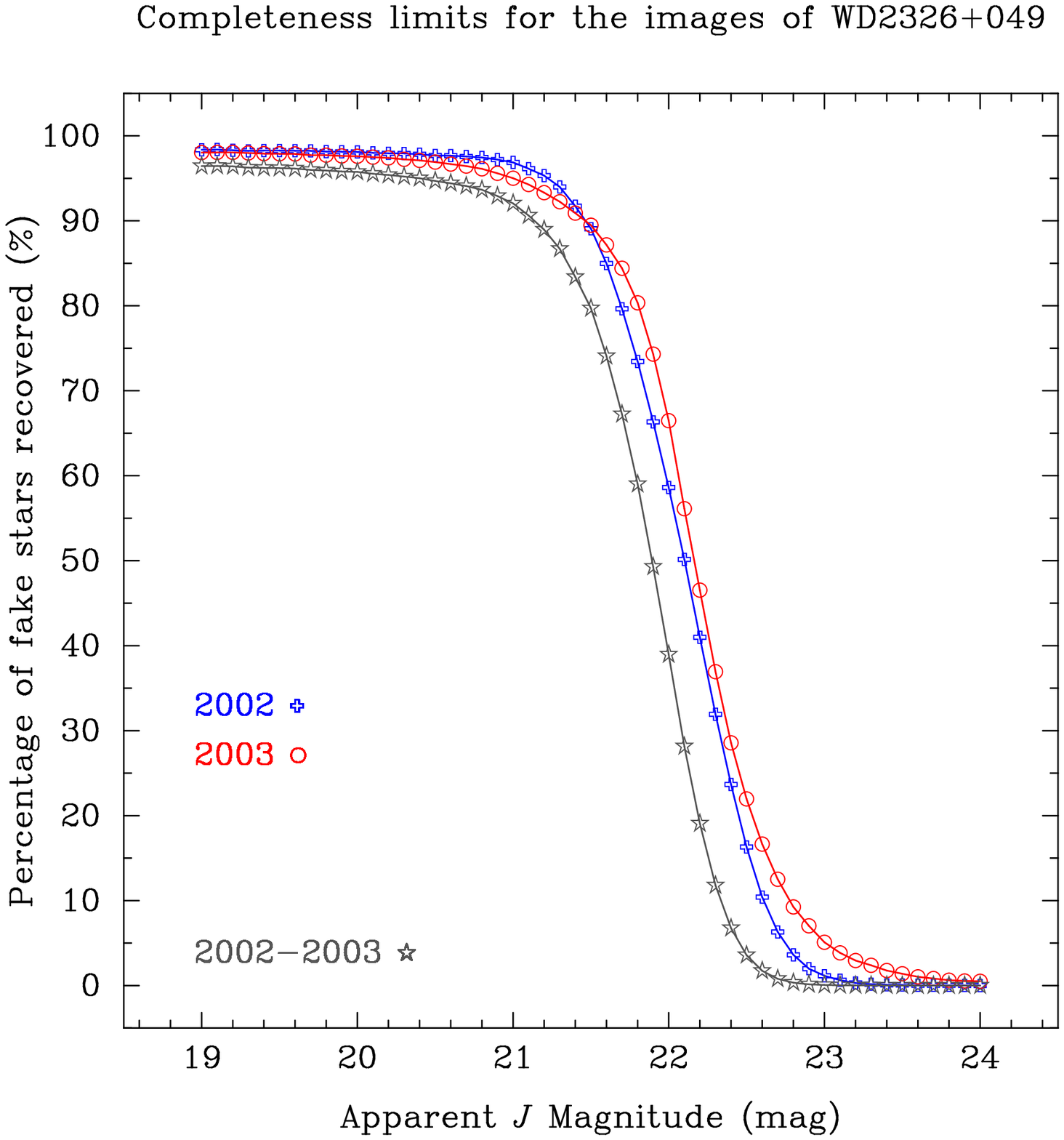}}
\mbox{\includegraphics[bb=45 130 579 616,clip,angle=270,scale=0.370]{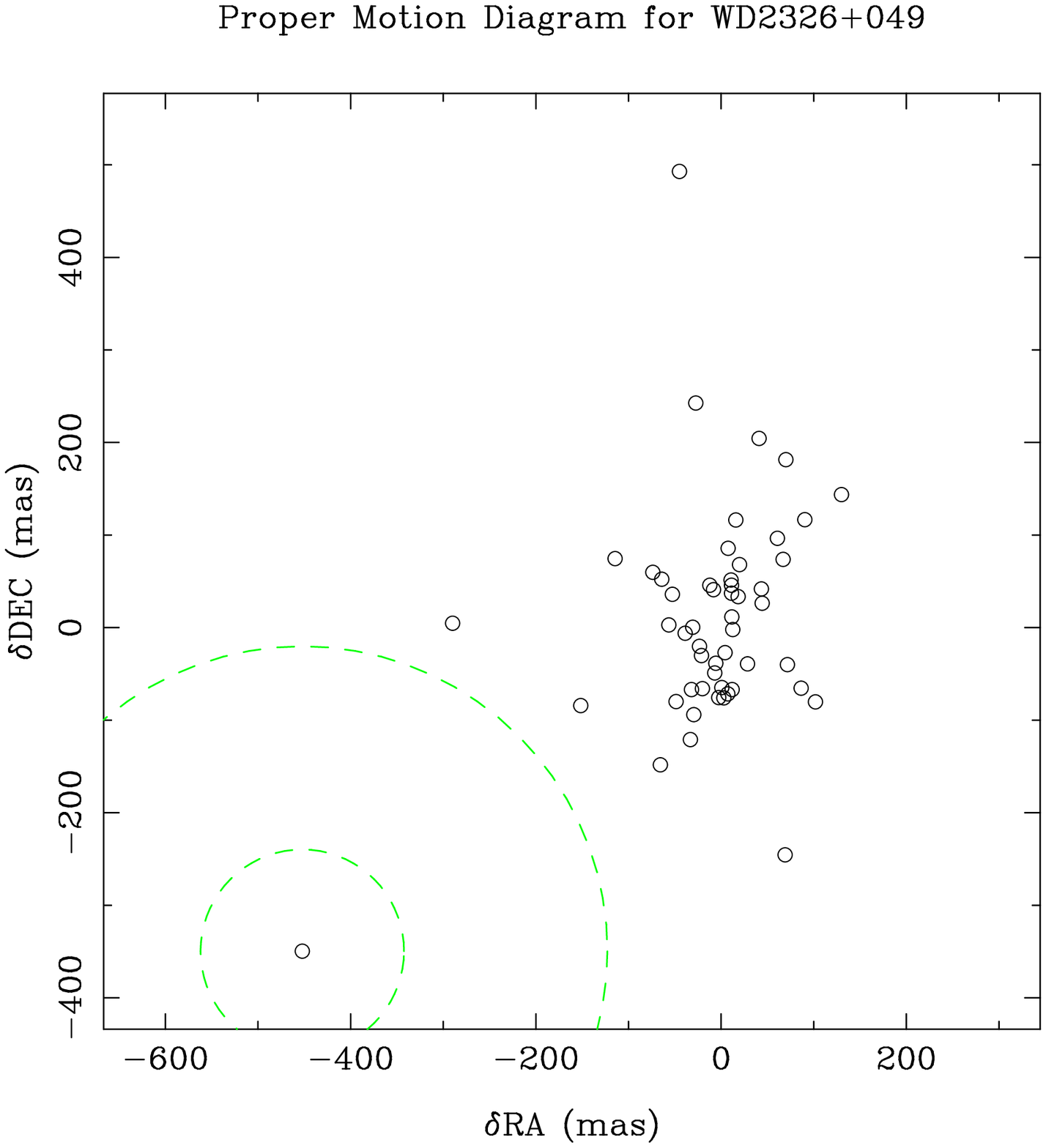}}
\caption{The completeness limit (left) and the proper motion diagram (right) for WD~$2326+049$.}
\label{WD2326_plots}
\end{figure*}
\end{center}
\clearpage
\begin{table*}
\begin{center}
\caption{Results for the 23 equatorial and northern hemisphere white dwarfs in the DODO survey\label{limits}}
\begin{tabular}{cccr@{$^{+}_{-}$}lccr@{$^{+}_{-}$}lcr@{ - }lr@{ - }lcl}
\hline
\hline
\multicolumn{1}{c}{White} & \multicolumn{1}{c}{$t_{\rm{tot}}$} & \multicolumn{1}{c}{$90\%$~$J$} & \multicolumn{2}{c}{$90\%$~$M$} & \multicolumn{1}{c}{$90\%$~$T$} & \multicolumn{1}{c}{$50\%$~$J$} & \multicolumn{2}{c}{$50\%$~$M$} & \multicolumn{1}{c}{$50\%$~$T$} & \multicolumn{2}{c}{WD Orbit} & \multicolumn{2}{c}{MS Orbit} & \multicolumn{1}{c}{$1\sigma$ error} &\\
\multicolumn{1}{c}{Dwarf} & \multicolumn{1}{c}{[Gyr]} & \multicolumn{1}{c}{[mag]} & \multicolumn{2}{c}{[$M_{\rm{Jup}}$]} & \multicolumn{1}{c}{[K]} & \multicolumn{1}{c}{[mag]} & \multicolumn{2}{c}{[$M_{\rm{Jup}}$]} & \multicolumn{1}{c}{[K]} & \multicolumn{2}{c}{[AU]} & \multicolumn{2}{c}{[AU]} & \multicolumn{1}{c}{[mas]} & \\
\hline
0115$\,+\,$159 & 1.7 & 21.0 & $\,\,10$ & $^{2}_{1}$ & 430 & 22.0 & $\,\,8$ & $^{1}_{1}$ & 380 & $\;\;$46 & 675 & $\;$11 & 160 & 94 &\\
0148$\,+\,$467 & 2.5 & 20.4 & $\,\,16$ & $^{5}_{3}$ & 480 & 21.9 & $\,\,10$ & $^{2}_{1}$ & 390 & $\;\;$48 & 457 & $\;$14 & 138 & 66 & 1,2\\
0208$\,+\,$396 & 2.6 & 20.5 & $\,\,16$ & $^{5}_{3}$ & 480 & 22.5 & $\,\,9$ & $^{1}_{1}$ & 360 & $\;\;$50 & 758 & $\;$13 & 194 & 91 &\\
0341$\,+\,$182 & 3.3 & 22.0 & $\,\,13$ & $^{3}_{2}$ & 400 & 22.9 & $\,\,10$ & $^{2}_{1}$ & 360 & $\;\;$57 & 801 & $\;$16 & 222 & 79 &\\
0435$\,-\,$088 & 4.1 & 21.2 & $\,\,13$ & $^{3}_{2}$ & 380 & 22.7 & $\,\,9$ & $^{1}_{2}$ & 320 & $\;\;$28 & 408 & $\;$9 & 124 & 94 &\\
0644$\,+\,$375 & 2.1 & 20.5 & $\,\,13$ & $^{3}_{0}$ & 460 & 22.4 & $\,\,8$ & $^{1}_{1}$ & 360 & $\;\;$46 & 652 & $\;$17 & 236 & 124 & 1\\
0738$\,-\,$172 & 2.4 & \multicolumn{1}{c}{$-$} & \multicolumn{2}{c}{$-$} & \multicolumn{1}{c}{$-$} & 22.0 & $\,\,7$ & $^{1}_{1}$ & 320 & $\;\;$27 & 379 & $\;$7 & 96 & 42 & 3\\
0912$\,+\,$536 & 3.0 & 20.9 & $\,\,13$ & $^{0}_{2}$ & 410 & 22.1 & $\,\,9$ & $^{1}_{2}$ & 350 & $\;\;$31 & 419 & $\;$7 & 93 & 171 & 1\\
1055$\,-\,$072 & 3.3 & 21.0 & $\,\,13$ & $^{3}_{2}$ & 400 & 22.6 & $\,\,9$ & $^{1}_{1}$ & 340 & $\;\;$36 & 503 & $\;$8 & 103 & 80 &\\
1121$\,+\,$216 & 2.3 & 21.2 & $\,\,10$ & $^{2}_{1}$ & 390 & 22.2 & $\,\,8$ & $^{2}_{1}$ & 350 & $\;\;$40 & 605 & $\;$9 & 134 & 117 & 1\\
1134$\,+\,$300 & 0.37 & 20.8 & $\,\,5$ & $^{0}_{1}$ & 440 & 21.9 & $\,\,3$ & $^{1}_{0}$ & 350 & $\;\;$46 & 664 & $\;$9 & 127 & 97 & 1,2\\
1344$\,+\,$106 & 2.5 & 20.8 & $\,\,16$ & $^{5}_{3}$ & 480 & 22.0 & $\,\,13$ & $^{0}_{2}$ & 440 & $\;\;$60 & 865 & $\;$14 & 208 & 114 & 1\\
1609$\,+\,$135 & 2.8 & 21.7 & $\,\,13$ & $^{3}_{2}$ & 420 & 22.5 & $\,\,10$ & $^{2}_{1}$ & 380 & $\;\;$55 & 642 & $\;$10 & 117 & 83 & 1\\
1626$\,+\,$368 & 2.2 & 22.1 & $\,\,9$ & $^{1}_{1}$ & 380 & 22.8 & $\,\,8$ & $^{1}_{1}$ & 360 & $\;\;$48 & 535 & $\;$13 & 141 & 72 & 1\\
1633$\,+\,$433 & 3.0 & 21.1 & $\,\,13$ & $^{3}_{0}$ & 410 & 22.3 & $\,\,10$ & $^{2}_{2}$ & 370 & $\;\;$45 & 533 & $\;$10 & 123 & 97 & 1,2,4\\
1647$\,+\,$591 & 0.91 & 19.6 & $\,\,9$ & $^{1}_{1}$ & 480 & 22.0 & $\,\,5$ & $^{0}_{1}$ & 350 & $\;\;$33 & 372 & $\;$7 & 77 & 127 &\\
1900$\,+\,$705 & 1.1 & 21.2 & $\,\,7$ & $^{1}_{1}$ & 400 & 22.2 & $\,\,5$ & $^{1}_{0}$ & 330 & $\;\;$39 & 452 & $\;$8 & 89 & 146 & 1,2\\
1953$\,-\,$011 & 2.1 & 19.2 & $\,\,16$ & $^{5}_{0}$ & 510 & 21.7 & $\,\,8$ & $^{1}_{1}$ & 360 & $\;\;$34 & 509 & $\;$7 & 111 & 60 & 1\\
2007$\,-\,$219 & 1.4 & 21.2 & $\,\,10$ & $^{0}_{2}$ & 450 & 22.4 & $\,\,7$ & $^{1}_{1}$ & 370 & $\;\;$55 & 831 & $\;$12 & 189 & 74 &\\
2047$\,+\,$372 & 0.89 & \multicolumn{1}{c}{$-$} & \multicolumn{2}{c}{$-$} & \multicolumn{1}{c}{$-$} & 21.8 & $\,\,6$ & $^{1}_{0}$ & 390 & $\;\;$54 & 202 & $\;$12 & 46 & 44 & 1\\
2140$\,+\,$207 & 4.4 & 20.0 & $\,\,21$ & $^{0}_{0}$ & 490 & 21.6 & $\,\,13$ & $^{3}_{0}$ & 370 & $\;\;$38 & 542 & $\;$13 & 181 & 73 & 1\\
2246$\,+\,$223 & 1.7 & 20.6 & $\,\,13$ & $^{3}_{0}$ & 490 & 22.0 & $\,\,9$ & $^{1}_{1}$ & 400 & $\;\;$57 & 835 & $\;$11 & 157 & 92 & 1,2\\
2326$\,+\,$049 & 1.1 & 21.1 & $\,\,7$ & $^{1}_{1}$ & 400 & 21.8 & $\,\,6$ & $^{1}_{1}$ & 370 & $\;\;$41 & 396 & $\;$9 & 89 & 110 & 1,5\\
\hline
\end{tabular}
\begin{tabular}{p{0.95\textwidth}}
Columns: $t_{\rm{tot}}$ is the ``COND'' evolutionary model age used; $90\%$ and $50\%$ gives the $90\%$ and $50\%$ completeness limits in terms of apparent $J$ magnitude, mass, $M$, measured in Jupiter masses, and effective temperature, $T$, measured in Kelvin, respectively; WD Orbit is the range of projected physical separations at which a companion of that mass could be found around the white dwarf, measured in AU; MS Orbit is the range of projected physical separations at which a companion of that mass could be found around the main sequence progenitor, measured in AU; $1\sigma$ error is the $1\sigma$ scatter of the distribution of the motions of all objects in the field, excluding the white dwarf, measured in milli arc seconds. Notes: (1) 60Hz signal present in at least one epoch image, (2) A third epoch image is required, (3) A common proper motion companion is known to orbit this white dwarf, (4) Streak across one epoch image due to a bright source just outside the field of view, (5) A dust disk is known to orbit this white dwarf.
\end{tabular}
\end{center}
\end{table*}

\noindent
epoch image is required to determine if any of these objects are genuine 
common proper motion companions.

\subsection{WD~{\boldmath$2007-219$}} 
\label{WD2007}
\begin{figure*} 
\begin{center}
\mbox{\includegraphics[bb=45 121 576 616,clip,angle=270,scale=0.370]{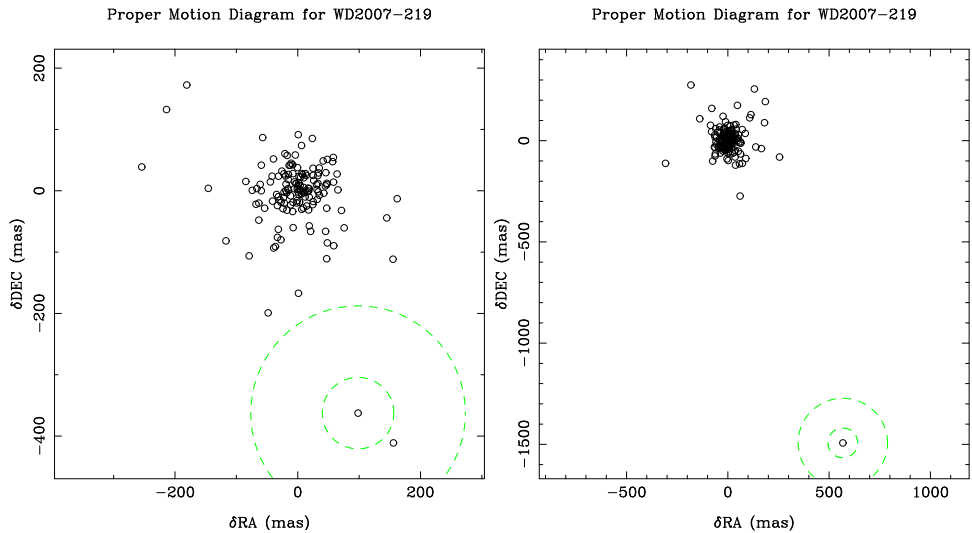}} 
\mbox{\includegraphics[bb=45 130 579 616,clip,angle=270,scale=0.370]{WD2007-219_pm.ps}}
\caption{The proper motion diagram (left) shows the motion of all objects in 
the field of WD~$2007-219$ between the first epoch and second epoch images. 
The dashed green circles represent the $1\sigma$ and $3\sigma$ scatter of the 
distribution of the motions of all objects excluding the white dwarf, centred 
on the white dwarf, to help determine possible common proper motion companions
to the white dwarf. A single object appears to have a motion similar to that 
of WD~$2007-219$. The latest proper motion diagram (right), utilising the 2008 
third epoch image, shows that there are no common proper motion companions to 
WD~$2007-219$.} 
\label{WD2007_old} 
\end{center}
\end{figure*}

A single object with a magnitude of $J\sim22$~mag appeared to have a motion 
similar to that of WD~$2007-219$ between the 2002 first epoch image and 2003 
second epoch images (Figure~\ref{WD2007_old}). This candidate companion 
was detected with a SNR of $\sim9$ and $\sim11$ in the first epoch and second 
epoch images, respectively. The magnitude of the motion of the candidate was 
$\sim440$~mas. In comparison, the rms of the magnitude of the motion of all 
point sources in the field with $21.5<J<22.5$~mag is $\sim130$~mas. The motion 
of the candidate was $>3$ times larger than this mean error and was clearly 
separated from the other faint objects in the field. Therefore, both the 
magnitude and direction of its motion was entirely consistent with that of the 
white dwarf, suggesting that the candidate was a true common proper motion 
companion. The candidate would have a mass of $7\pm1\,M_{\rm{Jup}}$ and a 
projected physical separation of $\sim980$~AU if it were confirmed to be a 
common proper motion companion to WD~$2007-219$. A third epoch image was 
recently acquired to investigate whether this apparently co-moving candidate 
was real. Unfortunately, the new observation showed that this object is in 
fact a non--moving background object. The \textit{NIRI} 2008 third epoch 
image is less noisy than the \textit{NIRI} 2003 second epoch image, which has 
improved the completeness limit and shows the candidate to have some 
structure. Therefore, it is likely to be a galaxy. In summary, these 
observations could have detected a companion with a mass of 
$7\pm1\,M_{\rm{Jup}}$, corresponding to an effective temperature of 
$\sim370$~K \citep{bcb2003}, between a projected physical separation of 
$55-831$~AU with a $50\%$ probability (Table~\ref{limits}). The motion of this 
white dwarf between the first epoch and third epoch images is large enough to 
confidently state that there are no common proper motion companions to 
WD~$2007-219$ within these limits. 

\subsection{WD~{\boldmath$2140+207$}} 
\label{WD2140}
The mass of this DQ white dwarf used throughout this paper ($0.49\,M_{\odot}$) 
was determined by including effects from the carbon present in its atmosphere 
\citep*{dbf2005}, while \citet{dgf2006} use a white dwarf mass of 
$0.62\,M_{\odot}$ \citep*{blr2001}, which was derived using a pure helium 
model. In contrast, a companion with a mass of $9^{+1}_{-2}\,M_{\rm{Jup}}$ 
could have been detected if the larger white dwarf mass was used to determine 
the total age of the white dwarf.

\subsection{WD~{\boldmath$0148+467$}, WD~{\boldmath$1134+300$}, WD~{\boldmath$1900+705$}, WD~{\boldmath$2246+223$}} 
\label{WD0148}
\begin{figure} 
\begin{center} 
\mbox{\includegraphics[angle=270,scale=0.316]{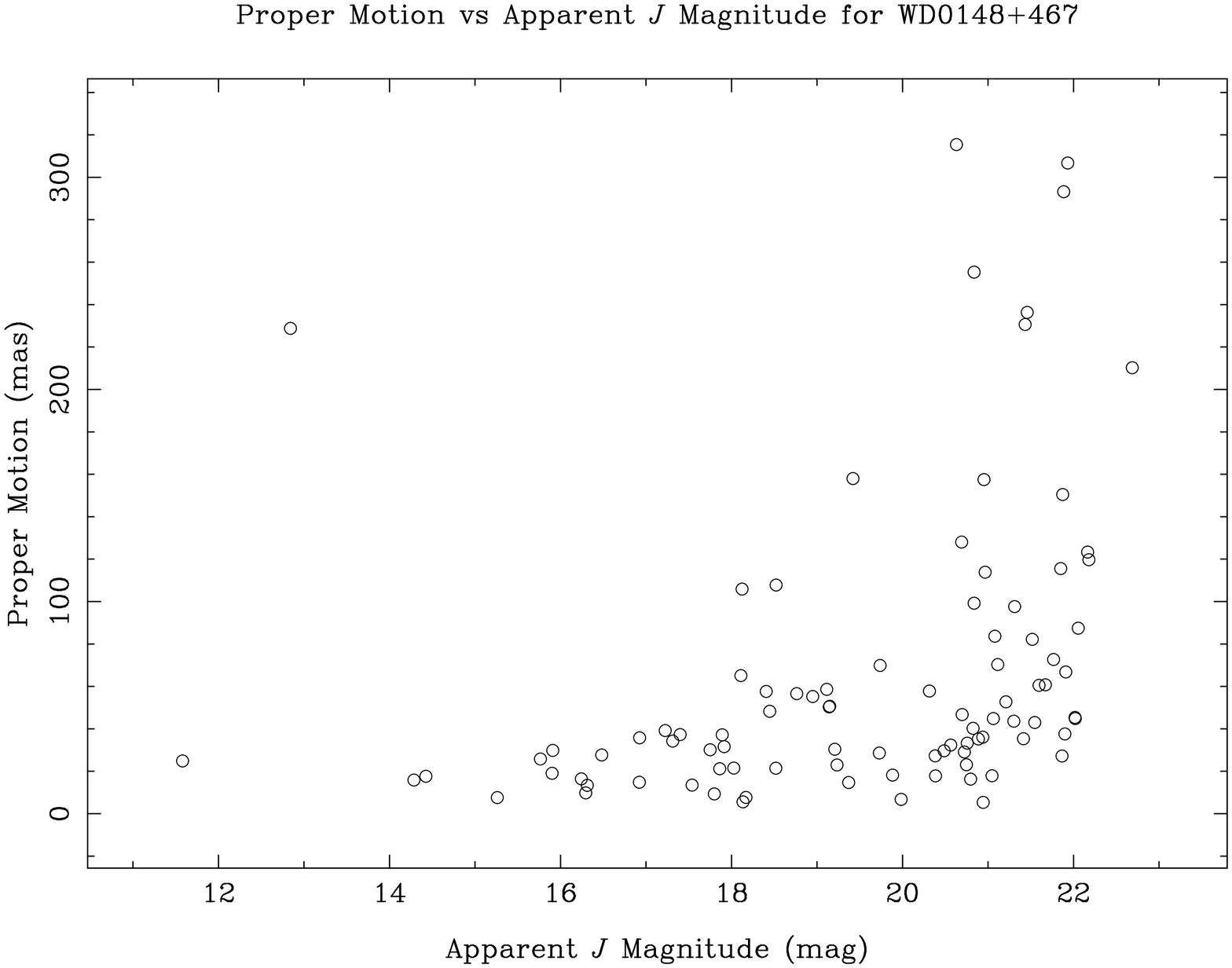}} 
\caption{The motion of the objects between the 2003 first epoch and 2005 second epoch images of WD~$0148+467$. The white dwarf has a magnitude of $J\sim13$~mag.} 
\label{WD0148_mag} 
\end{center} 
\end{figure}

\begin{figure} 
\begin{center} 
\mbox{\includegraphics[angle=270,scale=0.316]{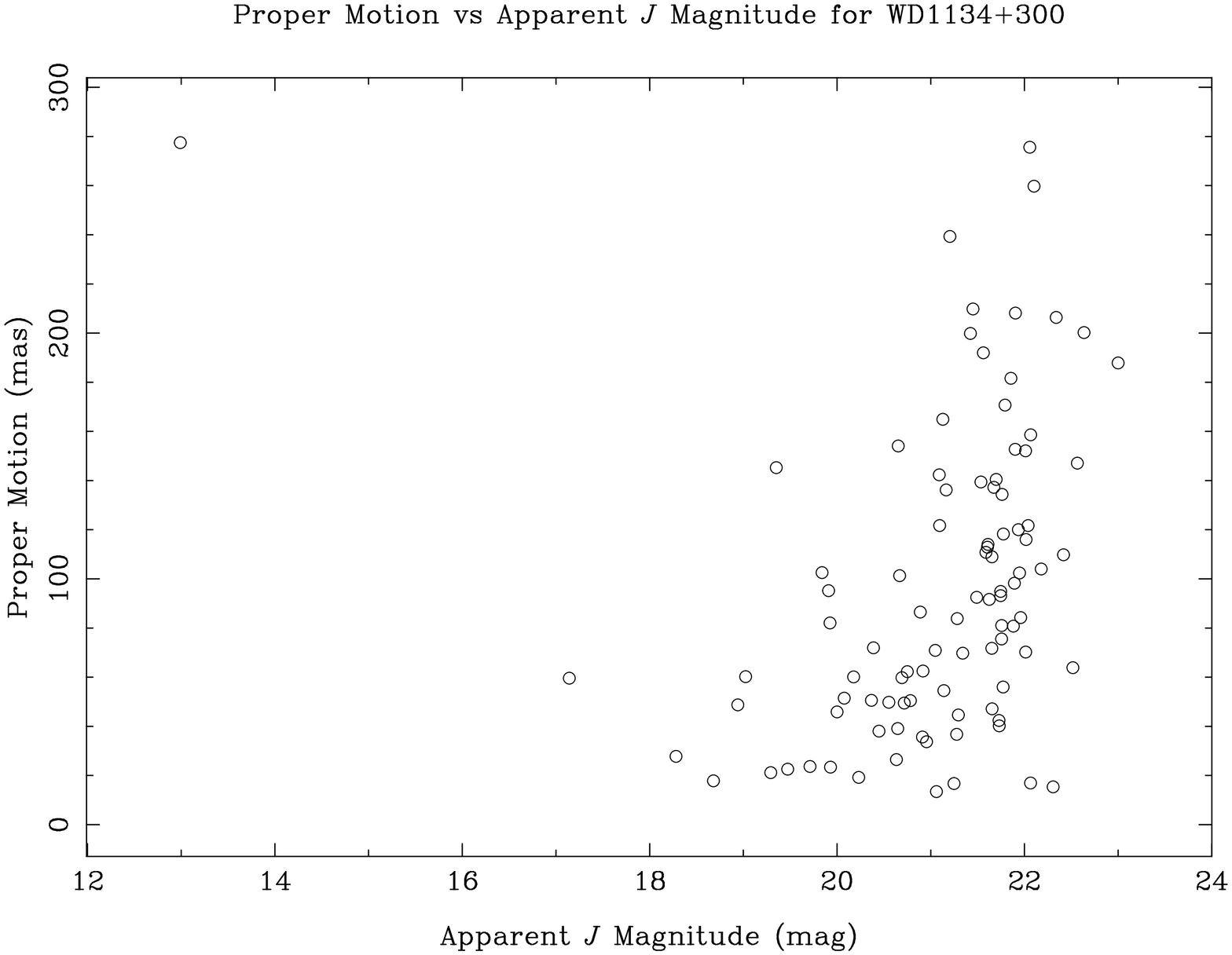}} 
\caption{The motion of the objects between the first epoch and second epoch images of WD~$1134+300$. The white dwarf has a magnitude of $J\sim13$~mag.} 
\label{WD1134_mag} 
\end{center} 
\end{figure}

\begin{center}
\begin{figure}
\mbox{\includegraphics[angle=270,scale=0.316]{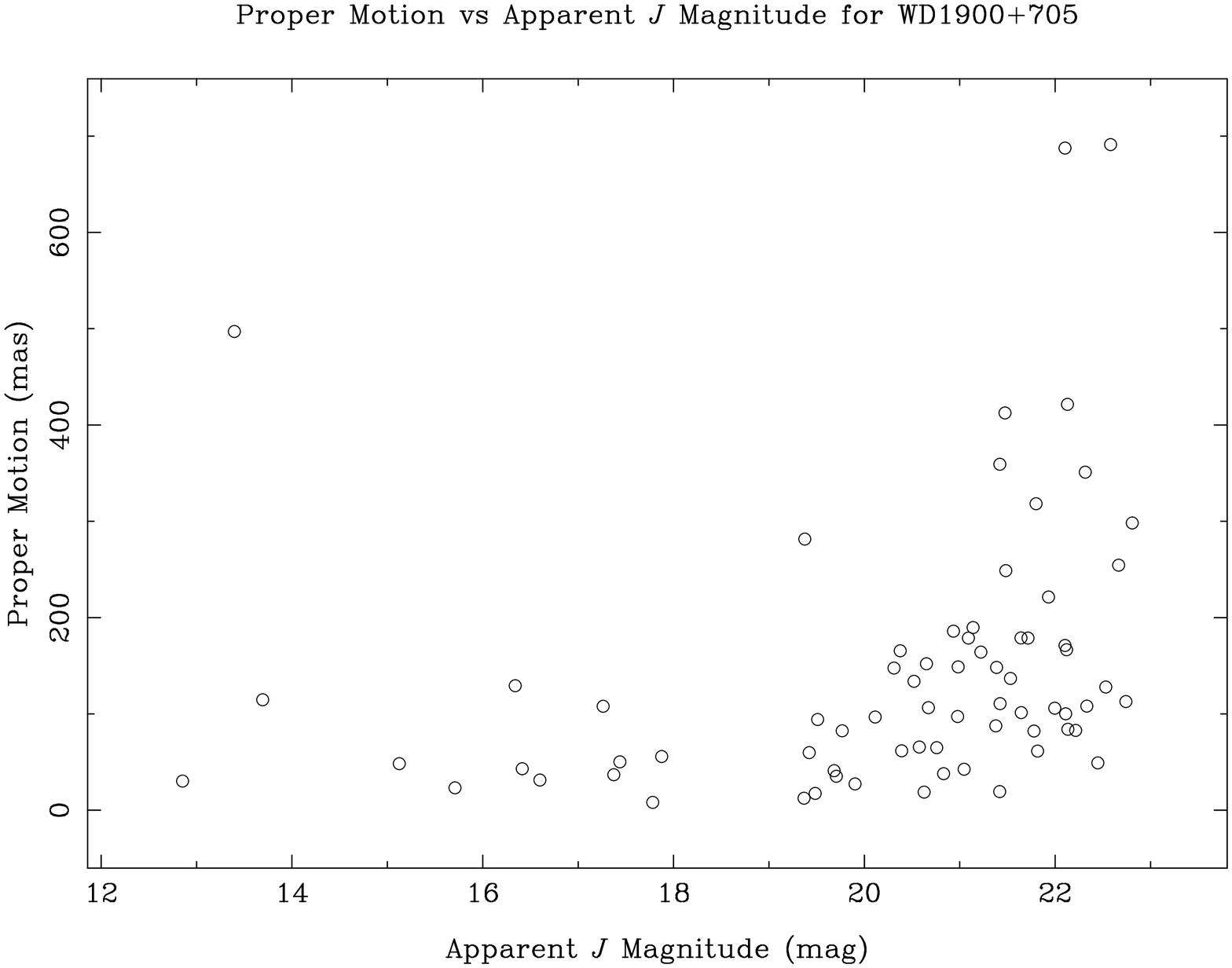}} 
\caption{The motion of the objects between the first epoch and second epoch images of WD~$1900+705$. The white dwarf has a magnitude of $J\sim13.5$~mag.}
\label{WD1900_mag}
\end{figure}
\end{center}

\begin{figure} 
\begin{center} 
\mbox{\includegraphics[angle=270,scale=0.316]{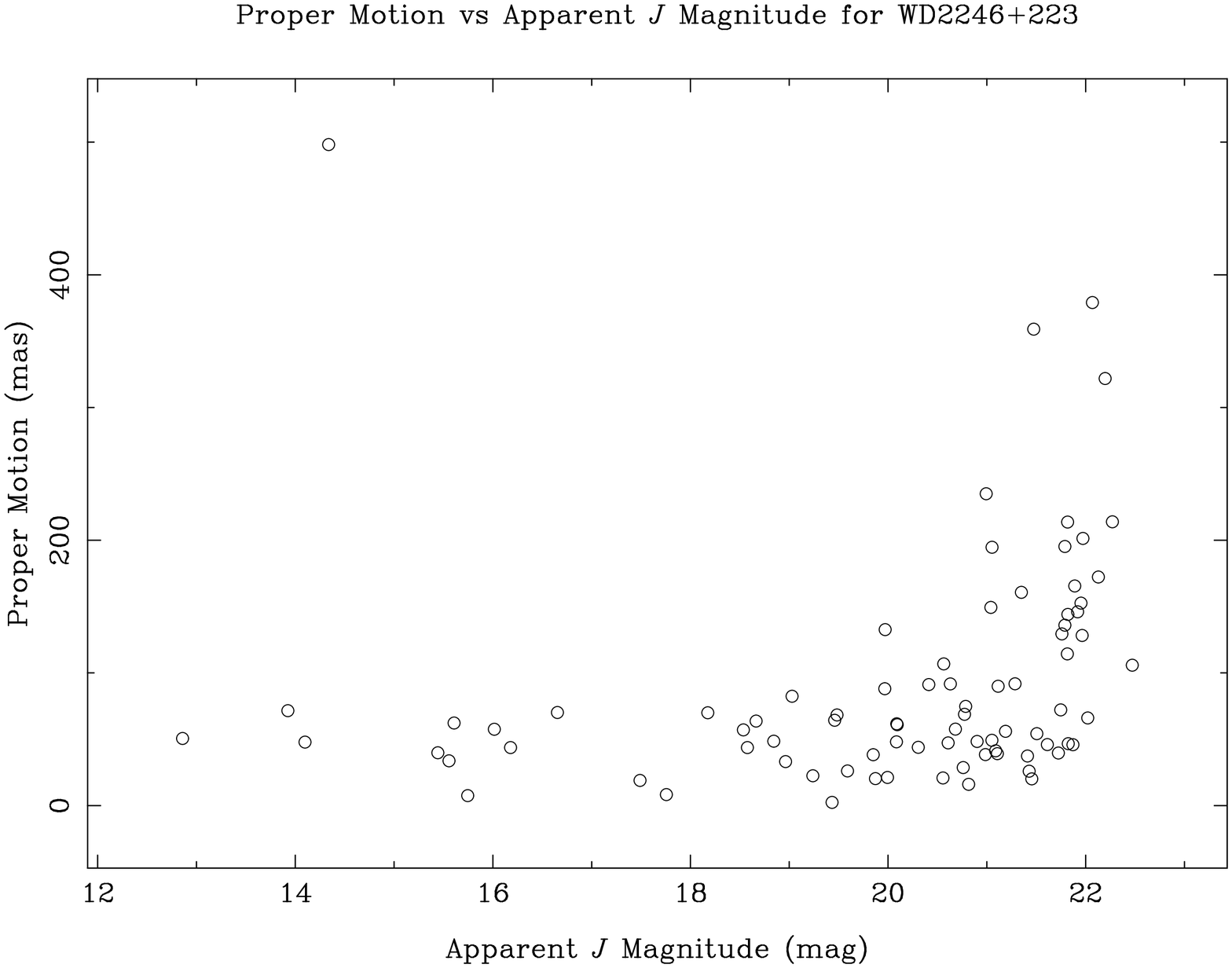}} 
\caption{The motion of the objects between the first epoch and second epoch images of WD~$2246+223$. The white dwarf has a magnitude of $J\sim14.3$~mag.} 
\label{WD2246_mag} 
\end{center} 
\end{figure}

The 2003 first epoch images of WD~$0148+467$, WD~$1134+300$, WD~$1900+705$ and 
WD~$2246+223$ were degraded by severe $60$~Hz interference 
(Table~\ref{observed}), which significantly decreased the completeness limit 
of these images. It is likely that this interference has introduced the large 
scatter in the motions of faint objects between the 2003 first epoch and 2004 
/ 2005 second epoch images (Figures~\ref{WD0148_mag}, \ref{WD1134_mag}, 
\ref{WD1900_mag} and \ref{WD2246_mag}). This suggests that the error on the 
motion of these faint objects is comparable to the motion of the white dwarfs. 
As a result, multiple objects appear to have motions similar to the motion of 
the white dwarfs. In the case of WD~$2246+223$, a single object appears to 
have a motion similar to the motion of WD~$2246+223$. The candidate common 
proper motion companion is detected with a SNR of $\sim5$ in both epoch 
images. If the candidate is confirmed to be a common proper motion companion 
to WD~$2246+223$, it would have a mass of $9\pm1\,M_{\rm{Jup}}$ and a 
projected physical separation of $\sim840$~AU. This corresponds a projected 
physical separation of $\sim160$~AU around the main sequence progenitor, 
assuming an expansion factor of 
$M_{\rm{MS}}/M_{\rm{WD}}\sim5.1\,M_{\odot}/0.97\,M_{\odot}\sim5.3$. However, 
due to the presence of $60$~Hz interference, the measurement of the motion of 
the candidate between the first epoch and second epoch images may be 
inaccurate. In addition, there are two other objects which appear to have the 
same magnitude of motion as the candidate, reducing the probability that the 
candidate is a genuine common proper motion companion. Therefore, a third 
epoch image is required, for all four of these white dwarfs, to determine if 
any of these objects are genuine common proper motion companions.
\begin{table}
\begin{center}
\caption{Previous searches for substellar companions around the 23 equatorial and northern hemisphere white dwarfs in the DODO survey\label{previous}}
\begin{tabular}{ccr@{ - }lr@{ - }lc}
\hline
\hline
\multicolumn{1}{c}{White} & \multicolumn{1}{c}{$M$} & \multicolumn{2}{c}{WD Orbit} & \multicolumn{2}{c}{WD Orbit} & \multicolumn{1}{c}{Ref}\\
\multicolumn{1}{c}{Dwarf} & \multicolumn{1}{c}{[$M_{\rm{Jup}}$]} & \multicolumn{2}{c}{[$^{\prime\prime}$]} & \multicolumn{2}{c}{[AU]} &\\
\hline
0115$\,+\,$159 & - & & & & &\\
0148$\,+\,$467 & - & & & & &\\
0208$\,+\,$396 & 10 & $\;\;$0.9 & 10 & $\;$15 & 167 & 1\\
0341$\,+\,$182 & - & & & & &\\
0435$\,-\,$088 & - & & & & &\\
0644$\,+\,$375 & - & & & & &\\
0738$\,-\,$172 & - & & & & &\\
0912$\,+\,$536 & 12 & $\;\;$1 & 7 & $\;$10 & 72 & 2\\
1055$\,-\,$072 & 14 & $\;\;$1 & 7 & $\;$12 & 85 & 2\\
1121$\,+\,$216 & 11 & $\;\;$1 & 7 & $\;$13 & 94 & 2\\
1134$\,+\,$300 & - & & & & &\\
1344$\,+\,$106 & 14 & \multicolumn{4}{c}{Unresolved} & 3\\
1609$\,+\,$135 & - & & & & &\\
1626$\,+\,$368 & 14 & $\;\;$1 & 7 & $\;$16 & 112 & 2\\
1633$\,+\,$433 & 14 & $\;\;$1 & 7 & $\;$15 & 106 & 2\\
	       & 14 & \multicolumn{4}{c}{Unresolved} & 3\\
1647$\,+\,$591 & - & & & & &\\
1900$\,+\,$705 & - & & & & &\\
1953$\,-\,$011 & 10 & $\;\;$1 & 7 & $\;$11 & 80 & 2\\
2007$\,-\,$219 & - & & & & &\\
2047$\,+\,$372 & - & & & & &\\
2140$\,+\,$207 & 10 & $\;\;$1 & 7 & $\;$13 & 88 & 2\\
2246$\,+\,$223 & 9 & $\;\;$1 & 7 & $\;$19 & 133 & 2\\
2326$\,+\,$049 & 6 & $\;\;$1 & 5 & $\;$14 & 68 & 4\\
\hline
\end{tabular}
\begin{tabular}{p{0.45\textwidth}}
Columns: $M$ is the minimum mass of a companion that could be found around the white dwarf, measured in Jupiter masses; WD Orbit is the range of projected physical separations at which a companion of that mass could be found around the white dwarf, measured in arc seconds and determined in AU using the distance to the white dwarf, measured in parsecs, taken from \citet*{vlh1995} (Table~\ref{parameters}); Two targets were part of a search for companions through the detection of an infrared excess and are denoted by ``Unresolved''; Ref = References: (1) \citet*{dswII2005}, (2) \citet*{dgf2006}, (3) \citet*{fbz2008}, (4) \citet*{dswI2005}.
\end{tabular}
\end{center}
\end{table}

\section{Discussion}
\label{discussion}
No common proper motion companions within the limits given in 
Table~\ref{limits} were discovered around 18 of the 23 equatorial and northern 
hemisphere white dwarfs in the DODO survey. Of these, 11 white dwarfs were 
subject to previous searches for substellar companions (Table~\ref{previous}). 
The DODO survey extends out to a much larger projected physical separation in 
each case and places a new lower, upper limit on the mass of any possible 
companion in orbit around 7 of these white dwarfs (Tables~\ref{limits} and 
\ref{previous}). For the remaining 5 targets, multiple objects in each field 
of view appear to have motions similar to the motion of the white dwarfs.

Two of these stars, WD~$0148+467$ and WD~$1134+300$, have the lowest proper 
motions in the survey, moving only $\sim225$~mas and $\sim286$~mas, 
respectively, between their 2003 first epoch and 2005 second epoch images. 
This motion is only $\sim1.7$ and $\sim2.0$ times larger than the rms of the 
magnitude of the motion of all the point sources in the fields of 
WD~$0148+467$ and WD~$1134+300$, respectively, with $21.5<J<22.5$~mag. In 
addition, their first epoch images were degraded by $60$~Hz interference. Due 
to the combination of the very low proper motions of these white dwarfs and 
the presence of $60$~Hz interference, third epoch images are required to 
clearly distinguish their proper motions from the motions of the background 
objects in the field.

WD~$1633+433$, WD~$1900+705$ and WD~$2246+223$ all move $>450$~mas between 
their 2003 first epoch and 2004 second epoch images. However, their first 
epoch images were degraded by $60$~Hz interference. In addition, the presence 
of a streak in the first epoch image of WD~$1633+433$ has most likely 
decreased the accuracy of the measurement of the motion of the faintest 
objects in the field. Again, third epoch images, without the presence of any 
interference, are required.

Using the 2008 third epoch image of WD~$2007-219$, the necessary minimum 
baseline to reliably distinguish real common proper motion companions from 
non--moving background objects was determined. For the faintest (and by 
implication, the most interesting) sources (SNR $\la10$, corresponding to 
$J\ga22$ for the majority of the images in the DODO survey) the white dwarf 
needs to have moved at least 4~pixels ($\sim470$~mas for the \textit{NIRI} 
data) between the two epoch images to conclusively rule out non--moving 
outliers at this magnitude.

\begin{figure} 
\begin{center} 
\mbox{\includegraphics[bb=80 122 580 624,clip,angle=270,scale=0.471]{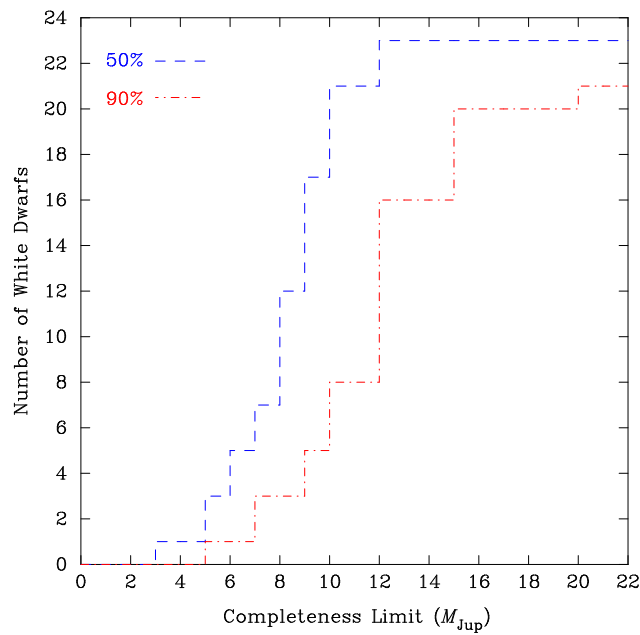}} 
\caption{The cumulative completeness limit, in terms of companion mass, for the 23 equatorial and northern hemisphere white dwarfs in the DODO survey. The red dotted--dashed line indicates the frequency of the completeness limit in $M_{\rm{Jup}}$ at which $90\%$ of companions with that mass could be detected, while the blue dashed line indicates the completeness limit in $M_{\rm{Jup}}$ at which $50\%$ of companions with that mass could be detected.} 
\label{histogram_mass} 
\end{center} 
\end{figure}

\begin{figure} 
\begin{center} 
\mbox{\includegraphics[bb=80 122 575 629,clip,angle=270,scale=0.471]{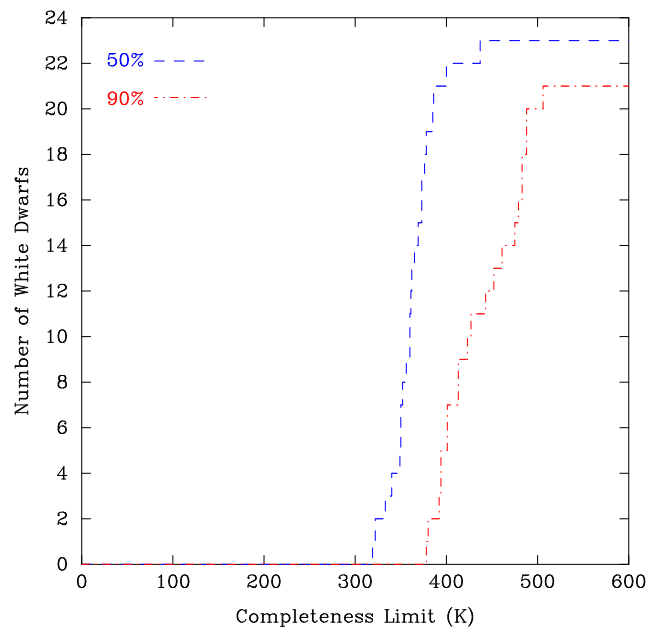}} 
\caption{The cumulative completeness limit, in terms of companion temperature, for the 23 equatorial and northern hemisphere white dwarfs in the DODO survey. The red dotted--dashed line indicates the frequency of the completeness limit in Kelvin at which $90\%$ of companions with that temperature could be detected, while the blue dashed line indicates the completeness limit in Kelvin at which $50\%$ of companions with that temperature could be detected.} 
\label{histogram_temp} 
\end{center} 
\end{figure}

\begin{figure}
\begin{center}
\mbox{\includegraphics[bb=80 122 575 616,clip,angle=270,scale=0.471]{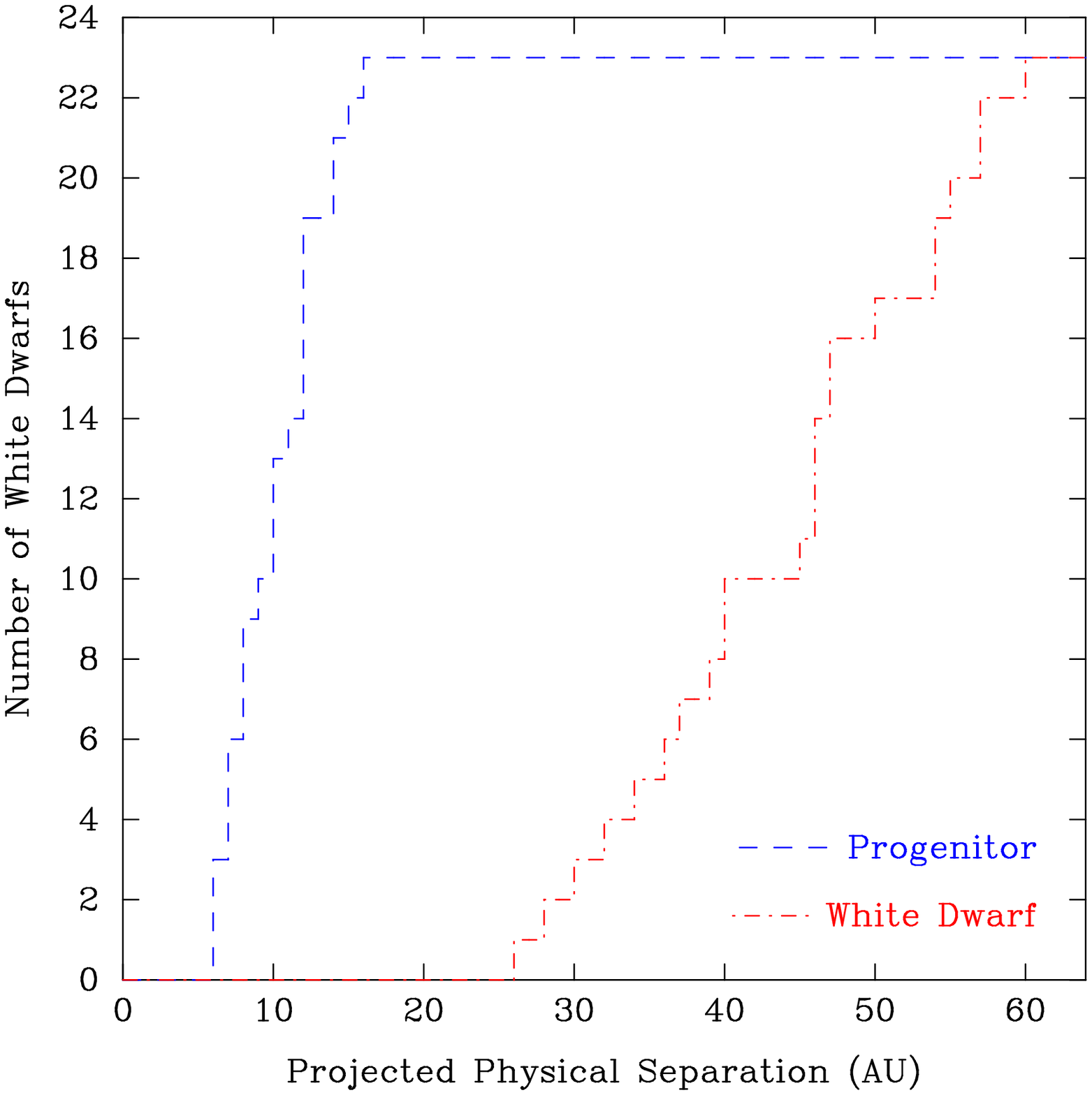}}
\caption{The cumulative minimum projected physical separations for the 23 equatorial and northern hemisphere white dwarfs in the DODO survey. The red dotted--dashed line indicates the minimum projected physical separations in AU at which a companion could be found around each white dwarf. The blue dashed line indicates the minimum projected physical separations in AU at which a companion could be found around the main sequence progenitor.}
\label{histogram_min}
\end{center}
\end{figure}

\begin{figure}
\begin{center}
\mbox{\includegraphics[bb=80 122 575 616,clip,angle=270,scale=0.471]{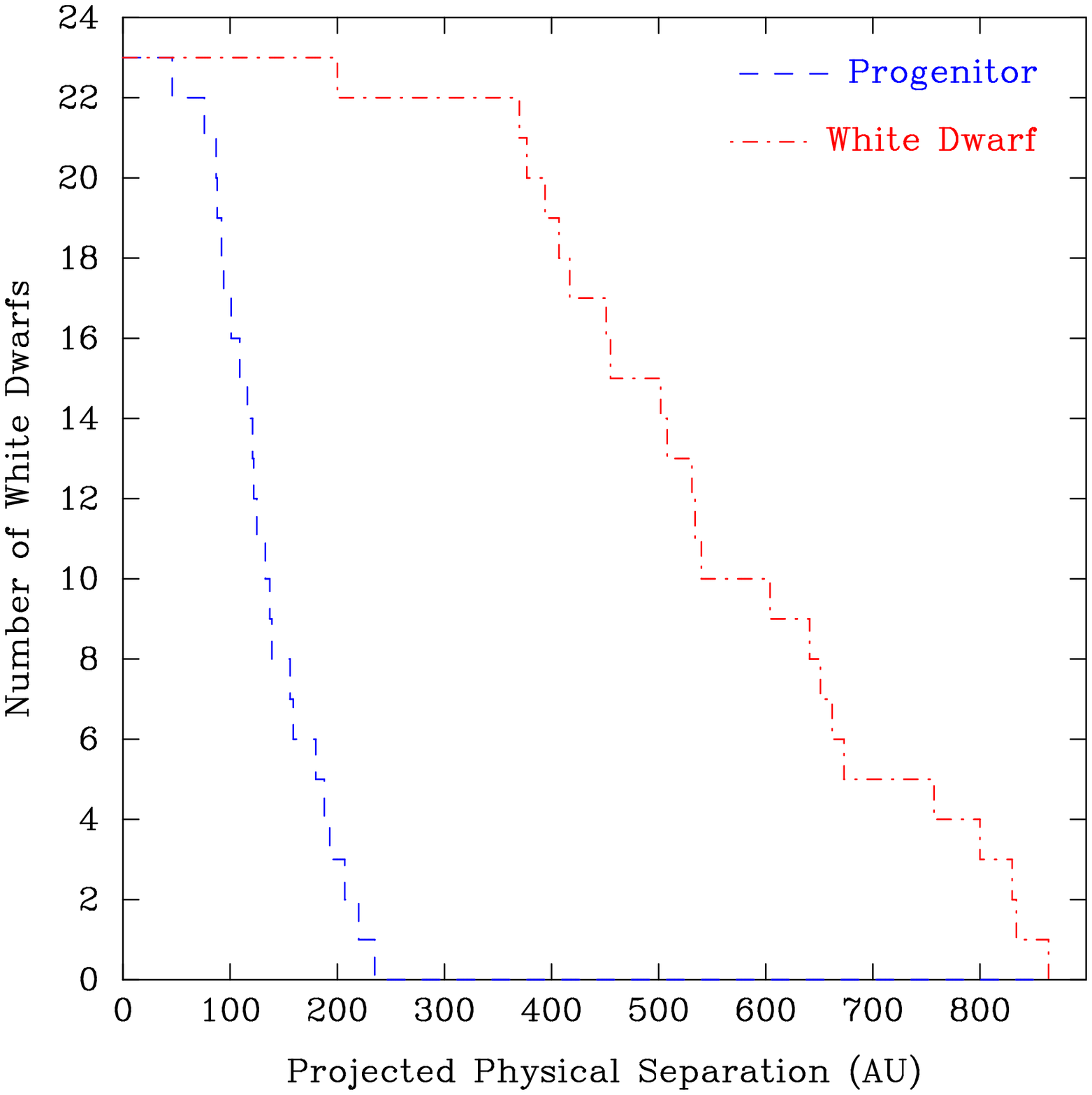}}
\caption{The cumulative maximum projected physical separations for the 23 equatorial and northern hemisphere white dwarfs in the DODO survey. The red dotted--dashed line indicates the maximum projected physical separations in AU at which a companion could be found around each white dwarf. The blue dashed line indicates the maximum projected physical separations in AU at which a companion could be found around the main sequence progenitor.}
\label{histogram_max}
\end{center}
\end{figure}

The cumulative completeness limits, in terms of mass and effective 
temperature, and the corresponding range of projected physical separations 
over which these limits apply have been determined for all 23 equatorial and 
northern hemisphere white dwarfs discussed in this paper (Figures~\ref{histogram_mass},~\ref{histogram_temp},~\ref{histogram_min},~\ref{histogram_max}).

From these results, tentative conclusions regarding the frequency of 
substellar and planetary mass companions to white dwarfs and their main 
sequence progenitors at wide separations can be made (we recognise that the 
DODO survey contains a relatively small number of targets). These conclusions 
assume that no common proper motion companions are confirmed around the 5 
white dwarfs requiring a third epoch image and include the non--detection of a 
companion around WD~$0046+051$ \citep{bch2008}. Firstly, using the $90\%$ 
completeness limits, the DODO survey can detect companions with effective 
temperatures $\ga500$~K around {\em{all}} targets. This is significantly below 
the currently coolest known brown dwarfs, ULAS~J$003402.77-005206.7$ 
\citep{wml2007} and CFBDS~J$005910.90-011401.3$ \citep{dda2008}, which have 
effective temperatures of $600<T_{\rm{eff}}<700$~K and spectral types of T8.5. 
In fact, these observations probe well into the hypothetical Y dwarf regime or 
at least into a significant extension of the T dwarf sequence \citep{k2005}. 
Therefore, we suggest that $\la5\%$ of white dwarfs have L, T and sub--T8.5 
(Y?) substellar companions with effective temperatures $\ga500$~K between 
projected physical separations of $60-200$~AU, although for many fields this 
applies to smaller ($\sim13$~AU for WD~$0046+051$; \citealt{bch2008}) and 
larger ($\sim800$~AU) projected physical separations. This corresponds to 
projected physical separations around their main sequence progenitors 
($1.5-8\,M_{\odot}$, i.e., spectral types F5--B5) of $20-45$~AU, although 
again for many fields these limits apply to smaller ($\sim3$~AU for 
WD~$0046+051$; \citealt{bch2008}) and larger ($\sim200$~AU) projected physical 
separations. For the same range of projected physical separations stated above 
and using the $50\%$ completeness limits, we suggest that $\la8\%$ of white 
dwarfs and their main sequence progenitors have companions with masses above 
the deuterium burning limit ($\sim13\,M_{\rm{Jup}}$), while $\la9\%$ have 
companions with masses $\ga10\,M_{\rm{Jup}}$.

These results can be compared to the results from other imaging surveys for 
wide substellar and planetary mass companions to white dwarfs and main 
sequence stars (Table~\ref{bddesert}). In particular, our results are 
consistent with those of \citet{mz2004} and \citet{ldm2007}. We note the 
recent claims of the directly imaged planetary mass companions to Fomalhaut 
\citep{kgc2008} and HR$8799$ \citep{mmb2008} and await the statistical 
analyses of those surveys for comparison with the DODO survey results. The 
DODO survey results can also be compared to complimentary recent MIR searches 
for {\em{unresolved}} substellar and planetary mass companions to white dwarfs 
(e.g., \citealt{mkr2007}). A recent MIR photometric survey of 27 white dwarfs 
using the Spitzer Space Telescope and IRAC was sensitive to the entire known T 
dwarf sequence \citep{fbz2008}. Their observations place similar limits 
($\la4\%$) on the frequency of such companions to white dwarfs, but at smaller 
separations (with some overlap) compared to the DODO survey. 
\begin{table*}
\begin{center}
\caption{Recent imaging searches for wide companions\label{bddesert}}
\begin{tabular}{llccr@{ - }lc}
\hline
\hline
\multicolumn{1}{l}{Survey} & \multicolumn{1}{l}{Targets}  & \multicolumn{1}{c}{Number} & \multicolumn{1}{c}{Limit} & \multicolumn{2}{c}{Separation} & \multicolumn{1}{c}{Frequency of}\\
& & \multicolumn{1}{c}{of Targets} & \multicolumn{1}{c}{[$M_{\rm{Jup}}$]} & \multicolumn{2}{c}{[AU]} & \multicolumn{1}{c}{Companions}\\
\hline
\citet{mz2004} & G K M & 102 & $>$~12 & 75 & 300 & $1\%\pm1\%$\\
& & 178 & $>$~30 & 140 & 1200 & $0.7\%\pm0.7\%$\\
& & & 5-10 & 75 & 300 & $<3\%$\\
\citet{fbz2005} & White Dwarfs & 261 & $>$~52 & 100 & 5000 & $<0.5\%$\\
& & 86 & $>$~21 & 50 & 1100 & $<0.5\%$\\
\citet{akm2007} & M7-L8 & 132 & $>$~52 & 40 & 1000 & $<2.3\%$\\
\citet{ldm2007} & F G K M & 85 & 13-40 & 25 & 250 & $<5.6\%$\\
\citet{ncb2008} & A F G K M & 60 & $>$~4 & 20 & 100 & $<20\%$\\
\hline
\end{tabular}
\end{center}
\end{table*}

At this stage, we prefer to refrain from speculating on the reasons for the 
negative results so far. For example, from radial velocity measurements of 
evolved giant stars, \citet{lm2007} estimate that at least $3\%$ of stars with 
$M\ga1.8\,M_{\odot}$ host $M_{\rm{p}}\,{\rm{sin}}\,i>5\,M_{\rm{Jup}}$ 
companions, including brown dwarfs. Therefore, it is not unreasonable to have 
expected that at least one of our target white dwarfs would have had a 
detectable companion. It may simply be a case of observing more targets. 

A more substantial comparison with similar surveys, along with a more thorough 
statistical analysis of our results, will be presented in a forthcoming paper 
on the southern hemisphere white dwarfs in the DODO survey.

\section{Acknowledgements}
\label{acknowledgements}
EH acknowledges the support of a PPARC Postgraduate Studentship. MRB 
acknowledges the support of a STFC Advanced Fellowship. Based on observations 
obtained at the Gemini Observatory, which is operated by the Association of 
Universities for Research in Astronomy, Inc., under a cooperative agreement 
with the NSF on behalf of the Gemini partnership: the National Science 
Foundation (United States), the Particle Physics and Astronomy Research 
Council (United Kingdom), the National Research Council (Canada), CONICYT 
(Chile), the Australian Research Council (Australia), CNPq (Brazil) and 
CONICET (Argentina). IRAF is distributed by the National Optical Astronomy 
Observatories, which are operated by the Association of Universities for 
Research in Astronomy, Inc., under cooperative agreement with the National 
Science Foundation. This publication makes use of data products from the Two 
Micron All Sky Survey, which is a joint project of the University of 
Massachusetts and the Infrared Processing and Analysis Center/California 
Institute of Technology, funded by the National Aeronautics and Space 
Administration and the National Science Foundation. This research has made use 
of the SIMBAD database, operated at CDS, Strasbourg, France.

\bibliographystyle{mn2e}
\bibliography{references}

\end{document}